\documentclass[twocolumn]{aastex631} 

\usepackage{amssymb,amsmath,graphicx,natbib,hyperref}
\usepackage{multirow}
\usepackage{subfigure}
\usepackage{float}
\usepackage{footmisc}
\usepackage{color}
\usepackage[flushleft]{threeparttable}
\usepackage{enumitem}
\usepackage{natbib}
\usepackage{tabularx}
\usepackage{booktabs}

\newcommand{\kms}{\,km\,s$^{-1}$}

\newcommand{\um}{\,$\mu$m}

\newcommand{\s}		  {$\sim$}

\newcommand{\tex}		  {$T_\mathrm{ex}$}

\begin{document}

\title{On the Interpretation of Mid-Infrared Absorption Lines of Gas-Phase H$_2$O as Observed by JWST/MIRI}

\author[0000-0003-0665-6505]{Jialu Li}
\affiliation{Department of Astronomy, University of Maryland, College Park, MD 20742, USA}

\author[0000-0001-9344-0096]{Adwin Boogert} 
\affiliation{Institute for Astronomy, University of Hawaii, 2680 Woodlawn Drive, Honolulu, HI 96822, USA}

\author[0000-0003-0306-0028]{Alexander G. G. M. Tielens}
\affiliation{Department of Astronomy, University of Maryland, College Park, MD 20742, USA}
\affiliation{Leiden University, Niels Bohrweg 2, 2333 CA Leiden, The Netherlands}

\correspondingauthor{Jialu Li}
\email{jialu@astro.umd.edu}

\shortauthors{Li et al.}

\shorttitle{Interpreting MIR H$_2$O Absorption Spectra in MIRI/JWST}



\begin{abstract}



Ro-vibrational absorption lines of H$_2$O in the 5--8~$\mu$m wavelength range selectively
probe gas against the mid-infrared continuum emitting background of the inner regions of YSOs
and AGN and deliver important information about these warm, dust-obscured environments. 
JWST/MIRI detects these lines in many lines of sight at a moderate spectral
resolving power of $R\sim3500$ (FWHM of 85 \kms).  Based on our analysis of high-resolution SOFIA/EXES observations, we find that the interpretation of JWST/MIRI absorption spectra can be severely hampered by the blending of individual transitions and the lost information on the intrinsic line width or the partial coverage of the background continuum source. In this paper, we point out {problems such as degeneracy} that arise in deriving physical properties from an insufficiently resolved spectrum. This can lead to differences in the column density by two orders of magnitude. We emphasize the importance of weighting optically thin and weak lines in spectral analyses and provide recipes for breaking down {the coupled parameters}. We also provide an online tool to generate the H$_2$O absorption line spectra that can be compared to observations.

\end{abstract}

\section{Introduction}

Water is one of the most abundant molecules in both the gas and ice phases in the regions associated with the protostars. Water forms on the grains in cold environments and then sublimates in warm conditions, and also forms easily in the gas phase when it's warm and dense. Therefore, water is a powerful diagnostic tool for studying the physical and chemical conditions in the protostellar environment~\citep{vds21}.



Due to the prevalence of water in the Earth's atmosphere, most studies of space gas-phase water were conducted via space or airborne telescopes at infrared or submillimeter wavelengths. Specifically, the rovibrational lines in the mid-infrared (MIR) are often seen in absorption against heavily obscured massive protostars~\citep[e.g.,][]{vh96, dartois98, boonman03, indriolo13, indriolo15, indriolo20, barr22, li23}. Because the disks of massive protostars are heated by accretion, they provide a pencil beam MIR continuum~\citep{barr20}, and water rovibrational lines selectively and directly probe gas in the innermost regions close to the protostars, such as the disk photospheres and winds. In contrast, submillimeter telescopes observe pure rotational emission lines which probe the protostellar environment on a larger scale, but the emission is still diluted in single dish submillimeter telescopes~\citep[e.g.,][]{snell00, wilson03, char10, karska14}. 

\begin{figure*}
    \centering
    \includegraphics[width=\linewidth]{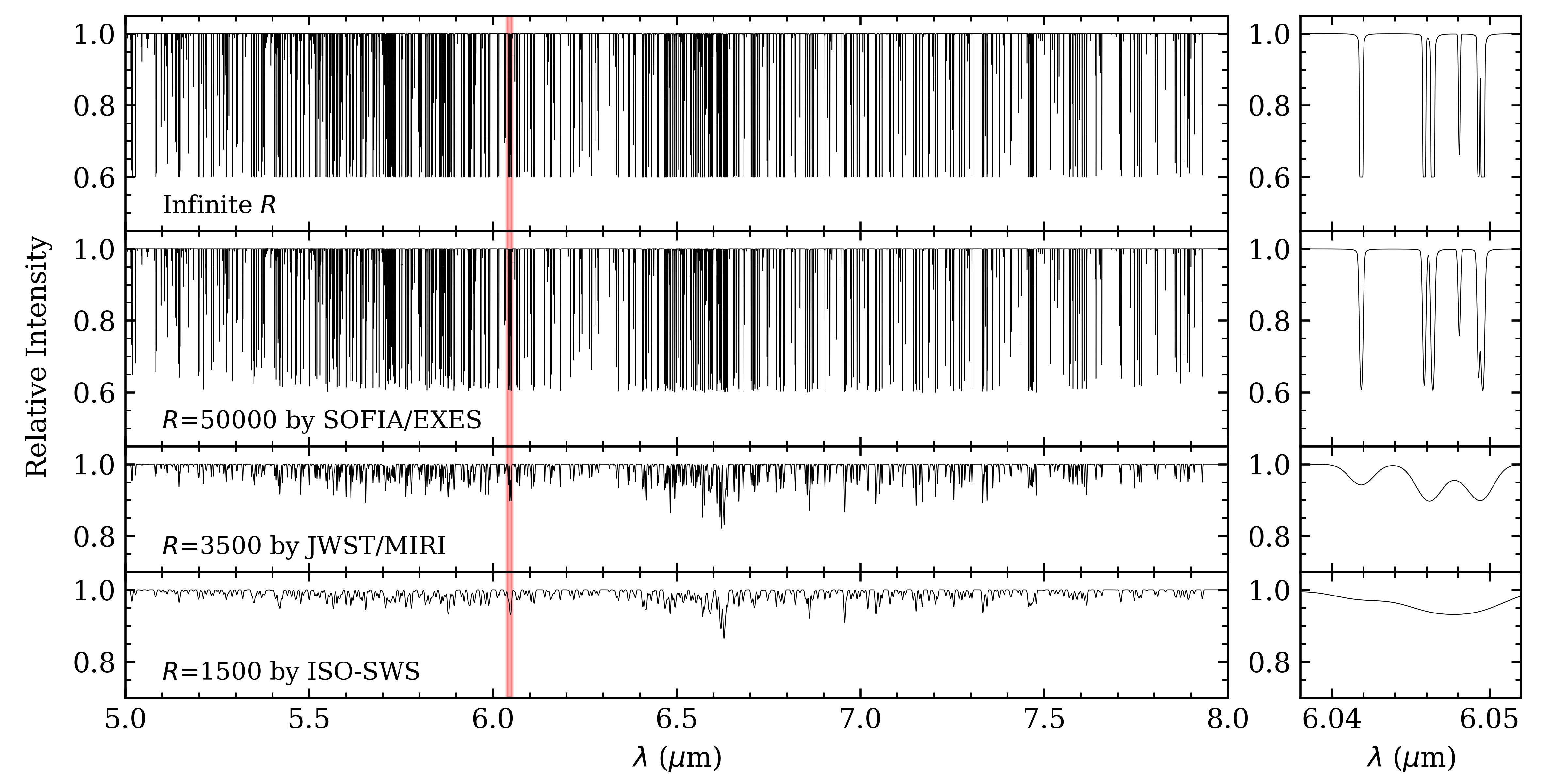}
    \caption{Modeled spectra of the $\nu_2$ vibrational band of H$_2$O of the massive protostar W3~IRS~5 observed by spectrometers onboard SOFIA/EXES ($R\sim$50,000), JWST/MIRI ($R\sim$3,500), and ISO-SWS ($R\sim$1,500). The model assumes that a hot ($T=471$~K) component with a water column density of $\sim2.5\times 10^{19}$~cm$^{-2}$ and a partial coverage of 40\% is in front of W3~IRS~5. The right panels present zoomed-in modeled spectra from the red regions in the left panels. At the limited resolution of MIRI, often, many transitions blend into a broad, unresolved spectral structure.}
    \label{fig:1}
\end{figure*}

JWST/MIRI~\citep[Mid-Infrared Instrument,][]{rieke15, labiano21} can detect the $\nu_2$ rovibrational band of gas-phase H$_2$O from 5--7.5\um\ with a spectral resolving power of $R\sim$3,500 (85\kms). MIRI has detected several tens to hundreds of gaseous water lines in a broad range of systems from merging galaxies~{\citep[e.g., VV~114 in][]{buiten23, ga24}} to hot cores associated with massive protostars~\citep{beuther23}, to brown dwarfs~\citep[e.g., VHS~1256~b in][]{miles23}. Depending on the astrophysical environment, the spectral resolution of MIRI may or may not be sufficient to resolve the lines individually (Figure~\ref{fig:1}). Therefore, the analysis of JWST/MIRI spectra can be hampered by the blending of individual transitions into broad absorption features, unresolved optical depth effects, and line depth reduction due to partial coverage of the source.

This paper aims to provide guidance to recover the physical conditions and abundance of the absorbing water molecules from JWST/MIRI spectra. While we recognize that the observed spectra with absorption lines can contain more complicated features such as emission due to the MIR pumping{~\citep[e.g., ][]{buiten23, gb24}}, this paper deals with pure absorption features. We present in Section~\ref{sec:2} the construction of modeled absorption spectra of gaseous water, and in Section~\ref{sec:3} the {coupled relation} of physical parameters in the crowded water lines based on the modeled spectra. We provide in Section~\ref{sec:recipe} recipes for JWST/MIRI spectra analysis. Certain problems regarding the application of the recipes are discussed in Section~\ref{sec:diss}. 

\section{Absorption Spectrum From a Foreground Slab}\label{sec:2}


The interpretation of an absorption spectrum seems straightforward if one considers the absorbing gas in front of the background MIR continuum with an isothermal dust-free slab model. {Considering one foreground cloud which does not cover the entire source (see Section~\ref{subsec:blend} the case for blended multi-components}), the observed intensity of the observed spectrum $I_\nu$ in frequency space is: 
\begin{equation}
    I_\nu = I_{\textrm{c}} (1 - f_c (1 - \textrm{e}^{-\tau_{\nu}})), \label{eq:ff}
\end{equation}
where $I_c$ is the continuum intensity, {$f_c$ is the fraction of the emission source covered by the absorbing slab ($0\leq f_c \leq1$)}, and $\tau_\nu$ is the optical depth, which is given by~\citep{tielens21}:
\begin{equation}
    \tau_\nu = \tau_p \phi_\nu = \sqrt{\pi} e^2/(m_e c)N_lf_{lu}(\lambda/b) \phi_\nu. \label{eq:tau}
\end{equation}
In the equation above, $e$ is the electron charge, $m_e$ is the electron mass, $c$ is the speed of light, $N_l$ is the column density in the lower state, and $f_{lu}$ is the oscillator strength. The Doppler parameter in velocity space, $b$, is related to the full width at half maximum (FWHM) of an optically thin line by $\Delta v_{\textrm{FWHM}} = 2\sqrt{\textrm{ln}2}b$; or is related to the dispersion in velocity space, $\sigma_v$, by $\sigma_v$=$\sqrt{2}b$. The normalized profile function, $\phi_\nu$, is a Voigt profile $H(a, v)$ that consists of a Doppler core and Lorentzian damp wings and is defined as~\citep[equation 9-34 in][]{mihalas78}:
\begin{equation}
    H(a, v) = \frac{a}{\pi}\int_{-\infty}^{+\infty} \frac{e^{-y^2}dy}{(v-y)^2 + a^2}. \label{eqn:voigt}
\end{equation}
The parameter, $v$, is defined as
\begin{equation}
    v = \frac{\nu - \nu_0}{\Delta \nu_D}
\end{equation}
and represents the shift from the line center in Doppler units. $\Delta \nu_D$ is the Doppler width in frequency space. The parameter $a$ is the damping factor of the Lorentzian profile. For water lines discussed in this paper, $a$ is in orders of 10$^{-8}$, and makes the Lorentzian line width negligible compared to the observed Doppler width~\citep{li23}. 

{In our analysis (Equation~\ref{eq:ff}), we do not account for re-emission, although it can be important in the MIR molecular bands. Observations show pure emission across the entire band, or alternatively, the P-branch in emission and the R-branch in absorption. These patterns are found in the 4.7~\um\ CO and 5--7~\um\ H$_2$O bands towards NGC 3256-S~\citep{ps24}, II Zw96~\citep{gb24}, Orion Peak 1 and Peak 2~\citep{ga02}, and Orion BN/KL~\citep{ga98}. Re-emission is generally expected toward line-of-sights where there is no strong MIR emission source behind the gas, but where the gas is laterally illuminated by a nearby MIR source. Therefore, our analysis applies primarily to sources where these line-of-sights are not present or important, i.e. where the spectrum is dominated by absorption towards an optically thick MIR emission source. Additionally, we assume $f_c$ to be constant. the parameter $f_c$ can vary with wavelength across the band if the beam includes PAH (Polycyclic Aromatic Hydrocarbons) emission, as the continuum absorbed by H$_2$O differs in shape from the observed continuum.}


Assuming the absorbing gas is in LTE, $N_l$ is set by the Boltzmann equation:
\begin{equation}
    \frac{N_l}{g_l} = \frac{N_\textrm{tot}}{Q(T_\textrm{ex})} \textrm{exp}\left(-\frac{\Delta E_l}{k_B T_\textrm{ex}}\right). \label{eq:bd}
\end{equation}
in which \tex\ is the excitation temperature, $N_\textrm{tot}$ is the total column density, {$\Delta E_l$ is the relative energy between upper and lower state. $Q(T_\textrm{ex})$ is the partition function, and $k_B$ is the Boltzmann constant. {We note that one \tex\ value is defined here and is applied to all lines regardless of their optical depth under the LTE assumption. According to the studies of W3~IRS~5 in \citet{li23}, the LTE assumption is valid: transitions from the vibrationally excited state ($\nu_2=1$) are detected and it is concluded that even the vibrational equilibrium is reached. Both the collisions due to the implied high density ($\sim 10^{10}$~cm$^{-3}$) and the strong radiation field contribute to the excitation condition. The collisions between gas and dust lead to gas kinetic temperatures coupled to but slightly lower than the dust temperature. At the same time, the radiative field drives the gas to the radiative temperature. }

The overall spectral profile, according to equations~\ref{eq:ff} and \ref{eq:tau}, is controlled by the total column density, the temperature, the intrinsic line width, and the covering factor of the absorbing gas. Specifically, the intensities of optically thick lines have flat bottoms due to the saturation in the cores, and the bottom level intensity of a normalized spectrum, $I_\nu/I_c$, equals $(1-f_c)$. 

The finite resolving power of a spectrometer, $R$, further regulates the spectral profiles via convolving an instrumental profile kernel (usually well represented by a Gaussian\footnote{{This assumption applies to instruments such as grating/echelle spectrometers. So the Gaussian profile is suitable for SOFIA/EXES and JWST/MIRI. However, there are notable exceptions such as Fourier Transform Spectrometers, which result in Sinc functions, like ISO-SWS or \textit{Herschel}/SPIRE.}}) with an FWHM of $c/R$ or a Doppler width of $c/(2\sqrt{\textrm{ln}2}R)$ in velocity space. For SOFIA/EXES~\citep[the Echelon Cross Echelle Spectrograph on board of the Stratospheric Observatory for Infrared Astronomy,][]{young12, richter18}, JWST/MIRI, and ISO-SWS~\citep[the Infrared Space Observatory, the Short Wavelength Spectrometer,][]{kessler96, richter18}, the corresponding Doppler widths are 3.6\kms~($R\sim50, 000$), 51.5\kms~($R\sim3, 500$), and 120\kms~($R\sim1,500$). As a comparison, the usual intrinsic Doppler widths of components found in massive protostars~\citep[e.g.,][]{barr22, li23} are a few \kms. SOFIA/EXES can barely resolve individual transitions, but JWST/MIRI and ISO/SWS will entirely reshape the profile of every single transition to the instrumental profile, even for the most optically thick lines, and blend many individual transitions into one feature (Figure~\ref{fig:1}). The information on the intrinsic line width $b$, as well as the line saturation level, or the covering factor $f_c$, are therefore lost after the kernel smoothing.

\subsection{An Online Absorption Spectrum Generator}

We provide an online tool to generate the MIR rovibrational absorption spectrum of water (website: \href{https://mirasg.astro.umd.edu/}{https://mirasg.astro.umd.edu/}) based on the slab model. The user can input a combination of parameters including the spectral resolving power $R$, the excitation temperature \tex, the total column density $N_\textrm{tot}$, the fractional coverage $f_c$, the line width $b$, and the relative velocity shift of the astronomical object to the Earth $\Delta v$. The generator will return both an image of the model spectrum and a \texttt{.txt} file of the spectrum. This online generator is ideal for understanding how different parameters influence the spectral profile and also helps to understand the degeneracy problem among parameters that we will illustrate in detail in \S~\ref{sec:3}. The users are encouraged to use the modeled spectrum to assist the analysis in a qualitative way (see \S~\ref{subsec:feature}), or to fit with the observed spectrum cautiously (see \S~\ref{subsec:de-obs}). All modeled spectra in this paper are generated via this tool. 

\section{The Degeneracy Problem Under Insufficient Spectral Resolution}\label{sec:3}

\begin{figure*}[!t]
    \centering
    \includegraphics[width=\linewidth]{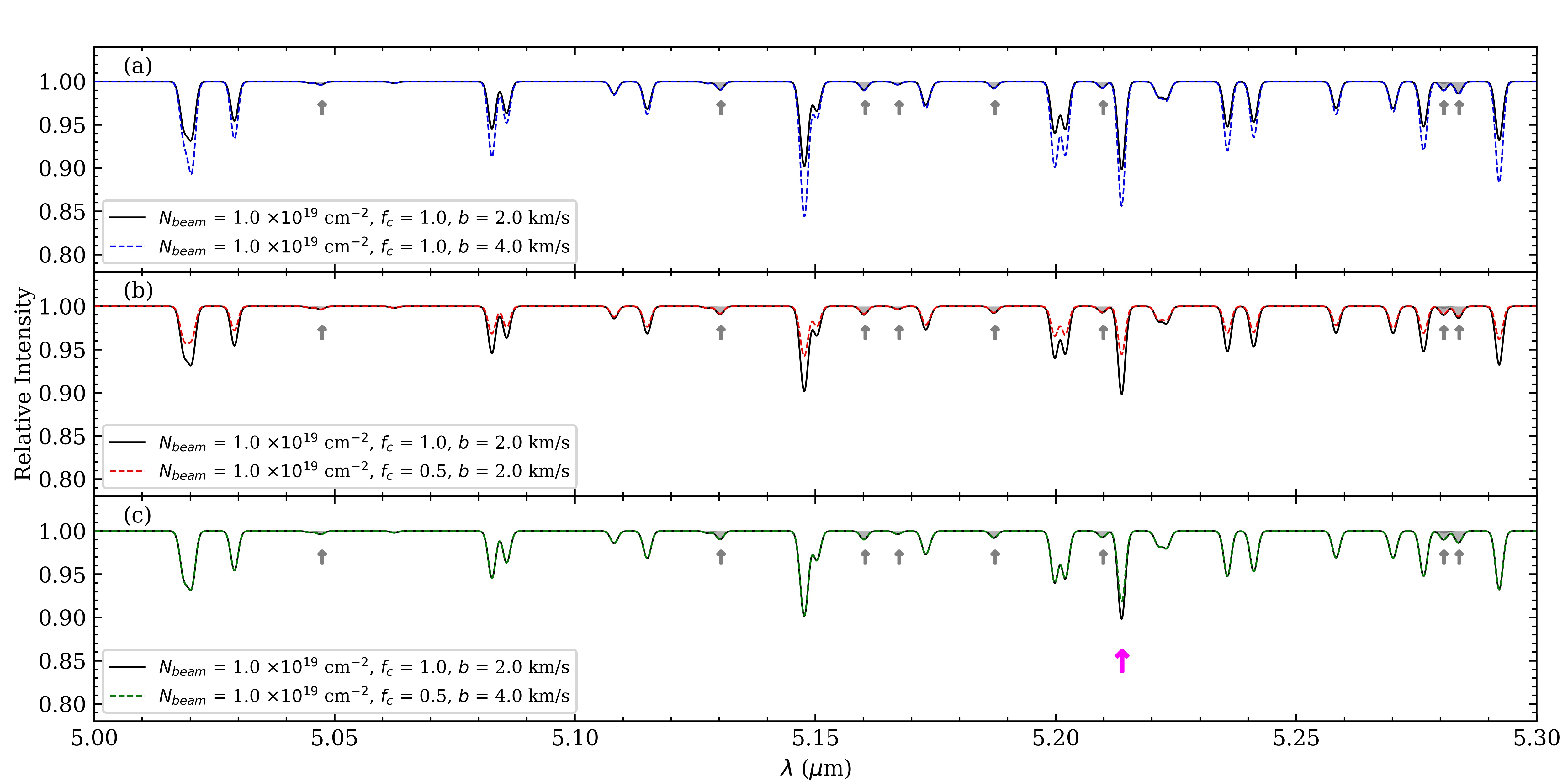}
    \caption{Modeled spectra of water rovibrational lines under different $N_\textrm{tot}$, $f_c$, and $b$ at $R=3,500$. All models have the same $N_\textrm{beam}$ ($=f_cN_\textrm{tot}$), and have the same line depths in particular in optically thin transitions which are colored in grey shades {and are indicated by grey arrows}. Specifically, panel (a) shows that optically thin lines overlap regardless of the line width, $b$. Panel (b) shows that $f_c$ ``dilutes" the spectral line depth. Panel (c) shows that under the same $N_\textrm{beam}$, if the product of $b$ and $f_c$, $bf_c$, is also a constant, line profiles are indistinguishable, except for regions where optically thick lines overlap (e.g., at $\sim$5.214~\um, indicated by a magenta arrow). In such an overlapping region, a spectrum with a larger line width would have a smaller absorption intensity. Moreover, both panels (a) and (b) imply that larger $b f_c$ results in deeper optically thick lines.}
    \label{fig:comp-dege}
\end{figure*}

The parameters that determine the observed spectrum are $T_\textrm{ex}$, $N_\textrm{tot}$, $f_c$, and $b$. In studies using high-resolution absorption spectra, in which each transition is spectrally resolved, the curve of growth is used to analyze simultaneously observed, individually identified transitions without prior assumptions about the $f_c$ and $b$-values~\citep[see][]{barr22, li22}. The curve of growth analysis derives the physical conditions of the absorbing gas properly, especially when some of the transitions are optically thick. While $f_c$ and $b$ are directly measurable under sufficient spectral resolution, the equivalent widths of each transition, which is defined as
\begin{equation}
  \begin{aligned}
   W_\nu &= \int (1 - I_\nu/I_c) d\nu \\ &= \int (1 - \textrm{e}^{-\tau(\nu)}) d\nu \\ &= \int  (1 - \textrm{e}^{-\tau_p \phi_\nu)}) d\nu, 
 \end{aligned}
\end{equation}
in frequency space, are measured and compared to those derived from the modeled curve of growth to determine the parameters $T_\textrm{ex}$, $N_\textrm{tot}$.

For spectra with limited spectral resolution, kernel smoothing reshapes the line profiles but keeps the equivalent widths of each individual transition unchanged. However, even if the equivalent widths can be measured from lines that are not blended with others, the curve of growth analysis cannot be used to derive the physical conditions {because $N_\textrm{tot}$, $f_c$, and $b$ are coupled}. As we will illustrate below, there are different coupling relations among $N_\textrm{tot}$, $f_c$, and $b$ under optically thin (\S~\ref{subsec:decouplebeam}) and optically thick (\S~\ref{subsec:bf}) conditions. Other than the intrinsic degeneracy, we also emphasize {the coupled nature} between $N_\textrm{tot}$ and $f_c$, which originates from the observational detection limit (\S~\ref{subsec:de-obs}). {We summarize how the coupled information influences the interpretation of MIR spectra under insufficient spectral resolving power in \S~\ref{subsec:2problems}.}

\subsection{Optically Thin Lines: Constant $N_{beam}$}\label{subsec:decouplebeam}

For an optically thin line, each absorbing molecule is able to remove photons from the radiation field, and the line intensity is proportional to the number of absorbers. An optically thin line is therefore on the \textit{linear} part of the curve of growth, and the equivalent width $W_\nu$ scales as follows,
\begin{equation}
    W_\nu \propto f_c \tau_p b \propto f_c N_\textrm{tot}. \label{equ:nbeam}
\end{equation}
If we define 
\begin{equation}
    N_\textrm{beam} = f_c N_\textrm{tot},
\end{equation}
equation~\ref{equ:nbeam} implies that the equivalent width of an optically thin line is proportional to $N_\textrm{beam}$, the column density within the observation beam, regardless of the value of $b$. In other words, $N_\textrm{beam}$ can be determined through the optically thin lines. Figure~\ref{fig:comp-dege} illustrates this point further, where modeled spectra are constructed with the same $N_\textrm{beam}$ but different $f_c$ and $b$. Even though these spectra have different $N_\textrm{tot}$ and $b$, the equivalent widths are the same when $N_\textrm{beam}$ is constant. 

\begin{figure*}[!t]
    \centering
    \includegraphics[width=\linewidth]{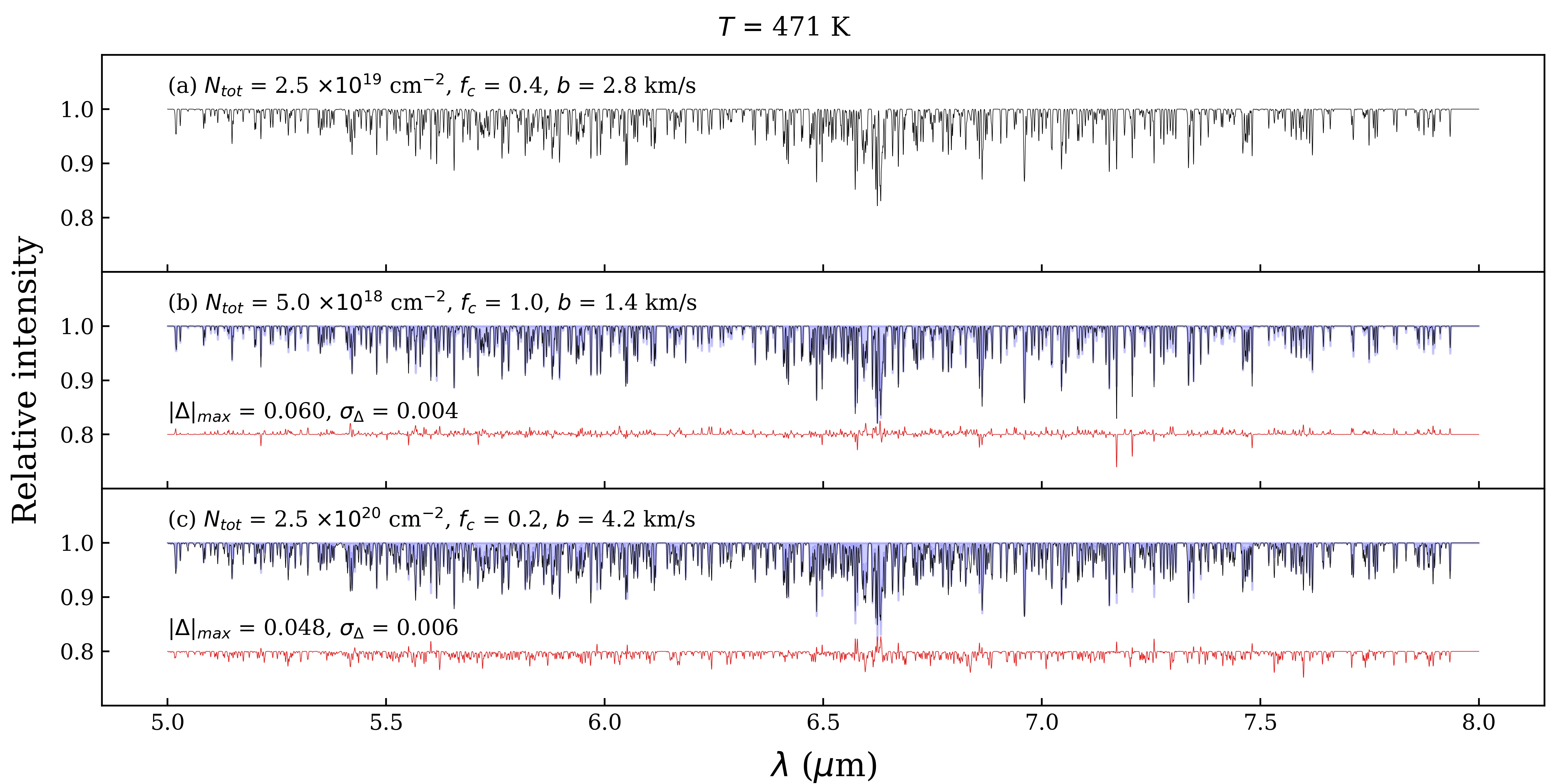}
    \caption{Modeled water rovibrational spectra in three sets of $N_\textrm{tot}, f_c$, and $b$ at a spectral resolving power 3,500. The physical conditions adopted in the model (a) follow the parameters derived from W3~IRS~5 used in Figure~\ref{fig:1}. In panels b and c, the model in panel (a) is plotted in blue, and their differences with model (a) are plotted in red. The maximum difference and the corresponding standard deviation between the spectrum in a and b or c are labeled in the lower left of each panel. We note that the total column density in b and c differ by a factor of 50.}
    \label{fig:three-models}
\end{figure*}

\subsection{Optically Thin \& Thick Lines: Degenerate $b$ and $f_c$ Under Constant $N_{beam}$}\label{subsec:bf}

For lines with the same $N_\textrm{tot}$, the Doppler parameter, $b$, redistributes the column density in frequency space. While $\int \tau_\nu d\nu$ is a constant under different $b$, the peak optical depth varies according with $b$ (equation~\ref{eq:tau}): $\tau_p\propto N_l/b$. The dependence of the equivalent width $\int (1-e^{-\tau_\nu}) d\nu$ on $b$ can be determined by the line opacity. This is illustrated in Figure~\ref{fig:comp-dege}a: for optically thin lines, $(1-e^{-\tau_\nu/b})$ is approximated by $\tau_\nu/b$, and the equivalent width, therefore, remains constant. For optically thick lines, in contrast, the equivalent width increases as $b$ increases, and this is reflected as an increment in the line intensity after the kernel smoothing.

A fractional coverage $f_c$ smaller than one ``dilutes" the line intensity (Figure~\ref{fig:comp-dege}b) and has an inverse effect on the equivalent width compared to an increasing $b$. Specifically, the effect cancels out when the product of $f_c$ and $b$ is a constant for the same $N_\textrm{beam}$. The curves of growth corresponding to the two sets of $f_c$ and $b$ will overlap; and as is illustrated in Figure~\ref{fig:comp-dege}c, the profiles of individual lines are indistinguishable. 

The degeneracy between $b$ and $f_c$ under the same $N_\textrm{beam}$ implies that when MIRI's spectral resolution is insufficient to spectrally resolve the line, one can only get the product of $N_\textrm{tot}f_c$ (=$N_\textrm{beam}$) and $bf_c$ from isolated individually identified lines. 
One cannot thereby derive a unique combination of $N_\textrm{tot}$, $f_c$, and $b$ from the curve of growth analysis. 

Parameters $f_c$ and $b$ can be decoupled under specific conditions. As Figure~\ref{fig:comp-dege}c shows, profiles of spectra with the same product of $f_c$ and $b$ do differ in certain regions (marked by the magenta arrow). This is because absorption intensities in such regions are contributed by two lines: $\int[1-e^{-(\tau_{1, \nu} + \tau_{2, \nu})}]d\nu$, where $\tau_{1, \nu}$ and $\tau_{2, \nu}$ represent optical depths of different lines at the same frequency. Therefore, counter-intuitively, regions with overlapping lines, rather than isolated lines can be used to decouple $f_c$ and $b$. More specifically, lines with larger $b$ will approach saturation faster than lines with smaller $b$ and have smaller equivalent widths. Therefore, to break down the $f_c$ and $b$ degeneracy is to look at line overlapping regions, and smaller line depth corresponds to larger $b$ (Figure~\ref{fig:comp-dege}c).

\subsection{Optically Thick Lines: Coupled $N_\textrm{tot}$ and $f_c$ Under Unknown $N_{beam}$}\label{subsec:de-obs}

The degenerated $N_\textrm{tot}f_c$ and $bf_c$ presented in \S~\ref{subsec:decouplebeam} and \S~\ref{subsec:bf} originate from ``intrinsic" coupling relations. However, in the actual analysis of JWST/MIRI spectra, the derived $N_\textrm{tot}$ and $f_c$ may also result in a large parameter space due to the limited signal-to-noise of the observations. Specifically, commonly, a least square fitting technique is used to derive the physical parameters involved by comparing the observed full spectrum to calculated synthetic spectra. As is shown in Figure~\ref{fig:three-models}, model spectra were constructed and compared in three sets of combinations of $N_\textrm{tot}$, $f_c$, and $b$. Although values of $N_\textrm{tot}$ differ by up to a factor of 50, the difference between the three spectra is small. 

Such a large parameter space of the physical conditions is derived because the strongest, or equivalently, optically thick lines dominate the fitting procedure. While the set of optically thick lines is concentrated on the logarithmic part of the curve of growth, their equivalent widths will fit well to different curves of growth that shift horizontally, or equivalently, with different combinations of $N_\textrm{tot}$, $f_c$, and $b$ {(see details in Appendix~\ref{app:0})}. Really, only the optically thin lines will carry information on the column density of absorbers in the beam but those are not weighted much because of the limited S/N that particularly affects weak lines in a least square fitting procedure. It is unlike the degeneracy problem discussed in \S~\ref{subsec:bf}: while the degeneracy residing in optically thick lines is under the condition where the considered $N_\textrm{beam}$ is a constant, in the reduced $\chi^2$ fitting described above, different models may have different $N_\textrm{beam}$. 

This problem has been realized in studies using ISO-SWS ($R\sim1500$) and SOFIA/EXES ($R\sim50,000$) toward the same target, a binary massive protostar, W3~IRS~5. While both studies derived consistent temperature in $\sim$400--500~K -- which is mainly set by the range of energy levels over which absorption is prominent -- the derived column densities differ by two orders of magnitude: ISO-SWS observations derived $3\times10^{17}$~cm$^{-2}$~\citep{boonman03}, two orders of magnitude less than SOFIA/EXES observations~\citep[$\sim3\times10^{19}$~cm$^{-2}$, the H1 component; see Table 5 in][]{li23}. ISO-SWS observations could not constrain the line width nor the partial coverage due to the insufficient spectral resolution, so both $f_c$ and $b$ were adopted with rough estimations.  {We do emphasize that in other similar objects, the difference in the estimation may not be as large: for AFGL~2136 and AFGL~2591,  the differences between the EXES results~\citep{barr22} and the ISO results~\citep{boonman03} are 3.6$\times10^{18}$/1.5$\times10^{18}\sim 2.4$ times 1.5$\times10^{19}$/3.5$\times10^{18}\sim 4.3$ times. Only W3~IRS~5 provides an example where the derived column density differs significantly between sufficient and insufficient resolving power}.

\subsection{Summary of the Problems}\label{subsec:2problems}


{In summary, there are two problems in interpreting a spectrum with insufficient resolving power and limited S/N. The first one (\S~\ref{subsec:de-obs}) relates from the higher S/N for the stronger optically thick lines. When optically thick lines dominate the fitting process, a wide range of parameter space parameter space will fit the data well. If the S/N is improved, including more optically thin lines can narrow the derived parameter space, specifically, in $N_\textrm{beam}$. However, even if $N_\textrm{beam}$ is constrained nicely, the second problem, the ``intrinsic degeneracy" still exists regardless of the noise level (\S~\ref{subsec:bf}). In the second problem, the product of $b$ and $f_c$ is a degenerated value, and can only be decoupled by indirectly fitting overlapping optically thick lines, or directly by obtaining ground-based observations at very high spectral resolution to determine $b$ directly.}



\section{Recipes}\label{sec:recipe}

\begin{figure*}[!t]
    \centering
    \includegraphics[width=0.92\linewidth]{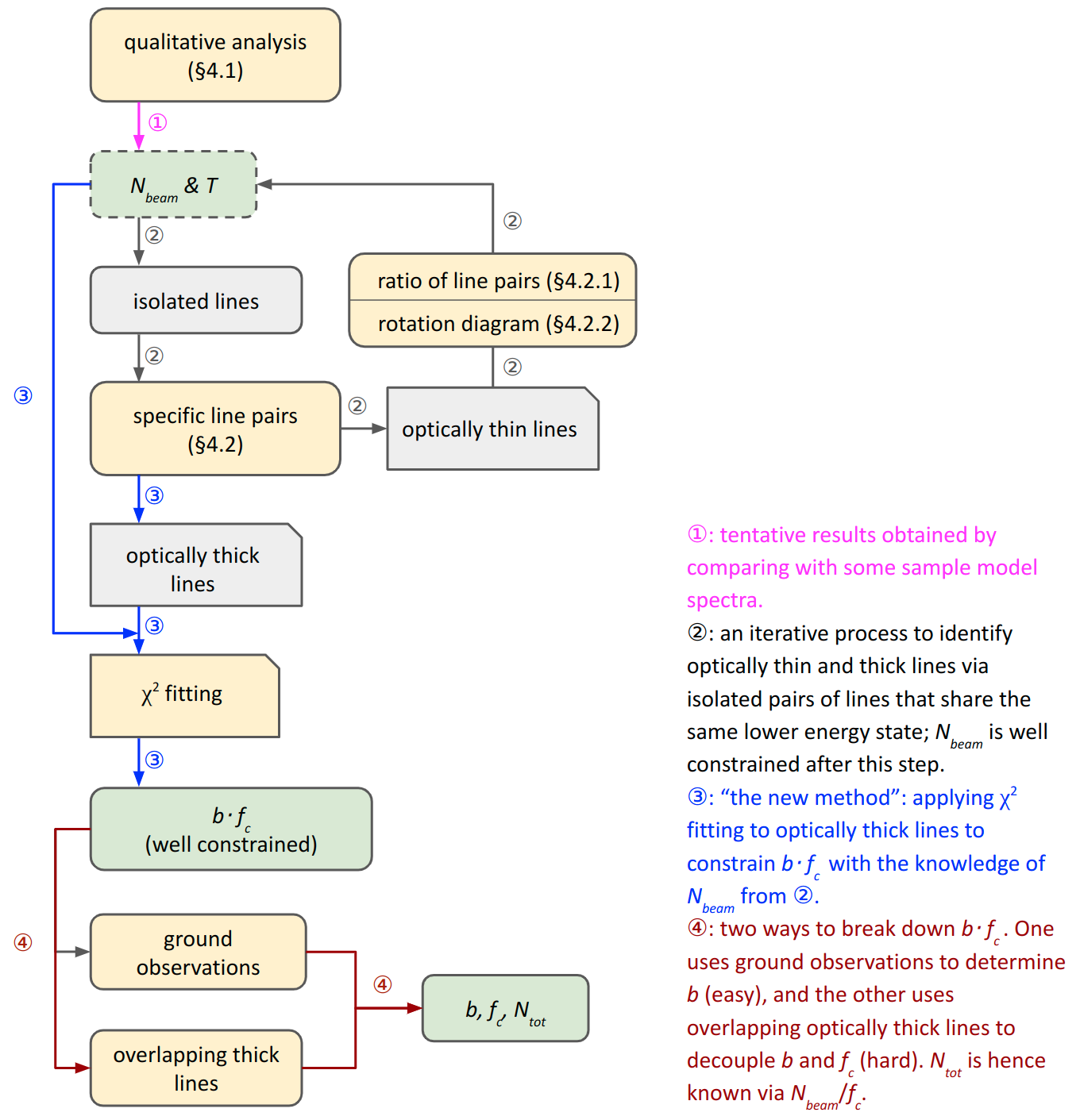}
    \caption{{Flow chart illustrating our recommended process in interpreting MIR spectrum for unresolved lines, such as $R\sim$3,000 for JWST. In this flow chart, yellow boxes represent analysis steps; green boxes represent parameters of interest; and grey boxes represent intermediate results of analysis. Compared to the ``old method" where the whole spectrum is used for the $\chi^2$ fitting to constrain $bf_c$, without knowing optically thick or thin lines (so as $N_\textrm{beam}$), the new recipe is iterative for initial identification of optically thin and thick lines. Once optically thin lines are selected, $N_\textrm{beam}$ and $T$ can be constrained via either the rotation diagram or curve of growth analysis, which are essentially the same for thin lines. Optically thick lines, with a good understanding of $N_\textrm{beam}$, can then be used to constraint the product $bf_c$. Whether $bf_c$ can be decoupled depends on whether regions of optically thick and overlapping lines exist. Otherwise, one has to refer to external information, such as ground observational results, to decouple $b$ and $f_c$.}}
    \label{fig:flowchart}
\end{figure*}

Our discussion in \S~\ref{sec:3} points out that for the MIR rovibrational band where several tens to hundreds of individual lines are observed simultaneously, a proper weighting of strong and weak lines is necessary. A blindly applied $\chi^2$ fitting method will overweight more optically thick lines which stand out better above the noise. Moreover, at lower instrumental resolving powers, the peak absorption intensities are suppressed even more. The derived result is then even more influenced by the degeneracy problem originating from optically thick lines and can be estimated incorrectly. 

We present in this section recipes that provide a more balanced weighting over the individual transitions for more accurate constraints on the parameters, $N_\textrm{tot}$, $f_c$, $b$, and \tex. The recipes include a visual comparison of the relative line depths to qualitatively estimate the lower limit of \tex\ and $N_\textrm{tot}$ (\S~\ref{subsec:feature}), and two ways to further quantitatively decouple $N_\textrm{tot}$, $f_c$, and $b$ (\S~\ref{subsec:decouple}). Specifically, \S~\ref{recipe1} suggests using line ratios of specific line pairs to pinpoint the parameter space, and \S~\ref{recipe2} introduces a more ``standard" curve of growth analysis.

\begin{figure*}[!t]
    \centering
    \includegraphics[width=\linewidth]{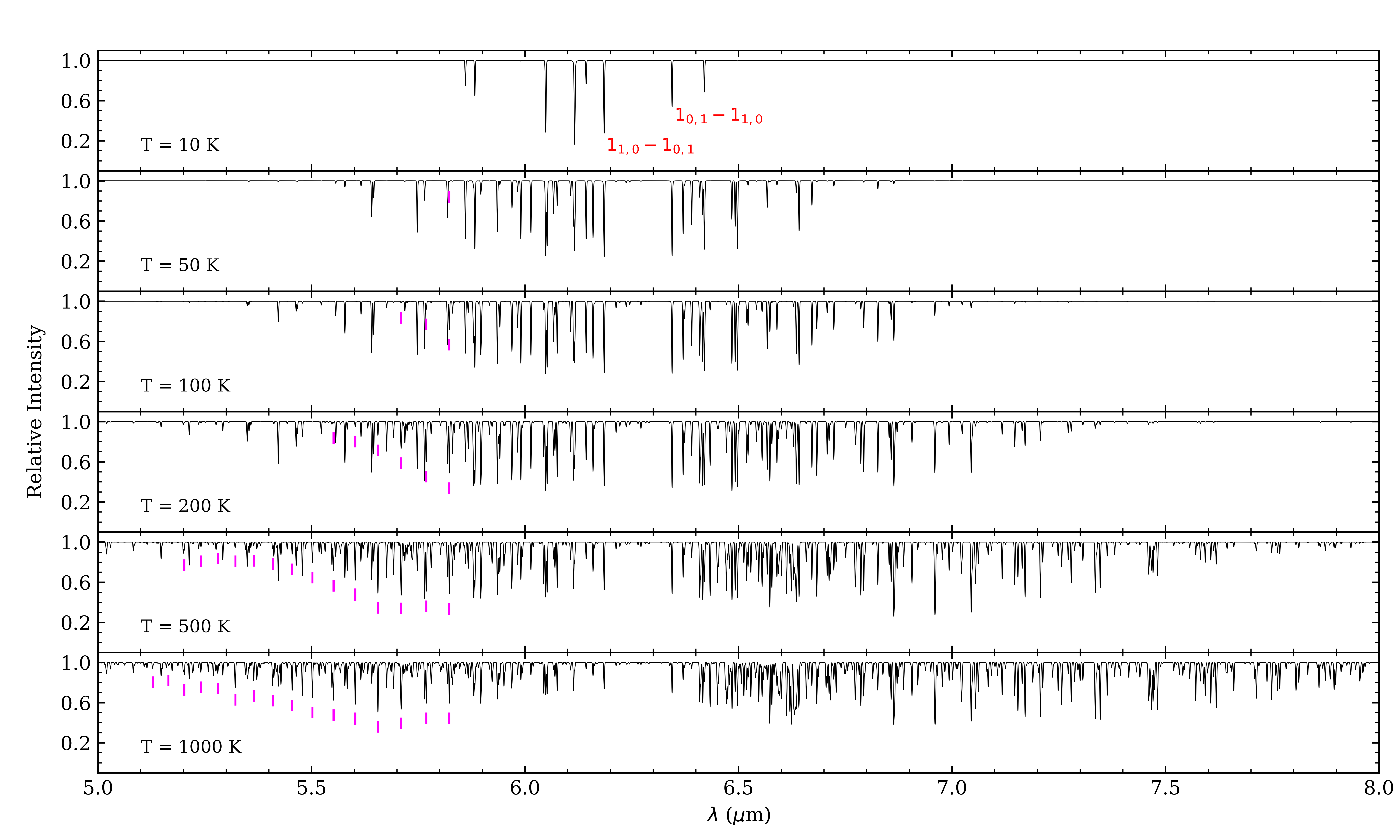}
    \caption{Modeled water rovibrational spectra at 10, 50, 100, 200, 500, 1000~K at a spectral resolving power of 3,500. All models have the same total column density ($N_\textrm{tot} = 5\times10^{18}$~cm$^{-2}$), velocity width ($b$ = 20~\kms), and fractional coverage ($f_c = 1$). We note that applying a different velocity width or a different fractional coverage would result in a similar spectral profile with different absorption depths (Fig~\ref{fig:comp-dege}).  The $R$ and the $P$ {branch} are roughly located on the left and the right side of $1_{1, 0}-1_{0, 1}$ and $1_{0, 1}-1_{1, 0}$ labeled in the first panel. Specifically, a set of seemingly evenly spaced spectral lines in the $R$ branch (marked with {magenta} ticks) that define the envelope of the spectrum stand out as \tex\ increases. These lines are composed of $J_{0, J}$--$(J-1)_{1, J-1}$ and $J_{1, J}$--$(J-1)_{0, J-1}$ transition when $J\geq10$ (see Figure~\ref{fig:r-branch}a for detail).}
    \label{fig:example}
\end{figure*}

\begin{figure}[!t]
    \centering
    \includegraphics[width=\linewidth]{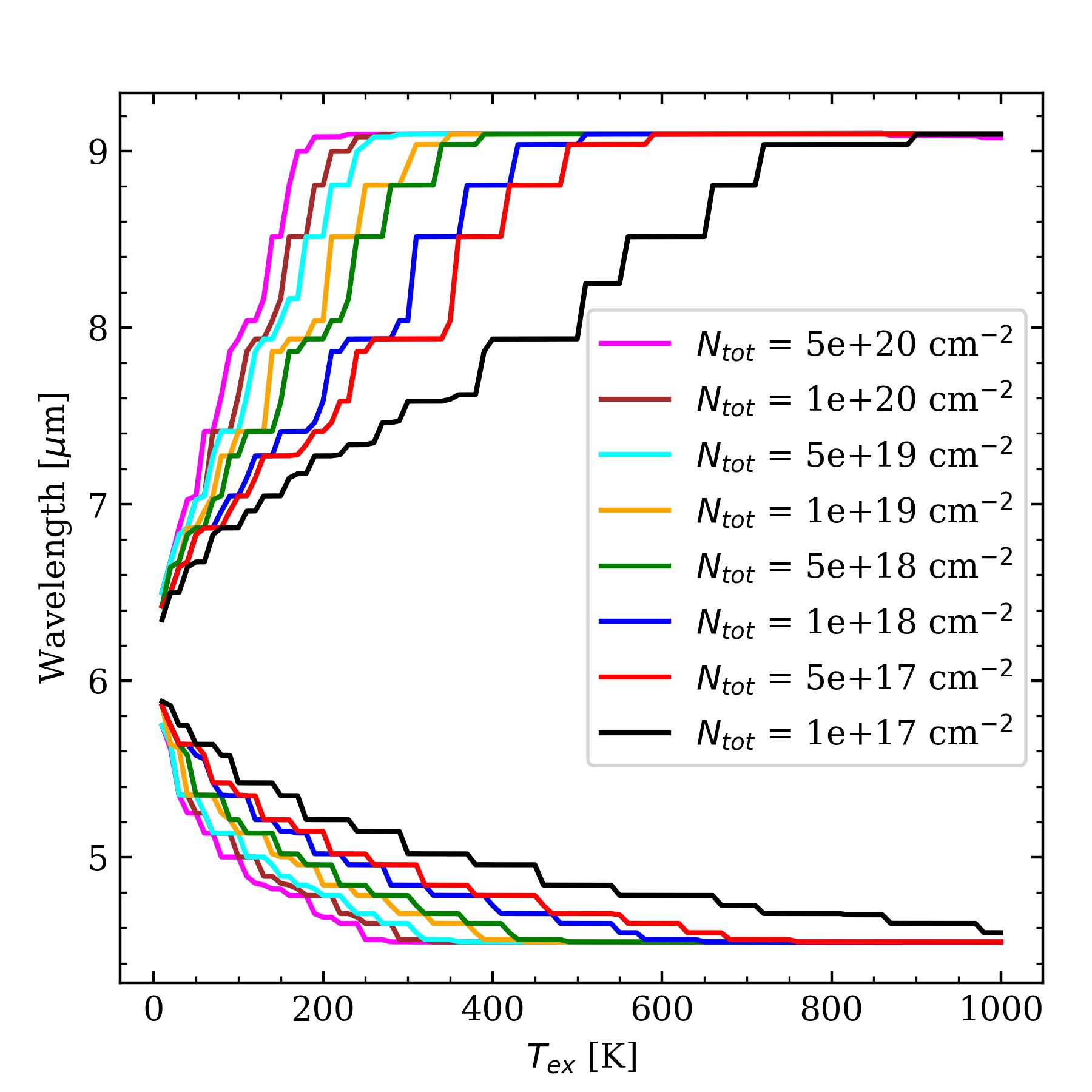}
    \caption{The lower and upper extremes of the wavelength range covered by the $\nu_2$ band above a 1\% absorption level under different \tex\ and $N_\textrm{tot}$. While higher \tex\ or $N_\textrm{tot}$ both increase the coverage range, a rough estimation of the upper and lower limits on the temperature can be made. The uncertainty of the estimation increases with higher \tex, though.}
    \label{fig:T}
\end{figure}

\subsection{Qualitative Recipe: Prominent Spectral Features for Quick Estimation on \tex\ and $N_\textrm{tot}$}\label{subsec:feature}


While the rovibrational band of water consists of crowded transitions with the $Q$ branch nested in the $R$ and $P$ branches, there are specific spectral features that are easy to identify and can deliver certain information about the excitation or the column density. Although these features provide a qualitative constraint, as we will show below, mostly lower limits to \tex\ and $N_\textrm{tot}$, being able to understand and recognize these features can help to establish a quick estimation. 


We present in Figure~\ref{fig:example}, as well as Figures~\ref{fig:e1} to \ref{fig:e3} in Appendix~\ref{app:1} a list of modeled water spectra in a grid of parameter space, in which the temperature ranges from 10 to 1000~K and $N_\textrm{tot}$ ranges from $10^{17}$ to 5$\times 10^{19}$~cm$^{-2}$. The line width $b$ is 20~\kms\ and cannot be resolved under MIRI's spectral resolution. 
 The chosen parameter space covers a variety of detectable spectral profiles from a few isolated lines in low \tex\ to tightly packed lines at high $N_\textrm{tot}$.

The parameter \tex\ is easy to constrain from the perspective that a higher temperature excites more transitions in a high $J$. To first order, for a higher \tex, the peak positions of the wing profiles of either the $R$ or the $P$ branch shift away from the ground state transition $1_{1, 1}-0_{0, 0}$ at $\sim$6.1~\um. As is shown in Figure~\ref{fig:example}, where the temperature of the modeled spectra ranges from 10 to 1000~K, the $R$ and the $P$ {branch} are roughly located on the left and the right side of $1_{1, 0}-1_{0, 1}$ and $1_{0, 1}-1_{1, 0}$ (labeled in Figure~\ref{fig:example}). More transitions get excited as the temperature increases, and a wider wavelength range is covered by the rovibrational lines. However, a higher $N_\textrm{tot}$ also excites lines with stronger depth (Figure~\ref{fig:T}). This relation, therefore, while it provides a rather good constraint for low \tex~($\lesssim$100~K), can be only used to constrain the lower limit of \tex\ unless $N_\textrm{tot}$ is constrained via other information.

A specific feature in the $R$ branch stands out for \tex$\gtrsim$300~K. As is shown in Figure~\ref{fig:example}, at 500 and 1000~K, the envelope of the $R$-branch seems to be defined by a set of seemingly evenly spaced spectral lines. These transitions involve the backbone levels and are labeled in Figure~\ref{fig:r-branch}a in Appendix~\ref{app:b-2}, and are $J_{0, J}$--$(J-1)_{1, J-1}$ and $J_{1, J}$--$(J-1)_{0, J-1}$ at the $\nu_2$ band. Specifically, for $J\geq10$, $J_{0, J}$--$(J-1)_{1, J-1}$ and $J_{1, J}$--$(J-1)_{0, J-1}$ overlap and blend to lines that define the envelope, and therefore form a wing-like profile as is seen in the $R$-branch of the rovibrational spectrum of a diatomic molecule. The set of blended lines is not evenly spaced, and the distance decreases from 0.055~\um\ ($J=10$) to 0.035~\um~($J=21$). The two transitions are separated at $J<10$, and the wing shape is thus broken at low $J$ (Figure~\ref{fig:r-branch}a). We also note that at low \tex\ when only low $J$ lines are excited, the wing-like profile does not exist, either. This wing-like feature is therefore helpful for recognizing high \tex\ conditions, although constrained $N_\textrm{tot}$ is also required for a good estimation of \tex~(Figure~\ref{fig:r-branch}b in Appendix~\ref{app:b-2}).

Certain spectral features help constrain $N_\textrm{tot}$ and we list a few that are instructive. Firstly, the absorption depth of the overall spectral profile constrains the lower limit of $N_\textrm{tot}$. Assuming $f_c = 1$, if most absorption lines are deeper than $\sim$50\%, $N_\textrm{tot}$ must be higher than $10^{18}$~cm$^{-2}$. Secondly, if the transitions between $1_{1, 0}-1_{0, 1}$ and $1_{0, 1}-1_{1, 0}$ are detected at levels $> 1\%$ (see Figure~\ref{fig:example}), this indicates a rather high $N_\textrm{tot}$.

\begin{figure*}[!t]
    \centering
    \includegraphics[width=\linewidth]{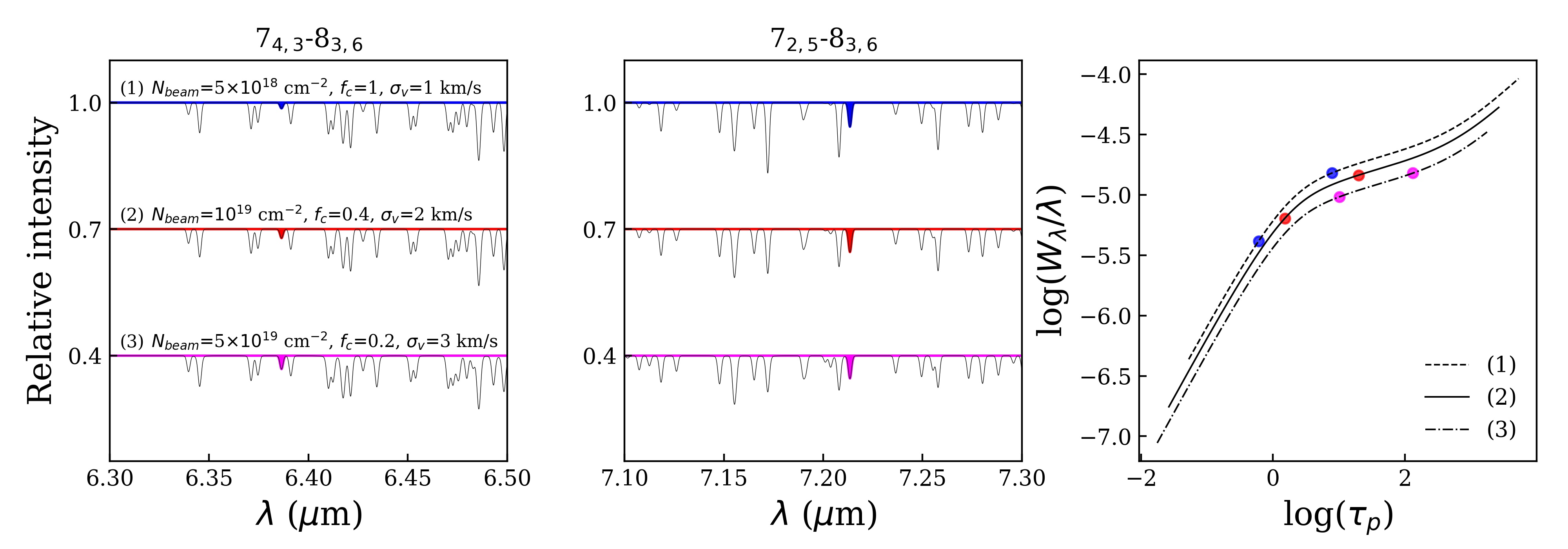}
    \includegraphics[width=\linewidth]{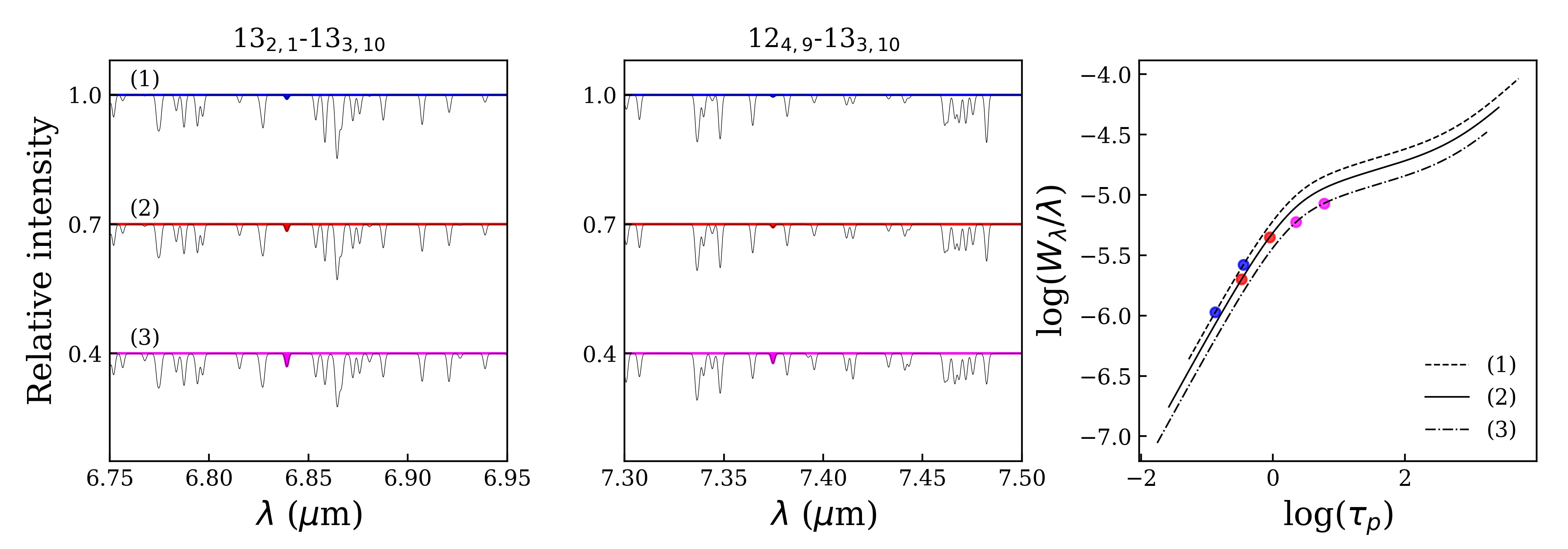}
    \caption{Two pairs of lines sharing the same lower energy state, 7$_{4, 3}$-8$_{3, 6}$, 7$_{2, 5}$-8$_{3, 6}$ and 13$_{2, 1}$-13$_{3, 10}$, 12$_{4, 9}$-13$_{3, 10}$ at an $R=3,500$ resolving power and their positions on the curve of growth. {Both line pairs are sufficiently isolated from nearby spectral lines.} The blue (model 1), red (model 2), and magenta (model 3) colors correspond to models (b), (a), and (c) adopted by Figure~\ref{fig:three-models}, with $N_\textrm{tot}$ = $5\times 10^{18}$, $2.5\times 10^{19}$, $2.5\times 10^{20}$~cm$^{-2}$, $N_\textrm{beam}$ = $5\times 10^{18}$, $10^{19}$, $5\times 10^{19}$~cm$^{-2}$, and $\sigma_v$ = 1, 2, 3~\kms\ (or $b$=1.4, 2.8, 4.2~\kms). On the curves of growth, the line pairs of models 1, 2, and 3 shift from the lower left to the upper right regions. Specifically, while other relatively deeper lines have similar depths in the three models, the color-filled lines with 8$_{3, 6}$ or 13$_{3, 10}$ as the lower level differ in the line depth as well as the opacity.}
    \label{fig:cog}
\end{figure*}

\begin{figure*}
    \centering
    \includegraphics[width=\linewidth]{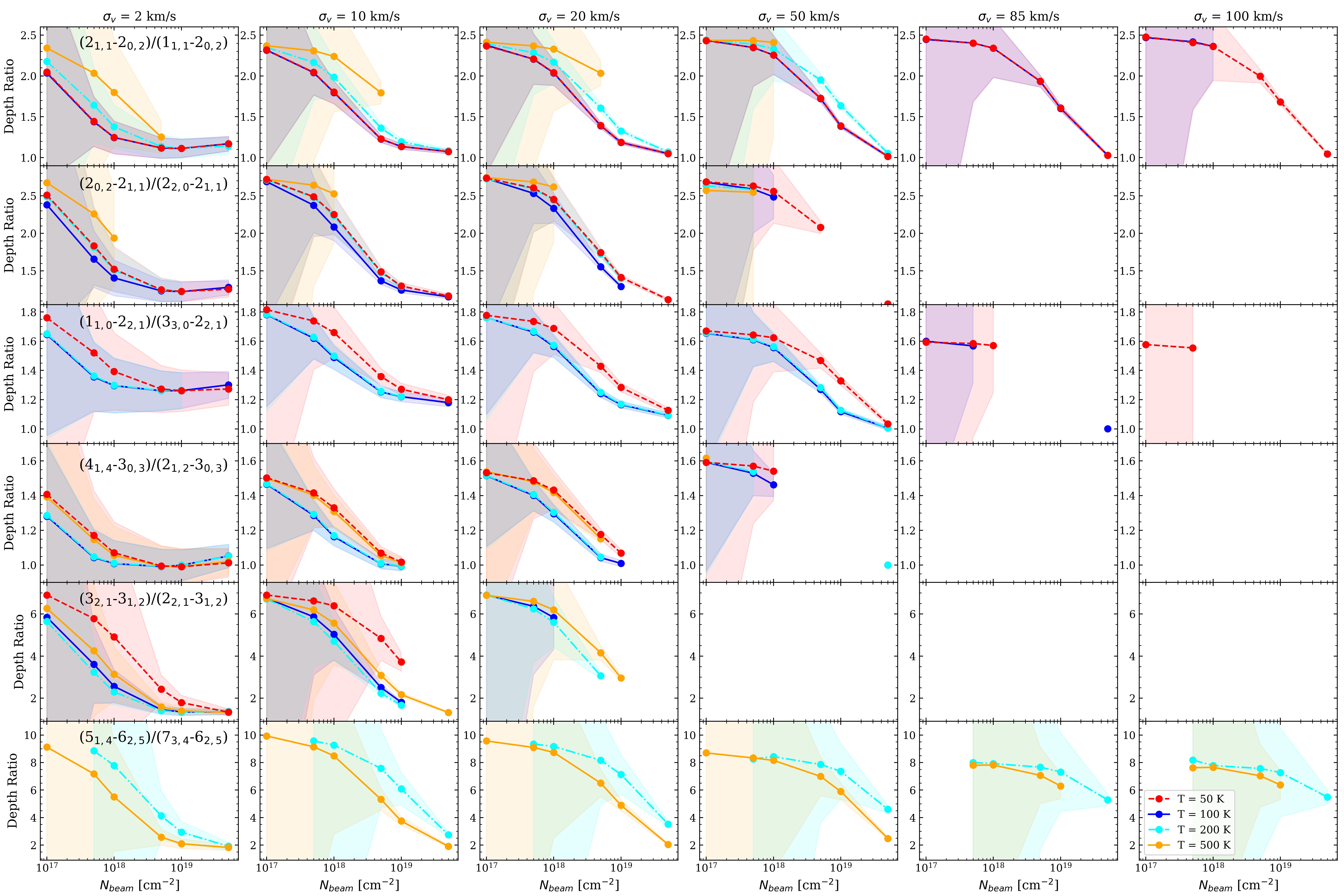}
    \caption{Ratios of six pairs of isolated lines under $\sigma_v$ of 2, 10, 20, 50, 85, and 100~\kms~(or $b$=2.8, 14, 28, 70, 120, 140~\kms), \tex\ of 50, 100, 200, 500~K, and $N_\textrm{beam}$ from $10^{17}$ to $5\times10^{19}$~cm$^{-2}$. The ratios are only available for specific temperatures and column densities. The shadowed regions correspond to the uncertainties of absorption depth or the depth ratios, adopting a continuum S/N of $\sim$100, or a noise level of 0.01 on a normalized continuum. The decrement in the ratios indicates that the line pairs are shifting from the linear part to the logarithmic part of the curve of growth and that the opacities are increasing from small to intermediate values. We note that some panels are empty because the targeted lines are either too weak or overlap with other lines under the designated physical conditions.}
    \label{fig:ratio}
\end{figure*}

\subsection{Quantitative Recipes}\label{subsec:decouple}

To quantitatively derive the physical conditions in the absorbing gas, optically thin lines deserve comparable weighting to optically thin lines. We provide below two recipes that balance the usage of optically thin and thick lines. {The first recipe focuses on a few lines that share the same lower energy state to derive the physical conditions. It specifically helps to determine whether a line (in the pair) is optically thin or thick, avoiding the degeneracy problem of $bf_c$.} {This method essentially treats these lines with equal weights in the curve of growth fitting.}
The second recipe uses a more traditional curve of growth analysis, treating optically thin and thick lines separately.




\subsubsection{Recipe 1: Ratios of Specific Line Pairs}\label{recipe1}

{At moderate spectral resolution, insufficient resolving power often results in the blending of spectral lines with their closely spaced neighbors. Despite these challenges, there exists a limited set of isolated spectral lines that are distinctly separate from adjacent transitions. Among these isolated lines, certain pairs share the same lower energy state, and consequently, the same column density of the lower state, $N_l$. Such pairs of isolated lines can be thereby used as tools for inferring the physical conditions of the gas.}

Consider lines with the same lower energy state level from the perspective of the curve of growth analysis, those lines are located at different positions of the curve if they have different peak optical depths $\tau_p$, which is proportional to $N_l f_{lu} \lambda$. Specifically, the ratios of $\tau_p$ of those lines are a constant which is equal to the ratio of $f_{lu} \lambda$. Thus, their separation on the $x$-axis of the curve of growth, or equivalently, $\Delta$log($\tau_p$) is a constant.

The equivalent widths, $W_\lambda$, of these lines can therefore be used to measure the opacity, and therefore the column density of the absorbing gas. Figure~\ref{fig:cog} shows examples as line pairs with the same lower state that are located at different positions on the curve of growth according to the input parameters for the models. Specifically, for the pair with 13$_{3, 10}$ as the lower level, their positions on the curve of growth move from the linear part to the logarithmic part, indicating these lines are transiting from optically thin to rather thick condition; for the pair with 8$_{3, 6}$ as the lower level, the shifts are within the logarithmic part, indicating that the lines have intermediate optical depth. While $\Delta$log($\tau_p$) is a constant, $\Delta$log($W_\lambda$) varies. When one or even two lines move to the logarithmic part, $\Delta$log($W_\lambda$) decreases. $\Delta$log($W_\lambda$) would increase again if the line pairs move to the square root part, but that is never relevant for the physical environments we discuss in this paper due to the too-small damping factor (see \S~\ref{sec:2}).

For spectra obtained at a moderate spectral resolution such as that of MIRI, the ratio of $W_\lambda$ of the unresolved lines can be equivalently converted to the ratio of the line peak optical depth. This is because the reshaped lines are in Gaussian and the reshaped width is dominated by the kernel size of MIRI. Therefore, the absorption intensity ratio of lines that share the same lower energy state, is an ideal tool for constraining the physical condition of the gas. This tool specifically has advantages in that the partial coverage, $f_c$ is canceled out in the form of the ratio, so one may first focus on decoupling \tex, $N_\textrm{tot}$, and $b$, then constrain $f_c$.

No pair of lines can be applied universally. As is illustrated in \S~\ref{subsec:feature}, different lines are excited under different conditions. For example, for larger $b$, one line may no longer be available as it becomes broader and is blended with a nearby line. Besides, lines that are not detectable under a lower \tex\ or a low $N_\textrm{tot}$ may pollute a usable line when \tex\ or $N_\textrm{tot}$ increase. On the other hand, line pairs that are strong and good indicators of high \tex\ or $N_\textrm{tot}$ might be too weak under lower \tex\ or low $N_\textrm{tot}$ to be useful. 

We present in Figure~\ref{fig:ratio} a set of selected line pairs that can be used to probe specific parts of the parameter space for the said reasons. All the line ratios presented decrease as the $N_\textrm{beam}$ increases, indicating the shift of the line pairs from the linear to the logarithmic part on the curve of growth. More specifically, there are two types of decrement: drop-then-flat (e.g., $2_{1, 1}-2_{0, 2}$/$1_{1, 1}-2_{0, 2}$, $\sigma_v$=2~\kms\ or $b$=2.8~\kms) and flat-then-drop (e.g., $2_{1, 1}-2_{0, 2}$/$1_{1, 1}-2_{0, 2}$, $\sigma_v$=85~\kms\ or $b$=120~\kms). The first case indicates that the line pairs start from the ``knee" region of the curve of growth, so the weaker line is optically thin and the strong line has intermediate optical depth, and that the ratio (or equivalently, $\Delta$log$W_\lambda$) decreases as the column density increases. The second case, on the other hand, indicates that both lines start from optically thin conditions and gradually move to the ``knee" region of the curve of growth.

While the line ratios in Figure~\ref{fig:ratio} help to unambiguously determine the line opacity, determining the desired $N_\textrm{beam}$, \tex, and $b$ requires cross-comparing values of multiple line ratios. This is because the ratio curves for two sets of lines presented in Figure~\ref{fig:ratio} can be nearly identical (e.g., 50 and 100~K in $2_{1, 1}-2_{0, 2}$/$1_{1, 1}-2_{0, 2}$). Although $b$ can be determined in the cross-comparing process, we suggest using the information of $b$ from high-spectral resolution MIR ground observations of other species (e.g., CO) and only using the line ratios to determine $N_\textrm{beam}$ and \tex. Grids with finer temperature intervals of 10~K for the six line pairs are presented in Appendix~\ref{app:grids}.

The line ratio method has several caveats. Firstly, due to the noise in the observed spectra, the uncertainty associated with the ratio at small $N_\textrm{beam}$ can be rather large since the absorption depths are small. Therefore, for the currently assumed noise level of 1\% on a normalized spectrum (continuum S/N of 100), this method is applicable for $N_\textrm{beam}$ larger than $10^{18}$~cm$^{-2}$. Secondly, since different line pairs are sensitive to different parameter spaces, it is not easy to start from a random line pair in Figure~\ref{fig:ratio} to look for the solution. If prior information of the temperature or column density can be constrained from, for example, the qualitative way introduced in \S~\ref{subsec:feature}, the cross-comparing process may get more simplified. Thirdly, this method assumes that the MIR continuum originates from the same region or spans the same physical scale across different wavelengths. However, this scenario is improbable in an actual astrophysical environment, since a temperature gradient must exist. {Lastly, emission by PAHs within the observational beam have a potential influence on the observed MIR continuum and complicate the interpretation further.}






\subsubsection{Recipe 2: Curve of Growth Analysis in Two-Steps}\label{recipe2}

If a rough estimation of the parameter space is derived, a full list of isolated lines can be determined by comparing the depth of the line in the spectral profile to its expected absorption depth from the model: if a line overlaps with some insignificant lines, its absorption depth will be larger than that derived from the model. {With the set of isolated lines in hand, as discussed in \S~\ref{subsec:bf}, applying the curve of growth analysis to all observed lines will result in degenerated results of $N_\textrm{tot}f_c$ and $bf_c$: this is the ``first problem" illustrated in \S~\ref{subsec:2problems}.}

Applying the curve of growth analysis separately to optically thin and thick lines can break down the degeneracy. {As Figure~\ref{fig:three-models} has shown, fitting optically thin lines determines $N_\textrm{beam}$ unambiguously by fitting to the curve of growth (in the linear part), or equivalently by applying the rotation diagram (see also Appendix~\ref{app:0})}. Fitting optically thick lines will determine $f_c$ which is degenerated with $b$, and either using the overlapped region (see \S~\ref{subsec:bf}) or $b$ from ground observations can break down the $bf_c$ degeneracy. In other words, at sufficiently high resolving power and sufficiently low $b$, the opacity may be determined from the depth of nearby lines. Otherwise, the line ratios discussed in \S~\ref{recipe1} should be used.

{Compared to the line-pair-ratio method proposed in \S~\ref{recipe1}, which uses a few lines and is expected to be a ``quick recipe", this method attempts to get a solution based on as many isolated lines in the JWST/MIRI spectrum and may serve as a ``standard recipe". However, this recipe faces the difficulty of getting the prior information about whether a line is optically thin or thick. As we have shown in Figure~\ref{fig:flowchart}, getting such information is unfortunately an iterative process. If a line is found to have another line sharing the same lower energy state, determining the opacity can be understood with the line ratio method. One other experiment can be helpful after fitting a curve of growth regardless of the opacity and obtaining the coupled $N_\textrm{tot}f_c$ and $bf_c$: as is shown in Figure~\ref{fig:comp-dege}, one can assume the values of $b$ or $f_c$ and construct a spectrum such as Figure~\ref{fig:comp-dege}c firstly; then tuning either $b$ or $f_c$ will reveal the unchanged lines to be optically thin lines.}

\section{Discussion}\label{sec:diss}

\subsection{Unresolved Kinematical Components}\label{subsec:blend}

While the previous sections discuss in detail the properties of a water rovibrational spectrum of one absorbing component, multiple kinematic components, usually in different temperatures, have been widely detected in front of the MIR background in massive protostellar systems~\citep[e.g.,][]{barr18, li22} with high spectral resolution spectroscopy. In these studies, the velocity difference of the components is on the order of several tens of km/s. Unless their velocity centers differ by at least 85~\kms, the spectral resolution of MIRI, these components will be blended into one and cannot be distinguished. 

\subsubsection{Recipes to the First Order}

For simpler molecules such as the diatomic ones, CO or CS, the existence of multiple unresolved components can be identified with the rotation diagram. In a rotation diagram, straight lines with different slopes show up, and the inverse of the slope equals the negative of the temperature. However, such a procedure can only be tentatively applied to water lines because water has a complex spectrum with lines that are difficult to measure individually. Specifically, a large scatter exists in the rotation diagram that exceeds what the error bars can account for, and this is because lines that share the same $N_l$ have different Einstein $A$ coefficients. The temperature and total column density derived from the rotation diagram can be significantly higher and lower than that derived from the curve of growth analysis which takes care of the opacity issue.

The qualitative analysis in \S~\ref{subsec:feature} described representative spectral features of certain temperatures or/and column densities that are prominent at specific wavelengths. Therefore, identifying such features helps determine if more than one component exists and helps make a rough constraint on these parameters. As we will discuss below (\S~\ref{subsec:cccog}), ``blindly" conducting a curve of growth analysis toward all lines may also deliver information on the components. Afterward, based on the understanding of specific wavelength ranges, one may determine unresolved but isolated lines at these ranges. Then, to the first order, the line ratio method can be applied to infer the properties of different components, separately. 

\subsubsection{Implications from the Curve of Growth Analysis}\label{subsec:cccog}

The curve of growth analysis, as the single curve illustrates (e.g. Figure~\ref{fig:cog}), only corresponds to one specific physical condition, and can only be used to derive that of one physical component. However, investigating how the existence of two components may influence the curve of growth analysis can be insightful. In a curve of growth analysis, first, the equivalent widths of each spectral line are measured and are converted to log($W_\lambda/\lambda$). Second, $\tau_p$ (equation~\ref{eq:tau}) is calculated given a variety of physical conditions. The two steps establish the positions of each spectral line on the log($W_\lambda/\lambda$)-log($\tau_p$) space, and the physical condition that gives the best fit between the dataset and the corresponding curve of growth is the solution. Consider two components with different physical conditions that coexist at different velocities and cannot be distinguished by the spectrometer. Assuming that the two components are within one beam but do not overlap along the same line of sight (in other words, their optical depths contribute to the attenuation independently), their equivalent widths add up to the observed $W_\lambda$, and the values of $\tau_p$ depend on the input physical conditions.

We present in Figure~\ref{fig:sum-cog-upper} and \ref{fig:sum-cog-lower} the distribution of the dataset on the log($W_\lambda/\lambda$)-log($\tau_p$) space applied with the physical condition of each absorbing component. Each component contributes to the sum of $W_\lambda/\lambda$ with a variety of weights (from 20\% to 80\%), and the sum was calculated transition by transition. The parameter $\tau_p$ was calculated with the physical conditions of the first (Figure~\ref{fig:sum-cog-upper}b, f, and \ref{fig:sum-cog-lower}b, f), the second (Figure~\ref{fig:sum-cog-upper}c, g and \ref{fig:sum-cog-lower}c, g) component, or with an artificial third model (Figure~\ref{fig:sum-cog-upper}d, h and \ref{fig:sum-cog-lower}d, h). Specifically, we considered three sets of conditions: the two components have the same temperature or the same column density (Figure~\ref{fig:sum-cog-upper}), or differ for all the parameters (Figure~\ref{fig:sum-cog-lower}).

\begin{figure*}
    \centering
    \includegraphics[width=\linewidth]{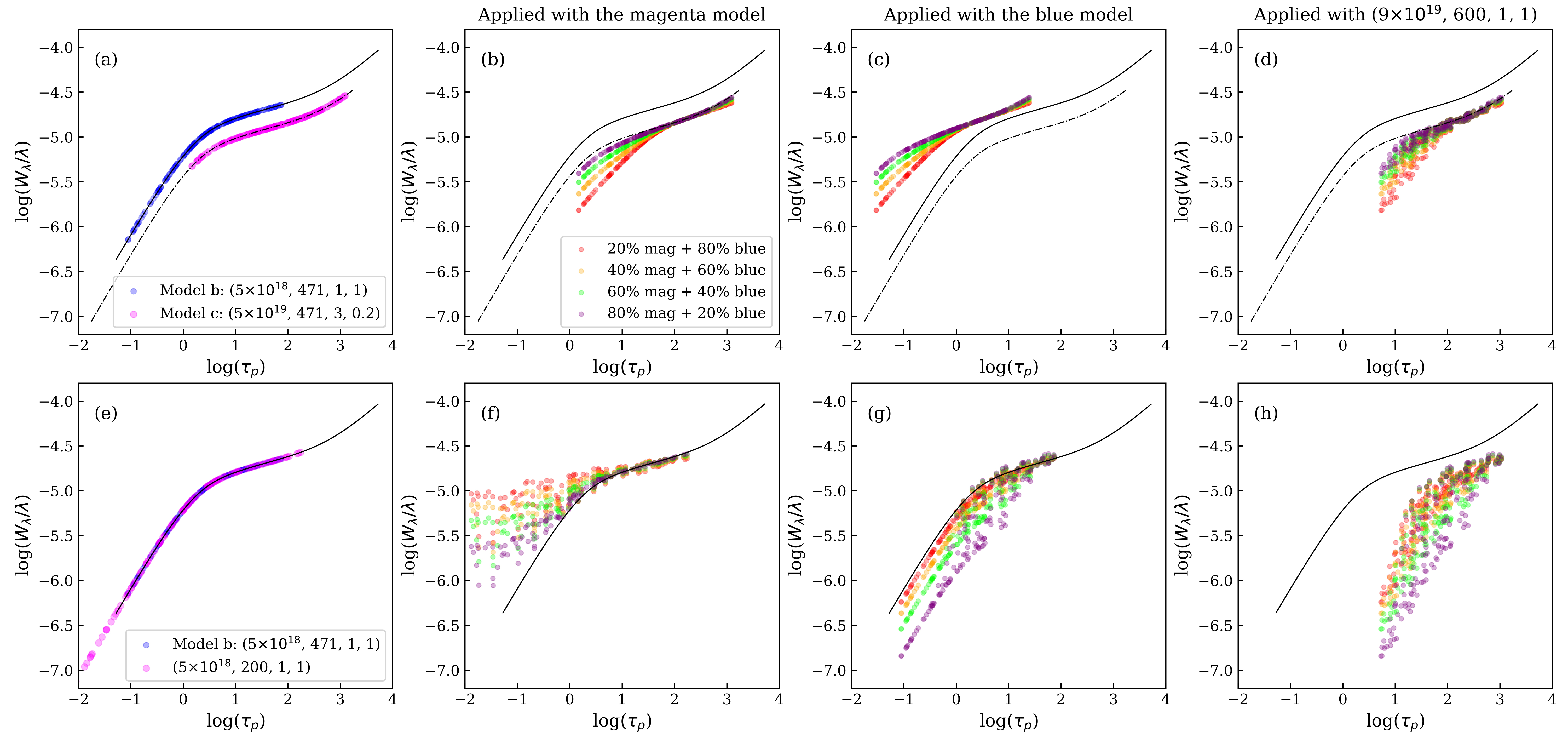}
    \caption{Locations of spectral lines that consist of two unresolved components, which have different velocities but the velocity components blend into one absorption feature at the low spectral resolution, on the space of curve of growth. Specifically, panels \textit{a}, \textit{e} present the curves of growth of absorbing gas under two physical conditions and the positions of individual isolated spectral lines on the curve. The physical parameters in the legend correspond to ($N_\textrm{beam}$, \tex, $\sigma_v$, $f_c$) in units of (cm$^{-2}$, K, \kms, 1). ``Model b" (and ``c") correspond to models b and c in Figures~\ref{fig:three-models} and \ref{fig:cog}. The rest panels present data from the weighted sum of the two models. Four sets of combinations of the two models are presented. While $W_\lambda/\lambda$ is directly observable, values of $\tau_p$ depend on the physical conditions of the model. The magenta (blue) models in \textit{a}, \textit{e} were applied to panels \textit{b}, \textit{f} (\textit{c}, \textit{g}). We note that in panels \textit{e}--\textit{h}, the curves of growth of the two models overlap because $N_\textrm{beam}$, $b$ (=$\sqrt{2}\sigma_v$) and $f_c$ are the same (see \S~\ref{subsec:bf}). }
    \label{fig:sum-cog-upper}
\end{figure*}

\begin{figure*}
    \centering
    \includegraphics[width=\linewidth]{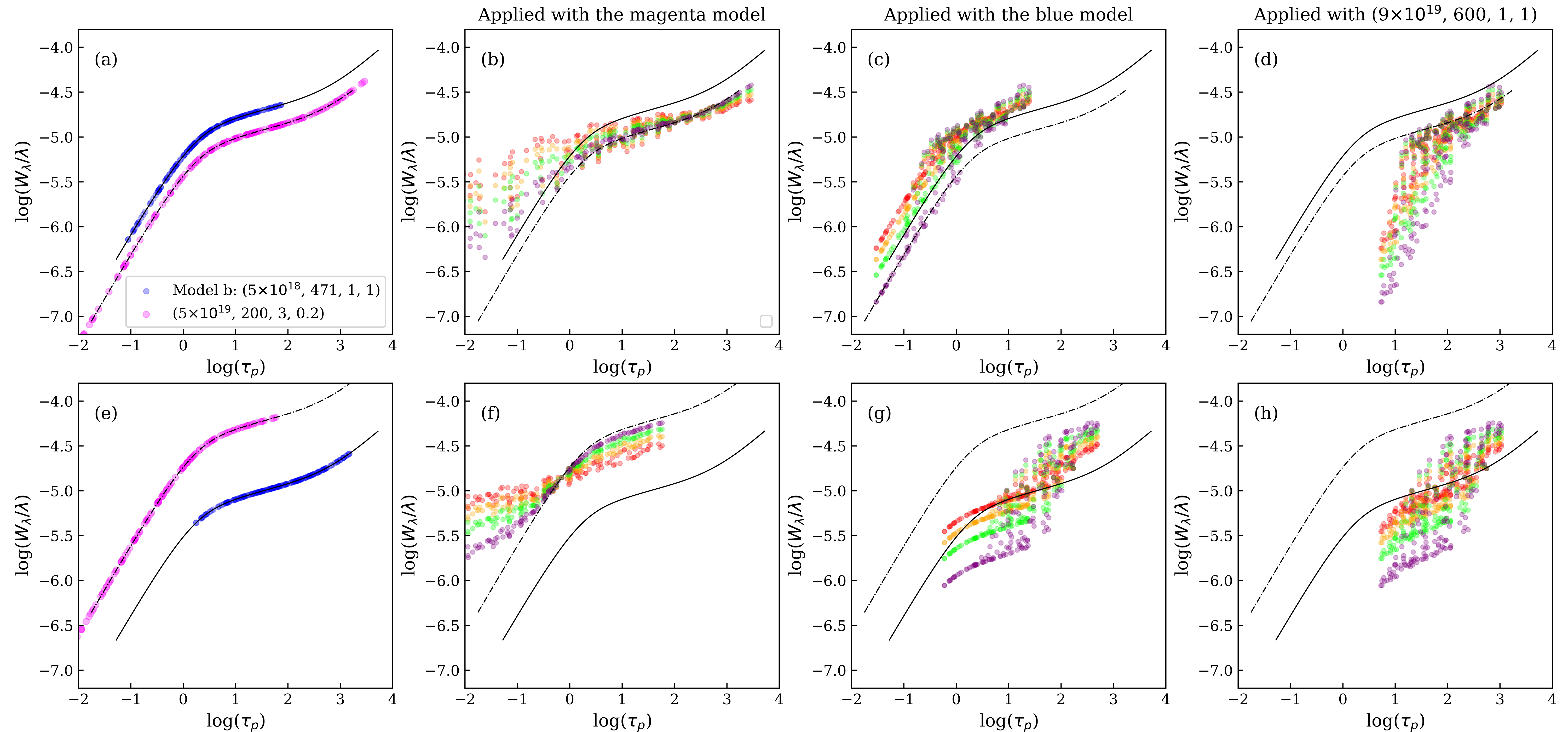}
    \caption{Similar to Figure~\ref{fig:sum-cog-upper}, with two components differ in both the temperature and the column density. }
    \label{fig:sum-cog-lower}
\end{figure*}


The distribution of the composite dataset on the log($W_\lambda/\lambda$)-log($\tau_p$) space is scattered with an artificial input parameter (Figure~\ref{fig:sum-cog-upper}d, h and Figure~\ref{fig:sum-cog-lower}d, h). However, the distribution has some representative features when the physical conditions of the absorbing component are applied. Figure~\ref{fig:sum-cog-upper}a--c and e--g show two ideal but potentially not practical cases as examples, where the temperatures or the column densities are the same. First, if the temperatures are the same, the sum of the curve of growth has no scatter. The existence of the other component is indicated by the smaller slope on the linear part since the component with a lower column density contributes more absorption for transitions at low opacities. Second, if the two components have the same column density but different temperatures, the scatter diminishes at the end of the curve of growth where one component dominates. More specifically, the composite dataset converges at the logarithmic part of the curve of growth that is associated with the cooler component (Figure~\ref{fig:sum-cog-upper}f) and converges at the linear part of the curve of growth that is associated with the hotter component (Figure~\ref{fig:sum-cog-upper}g). Figure~\ref{fig:sum-cog-upper}f shows that the properties of the cooler component can be well constrained with the logarithmic part. On the other hand, Figure~\ref{fig:sum-cog-upper}g shows that, depending on the weight of the hotter component, the converged end at low opacities may differ a lot from the theoretical curve of growth of the hotter component, and thus underestimate the column density if the cooler component weights more.

Figure~\ref{fig:sum-cog-lower} presents more realistic cases where the two absorbing components differ in both temperature and column density. Compared with the two ideal cases discussed above, Figure~\ref{fig:sum-cog-lower}b shows that a convergence seen in the logarithmic part to some extent gives a good constrain to the cooler component because the associated spectral lines dominate the low-$J$ lines that tend to have large opacities. The properties of the hotter component are hard to constrain even in the linear low-opacity part. As for examples shown in Figure~\ref{fig:sum-cog-lower}e--h, where one component dominates in the temperature as well as the column density, the convergence at one end is even easier to identify, but the distortion is also noticeable, depending on how the less dominant component distort the sum of the curve of growth. In short, constraining the properties of two unresolved components via a curve of growth is challenging, but Figure~\ref{fig:sum-cog-upper} and \ref{fig:sum-cog-lower} may provide hints to identify the second component.

\subsection{Dust Mixing with the Absorbing Gas}

While an isothermal slab model of dust-free gas was often considered as the absorbing component against a MIR continuum background, absorption in the continuum may happen when the gas is mixed with the dust that emits the continuum. For such a thermal continuum, to have the line intensity lower than the nearby continuum, the temperature is required to decrease toward the outer regions as the optical depth becomes smaller, analogous to the situation in a stellar photosphere~\cite [][Ch 10]{mihalas78}. Such an origin of the absorption line has been proposed in massive protostellar systems: the lines originate from the surface of the disk surrounding the protostar and the disk has an outward-decreasing temperature gradient~\citep{barr20}. {As we will show below, different parameters are adopted to describe the disk model (for example, the fractional coverage $f_c$ is no longer needed), and therefore one needs to interpret the spectrum observed under insufficient resolving power from a different perspective.}

\subsubsection{Line Transfer in the Photosphere Model}

In the photosphere model, the total opacity $k_\nu$ is contributed by both the continuum ($k_c$) and the line ($k_l\phi_\nu$), 
\begin{equation}
    k_\nu = k_c + k_l\phi_\nu.
\end{equation}
Assuming that the Planck function is linear in optical depth\footnote{This linear relation comes from the first-order Taylor expansion when the optical depth is in order of unity.}, {due to the dust temperature increasing with depth and because the Planck function depends solely on temperature,}
\begin{equation}
    B_\nu(\tau_c) = a + b\tau_c = a + \frac{b \tau_\nu}{1+\beta_\nu},
\end{equation}
where
\begin{equation}
    \beta_\nu = \frac{k_l}{k_c}\phi_\nu = \beta_0\phi_\nu,
\end{equation}
and $\beta_0$ is the line opacity relative to the continuum opacity at the line center. For a thermal continuum, the residual flux is~\citep[][eq 10-15]{mihalas78}:
\begin{equation}\label{eq:rv}
    R_\nu = \frac{I_\nu}{I_c} = \left[\frac{\sqrt{3\epsilon_\nu}a + b/(1+\beta_\nu)}{\sqrt{3}a+b} \right] \frac{2}{1+\sqrt{\epsilon_\nu}},
\end{equation}
where 
\begin{equation}
    \epsilon_\nu = \frac{1 + \epsilon \beta_\nu}{1+\beta_\nu}.
\end{equation}
The parameter $\epsilon$ is in the range of 0 to 1, corresponding to pure line scattering for $\epsilon=0$ and pure line absorption for $\epsilon=1$.

In the photosphere model, the equivalent width, $W_\nu$ and $\beta_0$ also form a curve of growth. For pure absorption line with $\epsilon=1$,  
 \begin{equation}\label{eq: mihalas}
  \begin{aligned}
W_\nu &= \int_0^{+\infty} (1-R_\nu) d\nu \\
     &= 2\Delta \nu_D A_0\int_0^{+\infty} \frac{\beta_\nu}{1+\beta_\nu}dv. 
 \end{aligned}
\end{equation}
where $\Delta\nu_D$ is the Doppler width of the line. The parameter $A_0$ is the absorption depth at the line center. It is dependent on the gradient of the Planck function, $a/b$, which ranges from \mbox{$\sim$ 0.5--0.9} from 900 to 100 K \citep[see Appendix~A in][]{barr20}. 

Equation~\ref{eq: mihalas} consists of three parts of the curve of growth: first, $W_\nu/\Delta \nu_D\sim \beta_0$ for small $\beta_0$, the linear part; then it grows slowly with $\sim\sqrt{\textrm{ln}\beta_0}$ in the logarithmic part. Finally $W_\nu/\Delta \nu_D\sim\sqrt{\beta_0}$ in the square root part as $\beta_0$ increases. We do not consider the square root part below for water lines due to the small damping factor. For a pure scattering line with $\epsilon=0$, the position of the logarithmic part differs because its line center intensity approaches 0 (versus $1-A_0$ for $\epsilon=1$), but the two curves of growth still share a similar shape because of the Voigt profile residing in $\beta_\nu$. 

\subsubsection{Comparison with the Slab Model}

The parameter $\beta_0$ in the photosphere model is analogous to $\tau_p$ in the slab model. Specifically, in the linear part of the curve of growth, for the slab model, there is~\citep[][eq 9-11]{draine11}
\begin{equation}
    \frac{W_\nu}{\Delta\nu_D}=\sqrt{\pi}f_c \tau_p\,\,\, (\tau_p \lesssim 1).
\end{equation}
For the photosphere model, there is~\citep[][eq 10-40]{mihalas78}
\begin{equation}
    \frac{W_\nu}{\Delta\nu_D}=\sqrt{\pi}A_0 \beta_0\,\,\, (\beta_0 \lesssim 1).
\end{equation}
The two curves of growth would overlap~\citep[Figure 2 in][]{li22} if one chooses $f_c$ equal to $A_0$ and $\tau_p$ equal to $\beta_0$. Since $\beta_0$ = $\tau_p/\tau_c$, this implies that we see the absorption line down to an order-unity optical depth of the continuum, which corresponds to the visible photosphere. 

In the logarithmic part of the curve of growth, for both models, the core gets saturated and the line profile becomes flat-topped. The depth of the line profile is fixed at $A_0$ or $f_c$, separately. $W_\nu/\Delta\nu_D$ has a weak dependence on $\beta_0$ but is more dependent on the line width. For the slab model, the line width corresponds to velocity $v_0$ (or equivalently, the frequency) where $\tau_p e^{-v_0^2} \approx 1$, since the additional absorption starts to remove the intensity through the line wing. For the photosphere model, $v_0$ occurs at $\beta_0 e^{-v_0^2} \approx 1$, corresponding to the depth where both the line and the continuum reach unit optical depth. More specifically, from a perspective of the actual physical depth, which increases from the disk surface toward the inner region, for $v < v_0$, the line reaches unit optical depth earlier than the continuum. However, the logarithmic parts of the two models do not overlap even if one adopts $f_c=A_0$ and $\tau_c=1$~\citep[Figure~2 in][]{li22}. For the slab model~\citep[][eq 9-17]{draine11},  
\begin{equation}
    \frac{W_\nu}{\Delta\nu_D} \sim 2f_c\sqrt{\textrm{ln}(\tau_p/\textrm{ln}2)}.
\end{equation}
For the photosphere model~\citep[][eq 10-43]{mihalas78}, 
\begin{equation}
    \frac{W_\nu}{\Delta\nu_D} \sim 2A_0\sqrt{\textrm{ln}\beta_0}.
\end{equation}

\subsubsection{Treatments to the Unresolved Spectrum}

The comparison above between the slab and the photosphere model reveals similar growth of lines as well as connections between physical parameters such as the pair $f_c$ and $A_0$, or the pair $\tau_p$ and $\beta_0$. Therefore, a similar degeneracy problem seems to exist in constraining the line absorption depth $A_0$ and the abundance, $N_\textrm{tot}/N_H$, of the absorbing gas that is mixed with the dust continuum from an unresolved spectrum. However, $A_0$ is dependent on the gradient of the Planck function ($a/b$, or intrinsically, the temperature $T_0$ at optical depth of unity) and is therefore not a free parameter as $f_c$ is. We note that the degeneracy problem would still exist because $A_0$ is defined specifically for $\epsilon=1$, and $\epsilon$ is unconstrained even if the lines are spectrally resolved~\citep[Table 3 in][]{li23}. As the determination of $\epsilon$ is beyond the scope of this paper, we suggest one designate the value of $\epsilon$ and apply the curve of growth analysis to derive the temperature and the abundance. As a reference, we also provide insufficiently resolved modeled spectra with arbitrary input parameters under the photosphere model on \href{https://mirasg.astro.umd.edu/}{https://mirasg.astro.umd.edu/}.

\section{Summary}

The rovibrational absorption spectrum of gaseous H$_2$O at MIR wavelengths consists of hundreds of lines and delivers valuable information about the innermost regions of dust-obscured continuum-emitting sources in systems like massive protostars or merging galaxies. Individual lines with velocity widths of several \kms, however, cannot be sufficiently resolved under JWST/MIRI's spectral resolution (FWHM of 85~\kms) and may even get blended with other lines. For insufficient resolution, the line profiles are smoothed to Gaussian. If the absorbing gas is assumed as a foreground cloud in front of the background continuum, information such as the partial coverage, $f_c$, and the intrinsic line width, $b$, are lost. 

The lost information about $f_c$ and $b$ hampers the spectrum analysis because the total column density $N_\textrm{tot}$ is degenerated with the two parameters. More specifically, isolated spectral lines with the same $N_\textrm{tot}f_c$ (i.e. $N_\textrm{beam}$) and the same $bf_c$ have exactly the same line profile. To break up the degeneracy, we suggest firstly constraining $N_\textrm{beam}$ from optically thin lines and then using the overlapped regions of optically thick lines to distinguish $b$ and $f_c$. While this is a ``standard" method for the analysis of the spectra, it requires a high signal-to-noise level in measuring the overlapped regions of optically thick lines. {Adopting $b$ derived from high spectral resolution spectroscopy in ground-based telescopes in certain spectral windows, when applicable, would simplify this process}.

We emphasize in this paper that blindly adopting a $\chi^2$ fitting between the modeled spectrum, which can be generated in the online tool we developed: \href{https://mirasg.astro.umd.edu/}{https://mirasg.astro.umd.edu/}, and the observed one may lead to a significant uncertainty on the derived column density for such insufficiently resolved spectra. This is because optically thin lines are under-weighted due to (1) their reduced absorption intensities after the convolution with the instrumental profile, and (2) the relatively lower signal-to-noise level compared to lines that are optically thick and strong. An improper estimation of $N_\textrm{beam}$ thus leads to further problems in estimating other parameters.

We provide both qualitative and quantitative recipes to provide feasible and quick guidance to the spectra analysis. Identifying representative spectral features helps to roughly constrain the lower limit of the temperature, and it is not very sensitive to the column density. Such a qualitative comparison may specifically stand out when the baseline subtraction is doubtful or when the noise level is not great. Cross-checking these different features may improve the constraints further. We encourage readers to compare the observed spectra to the models we present in Appendix~\ref{app:1}, or to compare with spectra one can generate in our online spectrum generator for the first step of the determination of the parameter space.

The quantitative recipes aim to provide a balanced usage of the optically thin and thick lines. One recipe uses pairs of isolated lines that share the same lower energy level. It provides grids of parameter space that corresponds to different ratios of these line pairs and can be used to specifically derive the line opacity information. This method is advantageous in that $f_c$ is a cancelled-out parameter in generating the grids.  The other recipe is based on the ``standard" method described above. If the opacities of lines are known from the previous recipe, one can use the curve of analysis firstly on optically thin lines to derive $N_\textrm{beam}$ and then on optically thick lines to decompose $b$ and $f_c$. 

We note that the recipes are for spectra that consist of single velocity components and are under the slab model. Spectra with multiple components are more challenging to understand, although rules of thumb can be applied. If the absorbing gas is mixed with the dust, we apply the photosphere model to solve the problem.

{The authors appreciate the valuable suggestions from the reviewers that greatly improve the manuscript. J. Li thanks the assistance from Marc Pound in making the online spectrum generator tool available. J. Li thanks Andy Harris, Alberto Bolatto, Weizhe Liu, Peizhi Du, and Jiali Lu for insightful discussions.} The authors acknowledge the support for the EXES Survey of the Molecular Inventory of Hot Cores (SOFIA \#08-0136) at the University of Maryland from NASA (NNA17BF53C) Cycle Eight GO Proposal for the Stratospheric Observatory for Infrared Astronomy (SOFIA) project issued by USRA.

\software{Astropy \citep{astropy13, astropy18}, Flask~\citep{grinberg18}, Matplotlib \citep{hunter07}, NumPy \citep{harris20}, SciPy \citep{virtanen20}.}

\appendix

\section{Curve-of-Growth Analysis on Optically Thin and Thick Lines}\label{app:0}


{To illustrate the importance of analyzing the weak, optically thin lines in addition to the strong saturated lines, we discuss here what would be the results of the curve-of-growth analysis if only optically thin or only optically thick lines, as shown in Figure~\ref{fig:app-1}, were used.}

{In Figure~\ref{fig:app-2}, the positions of optically thin lines and thick lines are plotted separately on the curve of growth. For optically thin lines,
the value of the fractional coverage, $f_c$, does not influence the equivalent widths (Figure~\ref{fig:app-2}b) if $N_\textrm{beam}$ stays the same.
Thus, optically thin lines constrain $N_\textrm{beam}$ very well.  But $N_\textrm{tot}$ scales inversely with $f_c$ ($N_\textrm{tot}$=$N_\textrm{beam}/f_c$), e.g., it changes by a factor of 3.5 for $f_c$ in the range of 0.2-0.7. In contrast, for optically thick lines the equivalent width of each line depends strongly on $f_c$ (Figure~\ref{fig:app-2}d) at constant $N_\textrm{beam}$. And, as shown in Figure~\ref{fig:app-2}f, for the same equivalent width, $N_\textrm{tot}$ differs by two orders of magnitude for $f_c$ in the range of 0.2-0.7 in the optically thick case.}

{Next, we perform a $\chi^2_\nu$ analysis to derive physical parameters from spectra in the presence of observational noise, i.e., we calculate}

\begin{equation}
    \chi^2_\nu = \frac{1}{N-2} \sum_i \frac{(I_i - M_i)^2}{\sigma_i^2},
\end{equation}
{where $N$ represents the channel number of the spectral regions to be compared, $I_i$ is the intensity of the observed spectrum under a combination of parameters $N_\textrm{beam}$, $f_c$, and $b$, and $M_i$ is the intensity of spectrum model (a) in Figure~\ref{fig:three-models} plus Gaussian distributed noise, $\sigma_i$. We select optically thick ($\tau_p>5$, or log$_{10}\tau_p>$0.7) and thin ($\tau_p<1$, or log$_{10}\tau_p<$0) lines to calculate $\chi^2_\nu$, separately, and each has a degree of freedom of 1034 and 1237 for assessing the values of $\chi^2_\nu$. According to \citet{press92}, the probability $Q$ that a value of $\chi^2$ as poor and poorer as the calculated one occur by chance is}

\begin{equation}
    Q = \texttt{gammq}\left( \frac{N-2}{2}, \frac{\chi^2}{2}\right),
\end{equation}
{where \texttt{gammaq} is the incomplete gamma function defined in \citet{press92}. Therefore, $Q$ greater than 0.1 indicates an acceptable model (or greater than 0.001 if the errors are moderately underestimated). We calculate the corresponding $\chi^2_\nu$ of 1.06 ($Q=0.1$) and 1.14 ($Q$=0.001) for the thick lines, and 1.05 ($Q=0.1$) and 1.13 ($Q$=0.001) for the thin lines to determine if the parameter space covers acceptable models.} 

{We present in Figure~\ref{fig:banana} the results of applying the $\chi^2_\nu$ analysis to optically thick and thin lines, separately. The results are consistent with our qualitative analysis above: for optically thick lines, a ``banana"-shape contour indicates that the acceptable combination of $N_\textrm{beam}$ and $f_c$ would result in a large range of $N_\textrm{tot}$. In contrast, the optically thin lines constrain $N_\textrm{beam}$ nicely. 
We note that this is not the case for the first panel of Figure~\ref{fig:banana}: the noise level is too high and therefore the weak, optically thin lines poorly constrain $N_\textrm{beam}$. We also emphasize that such results from the $\chi^2$ analysis are limited by the spectral region one chooses (thereby the degree of freedom), so as the contours used to determine the acceptable parameter space.}

\begin{figure}
    \centering
    \includegraphics[width=0.5\linewidth]{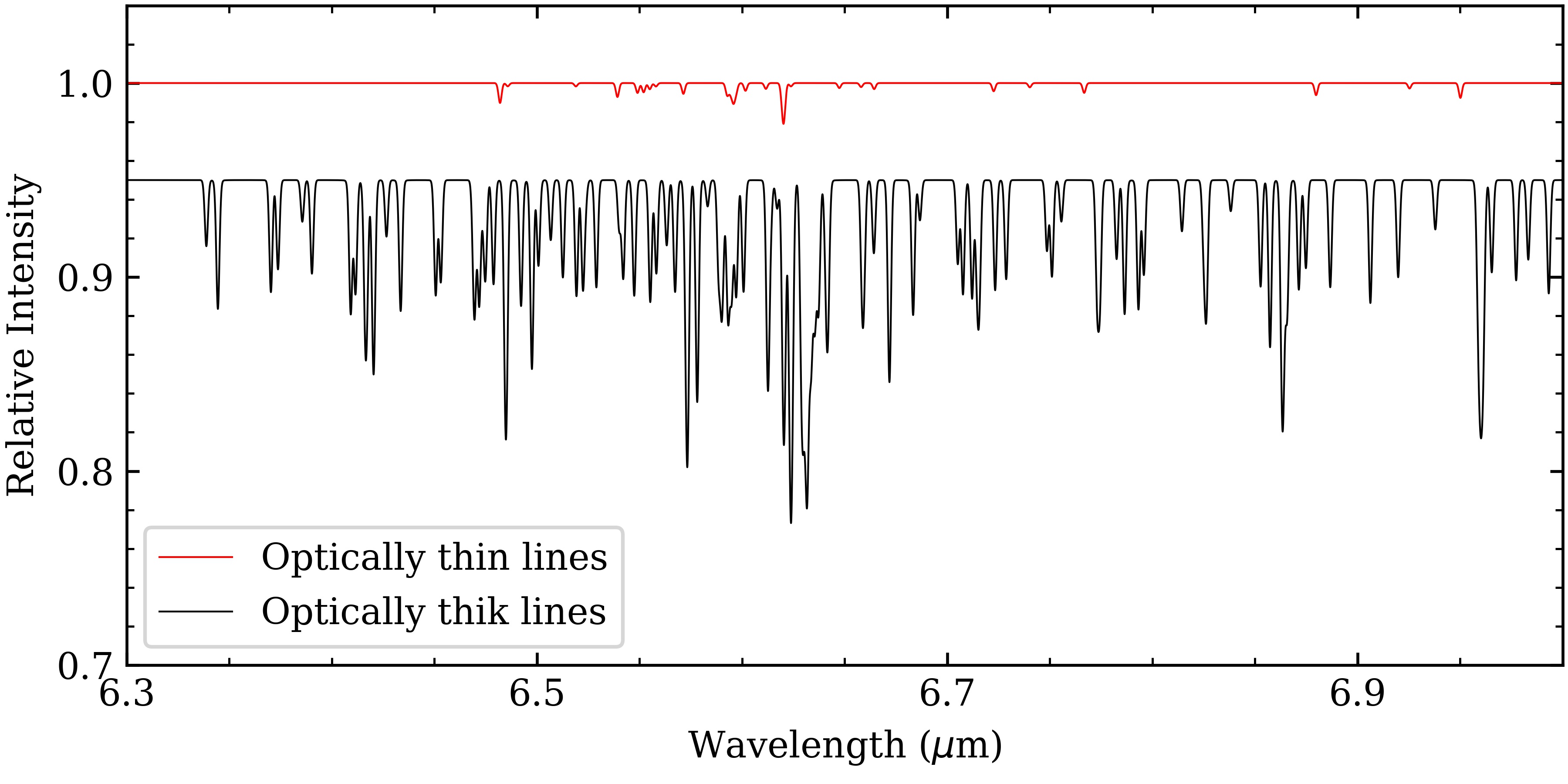}
    \caption{Optically thin (red) and thick (black) lines separated from part of the spectrum (a) in Figure~\ref{fig:three-models}.}
    \label{fig:app-1}
\end{figure}

\begin{figure*}
    \centering
    \includegraphics[width=\linewidth]{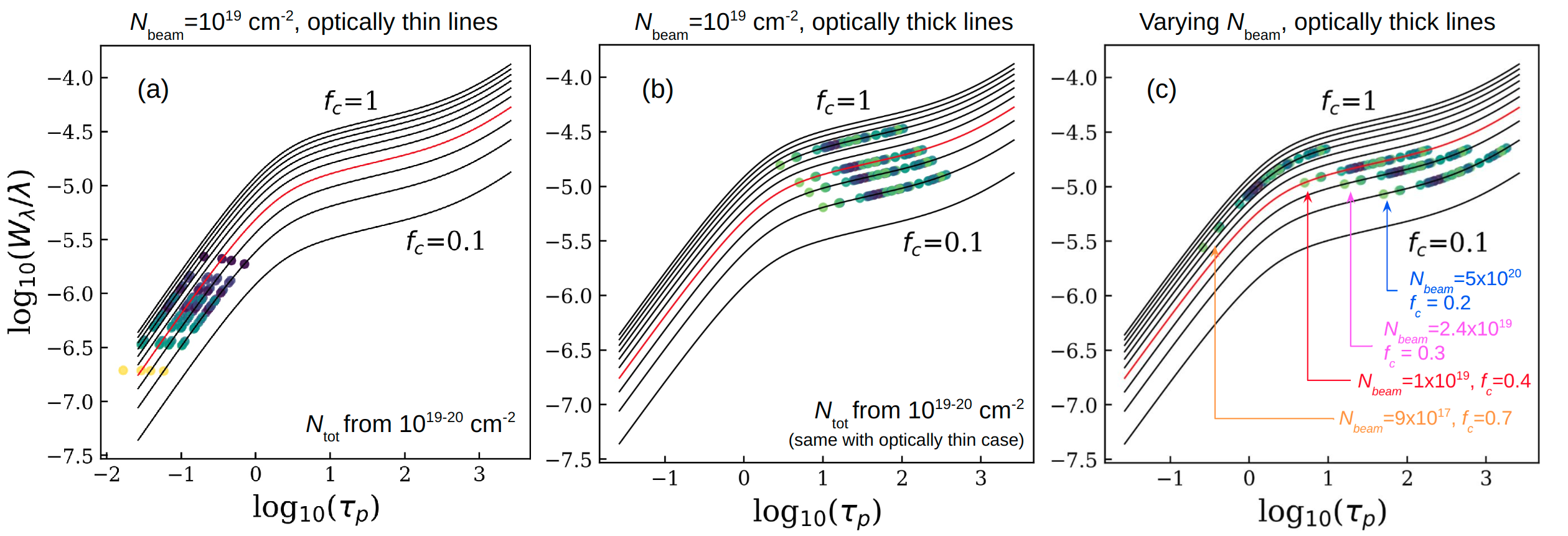}
    \caption{{Optically thin and thick spectral lines on the curve of growth. In all panels, the bullets represent specific spectral absorption lines. The red line represents the curve with the same physical parameters adopted from the spectrum in Figure~\ref{fig:three-models}a ($f_c$=0.4). The black lines represent curves of growth for a range of fractional coverage ($f_c$), increasing from 0.1, 0.2, ... to 1 in step of 0.1. Panels {(a) and (b)} illustrate how the locations of spectral lines on curves of growth change under different fractional coverage but with the same $N_\textrm{beam}$: panel~{(a)} shows that for optically thin lines, their measured equivalent widths do not change, so the column density $N_\textrm{beam}$ is well determined. This is consistent with the conclusion from Figure~\ref{fig:comp-dege}. In contrast, panel {(b)} indicates that the measured equivalent widths of optically thick lines change a lot, so $N_\textrm{beam}$ is not well determined unless $f_c$ is known (decoupled from $b$). Of course, changing the values of $N_\textrm{beam}$ can produce comparable equivalent widths, as is illustrated in panel {(c)}. Then, for optically thick lines,  the resulting $N_\textrm{tot}$ ($=N_\textrm{beam}/f_c$) can differ by two orders of magnitude, ranging from 2.5$\times10^{20}$ ($f_c$=0.2) to 1.2$\times10^{18}$~cm$^{-2}$ ($f_c$=0.7), consistent with the conclusions from Figure~\ref{fig:three-models}. But when only optically thin lines are fitted, the derived $N_\textrm{tot}$ differs by a factor of a few, depending on the range of $f_c$.}}
    \label{fig:app-2}    
\end{figure*}

\begin{figure*}
    \centering
    \includegraphics[width=\linewidth]{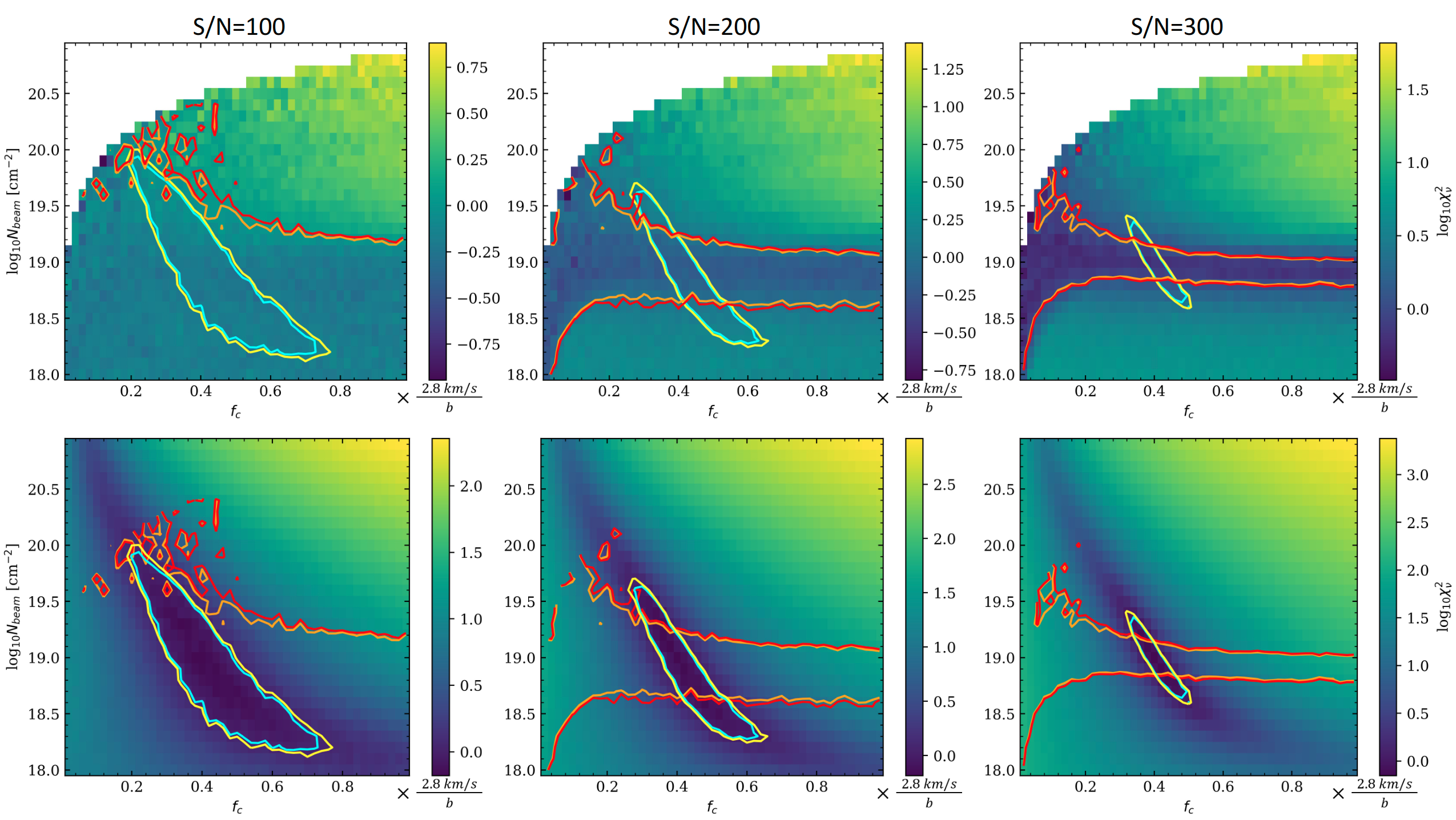}
    \caption{{$\chi^2_\nu$ analysis of the parameters space based only on the optically thin (upper panels) and thick lines (lower panels). For better comparison, the contours of the upper (lower) panels are copied onto the lower (upper) panels. Specifically, optically thick lines ($\tau_p>5$, or log$_{10}\tau_p>$0.7) and optically thin ($\tau_p<1$, or log$_{10}\tau_p<$0) lines are chosen from the 5 to 8~$\mu$m range, and each panel presents the calculated $\chi^2$ when adopting a noise level of 0.1, 0.05, 0.03 (or S/N=100, 200, 300) on a normalized spectrum. The cyan-yellow and orange-red contours indicate contours of $\chi_\nu^2$ corresponding to $Q$ values of 0.1 and 0.001 (see texts for the details). We note the scale of the $x$-axis, $f_c$, is scaled by the velocity width, as $b$ of 2.8~\kms\ is used in the presented models.}}
    \label{fig:banana}
\end{figure*}
\section{Graphic Collection of Models}\label{app:1}

\subsection{Overall Models}

{In this section, we present in Figures~\ref{fig:e1} to \ref{fig:e3} a list of modeled water spectra. The temperature ranges from 10 to 1000~K and $N_\textrm{tot}$ ranges from $10^{17}$ to 5$\times 10^{19}$~cm$^{-2}$. The line width $b$ is 20~\kms\ and cannot be resolved under MIRI's spectral resolution. We note that we have run models in widths of 2, 5, 10, and 20~\kms\ for which MIRI's spectral resolution is insufficient to resolve. Because their spectral profiles look similar but only differ in absorption depth, we only present models with 20~\kms\ in this paper. Specifically, $N_\textrm{tot}$ should only be considered as a lower limit here because we are presenting models with $f_c = 1$. Since the spectral lines are smoothed by MIRI's kernel, applying a different $f_c$ would only re-scale the absorption depth by $f_c$.}

\begin{figure*}[!t]
    \centering
    \includegraphics[width=\linewidth]{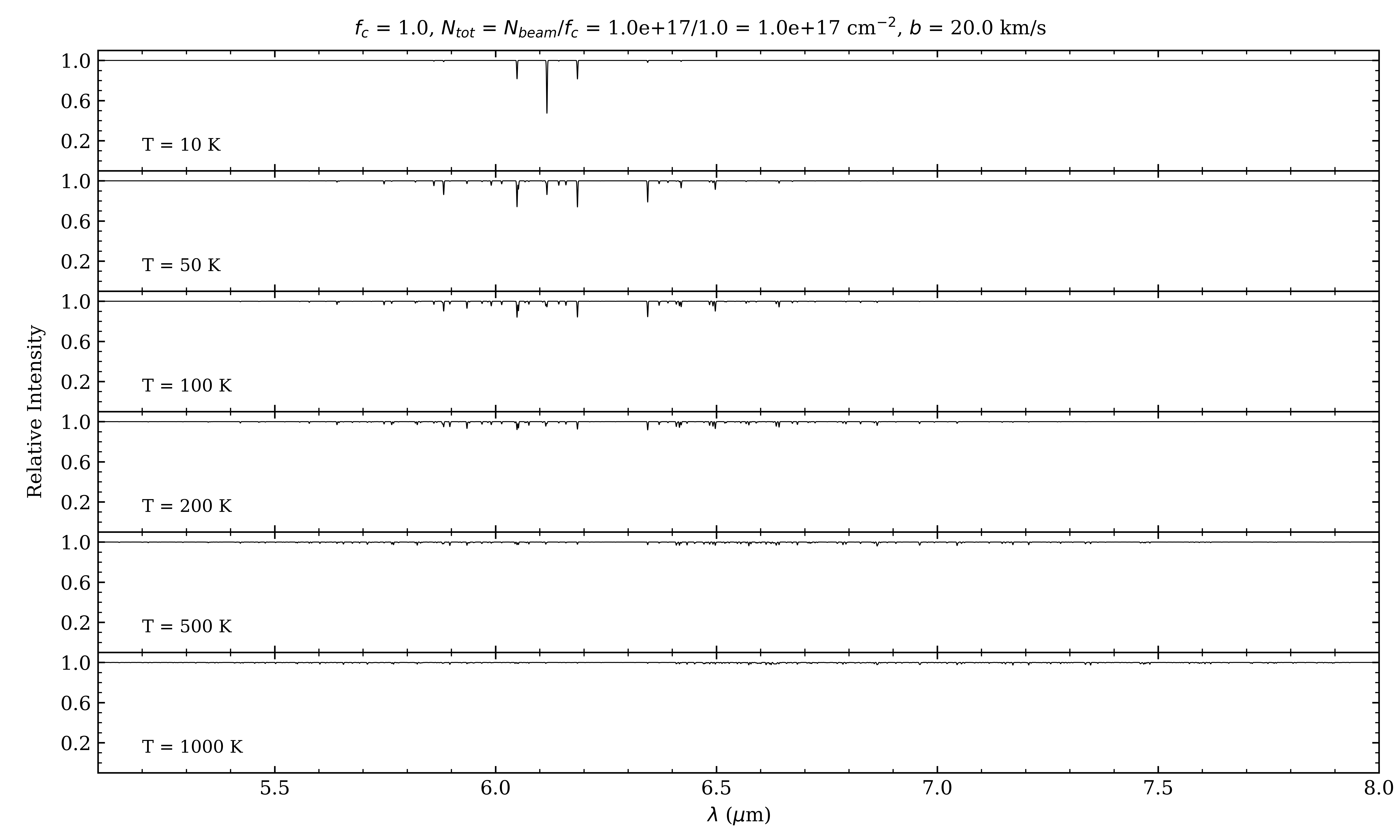}
    \includegraphics[width=\linewidth]{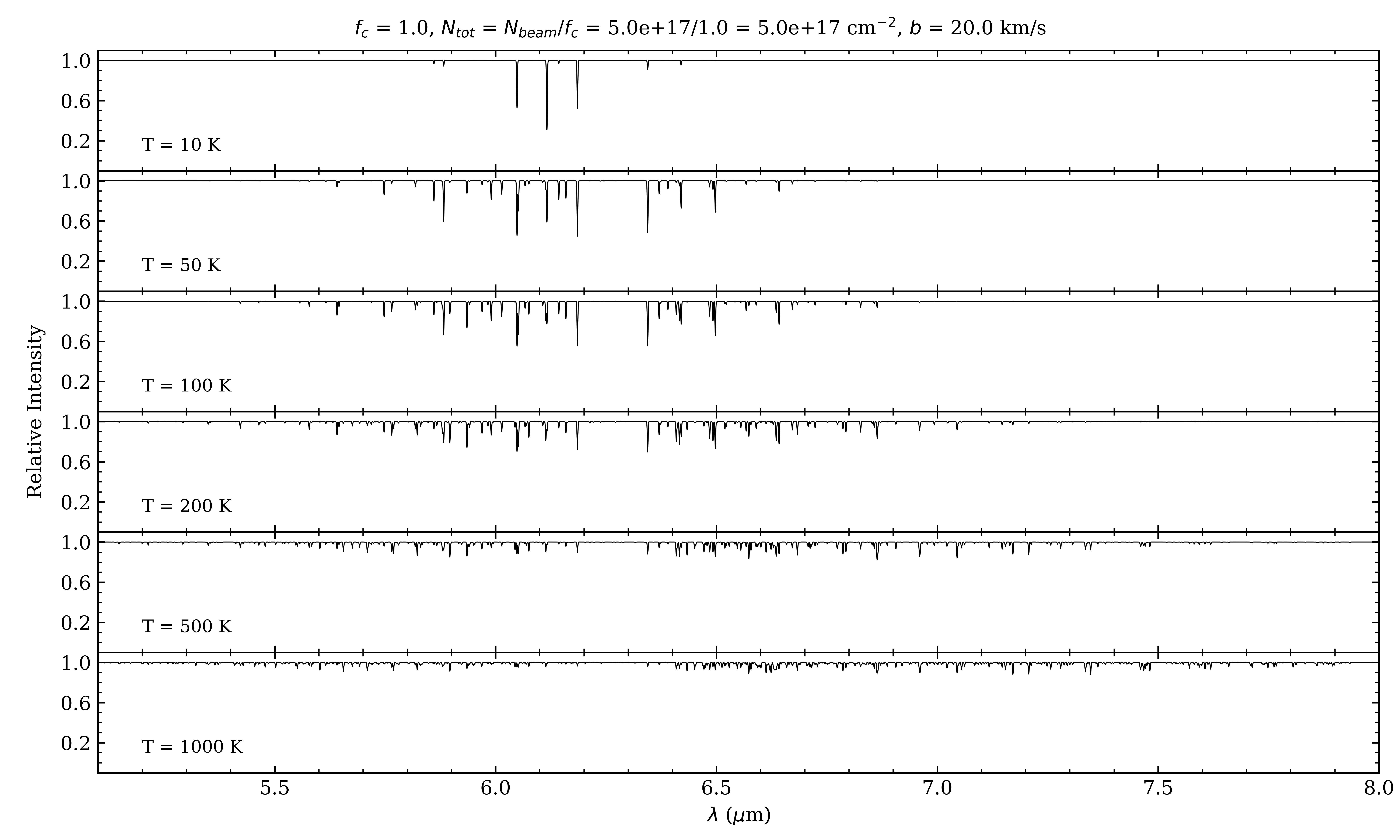}
    \caption{Modeled water rovibrational spectra similar to Figure~\ref{fig:example} with $N_\textrm{tot} = 1\times10^{17}, 5\times10^{17}$~cm$^{-2}$. }
    \label{fig:e1}
\end{figure*}

\begin{figure*}[!t]
    \centering
    \includegraphics[width=\linewidth]{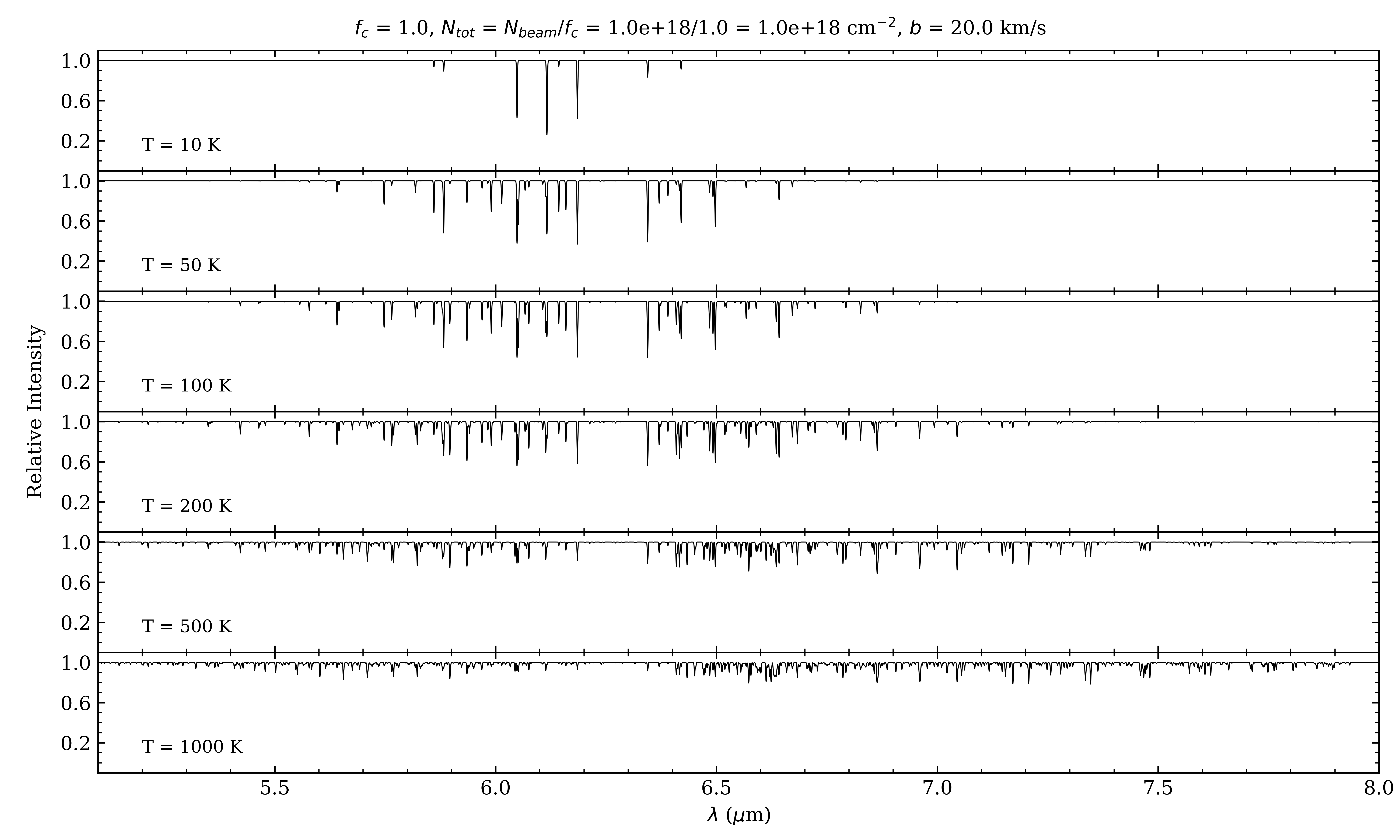}
    \includegraphics[width=\linewidth]{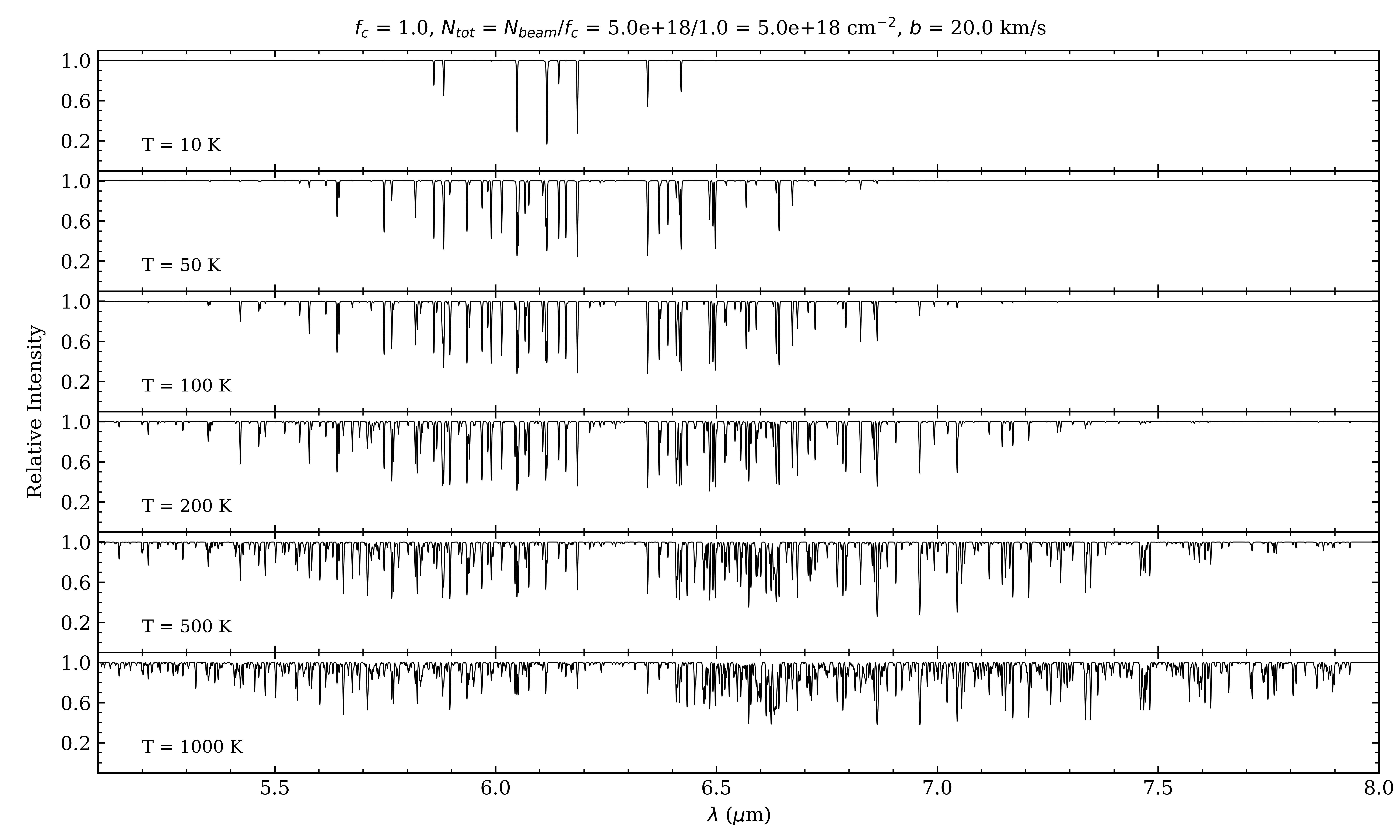}
    \caption{Modeled water rovibrational spectra similar to Figure~\ref{fig:example} with $N_\textrm{tot} = 1\times10^{18}, 5\times10^{18}$~cm$^{-2}$. }
    \label{fig:e2}
\end{figure*}

\begin{figure*}[!t]
    \centering
    \includegraphics[width=\linewidth]{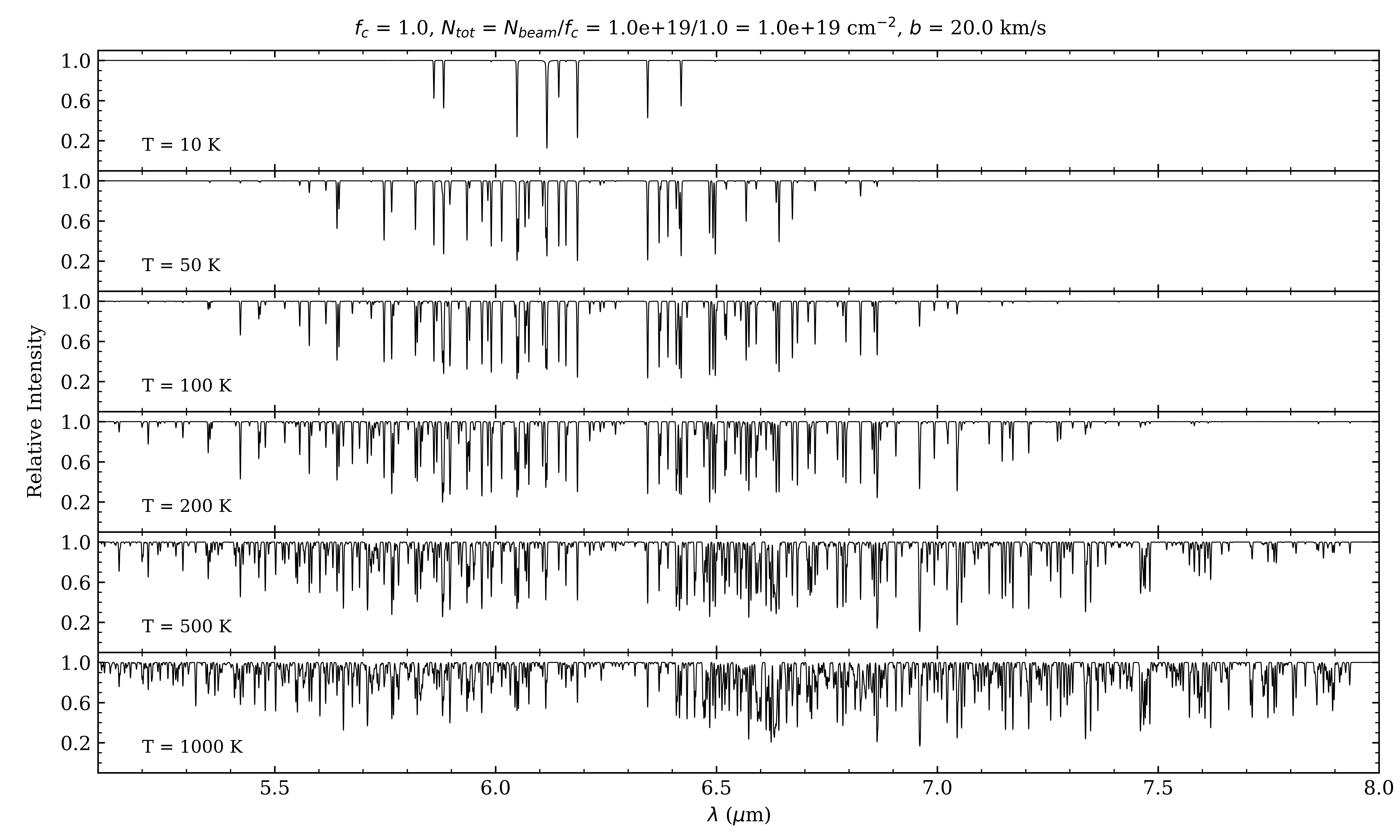}
    \includegraphics[width=\linewidth]{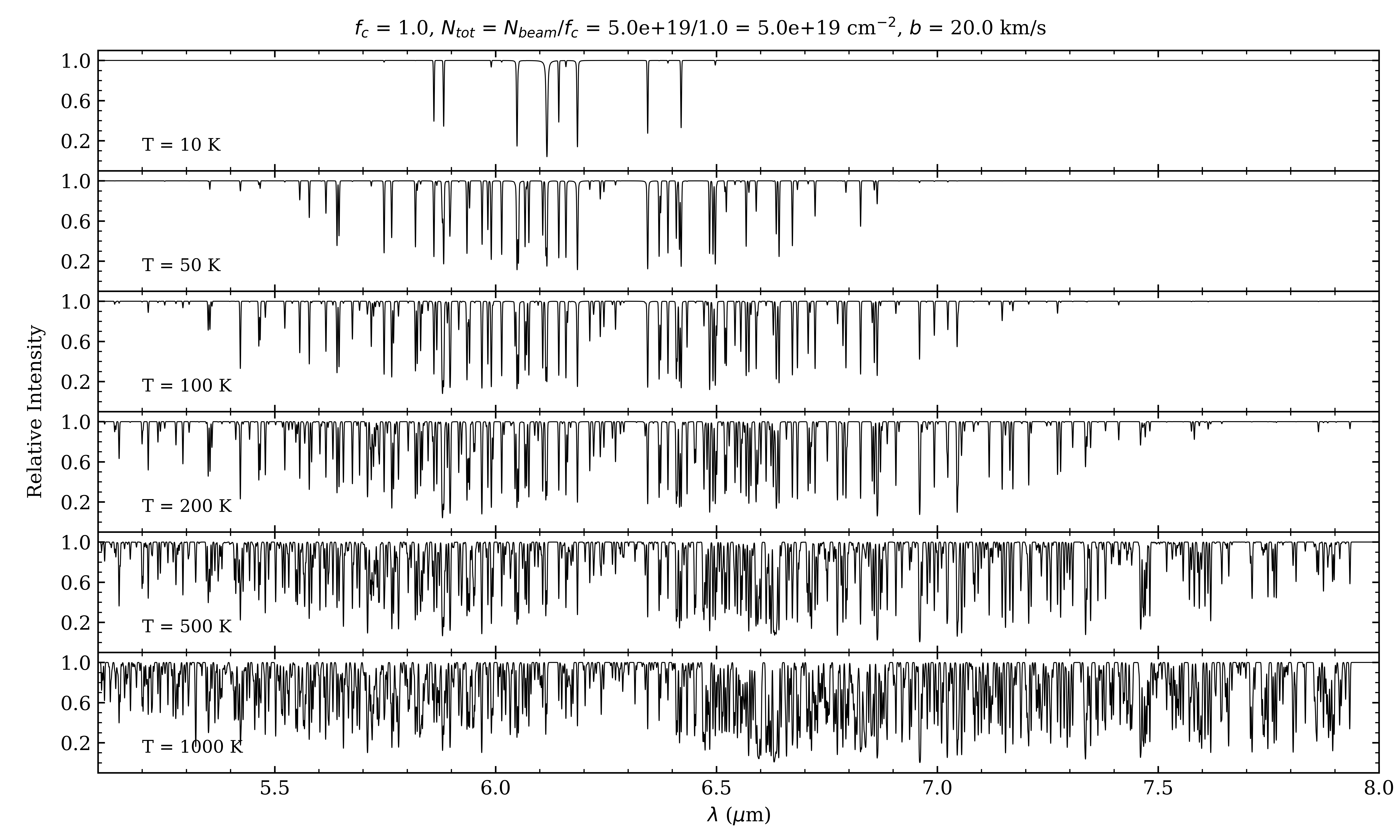}
    \caption{Modeled water rovibrational spectra similar to Figure~\ref{fig:example} with $N_\textrm{tot} = 1\times10^{19}, 5\times10^{19}$~cm$^{-2}$. }
    \label{fig:e3}
\end{figure*}

\subsection{Specific Features in the R-branch} \label{app:b-2}

{We present in Figure~\ref{fig:r-branch}a the detailed labels of the transitions involved in the backbone levels. For $J\geq10$, $J_{0, J}$--$(J-1)_{1, J-1}$ and $J_{1, J}$--$(J-1)_{0, J-1}$ overlap and blend to lines which define the envelope and form a wing-like profile as is seen in the $R$-branch of the rovibrational spectrum of a diatomic molecule.  The two transitions are separated at $J<10$ and the wing shape is thus broken at low $J$. The blended lines are not evenly spaced and the distance decreases from 0.055~\um\ ($J=10$) to 0.035~\um~($J=21$). Figure~\ref{fig:r-branch}b shows the $J_{0, J}$--$(J-1)_{1, J-1}$ and $J_{1, J}$--$(J-1)_{0, J-1}$ transitions under a range of column densities.}

\begin{figure*}[!t]
    \centering
    \includegraphics[width=0.48\linewidth]{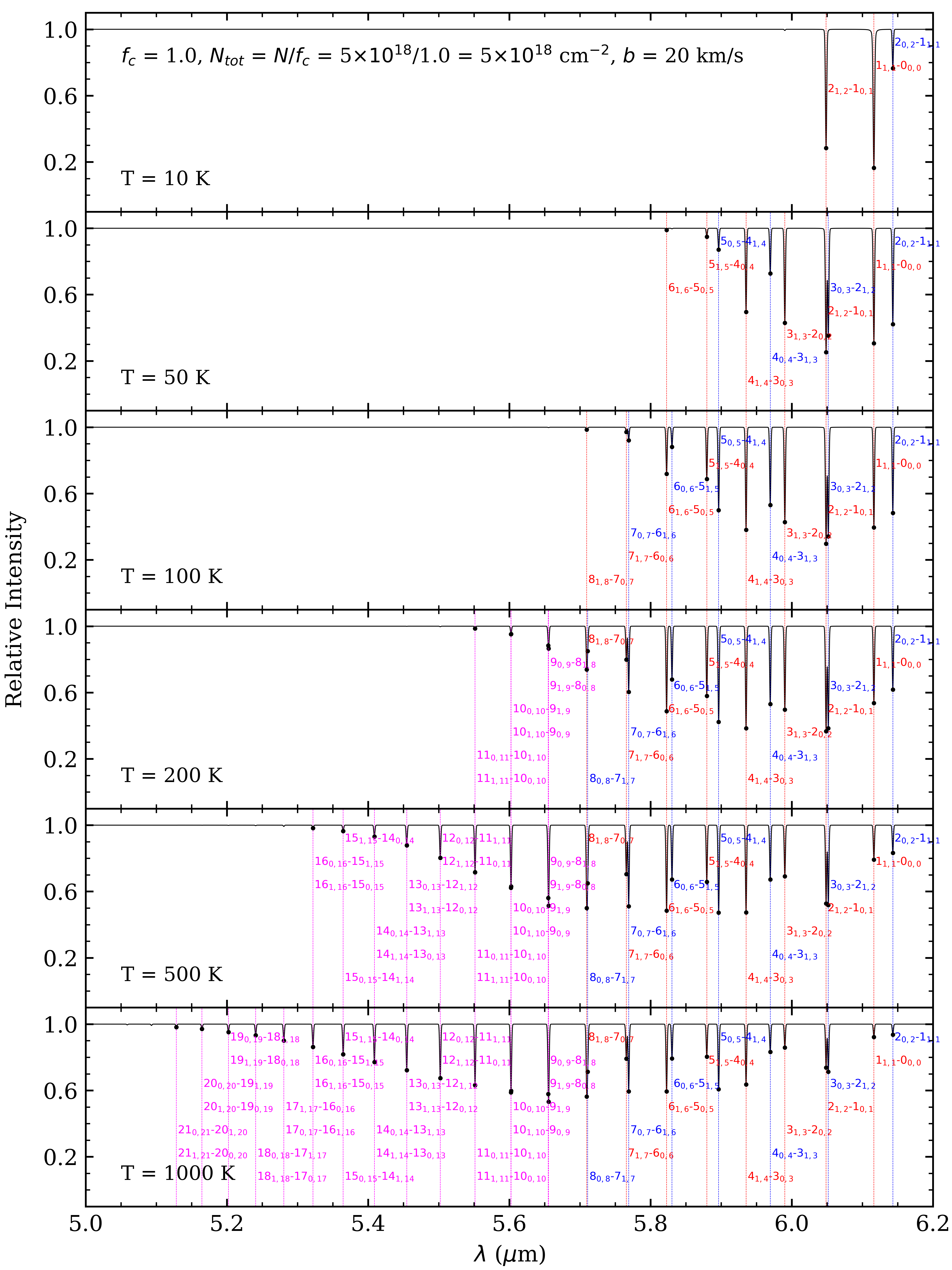}
    \includegraphics[width=0.48\linewidth]{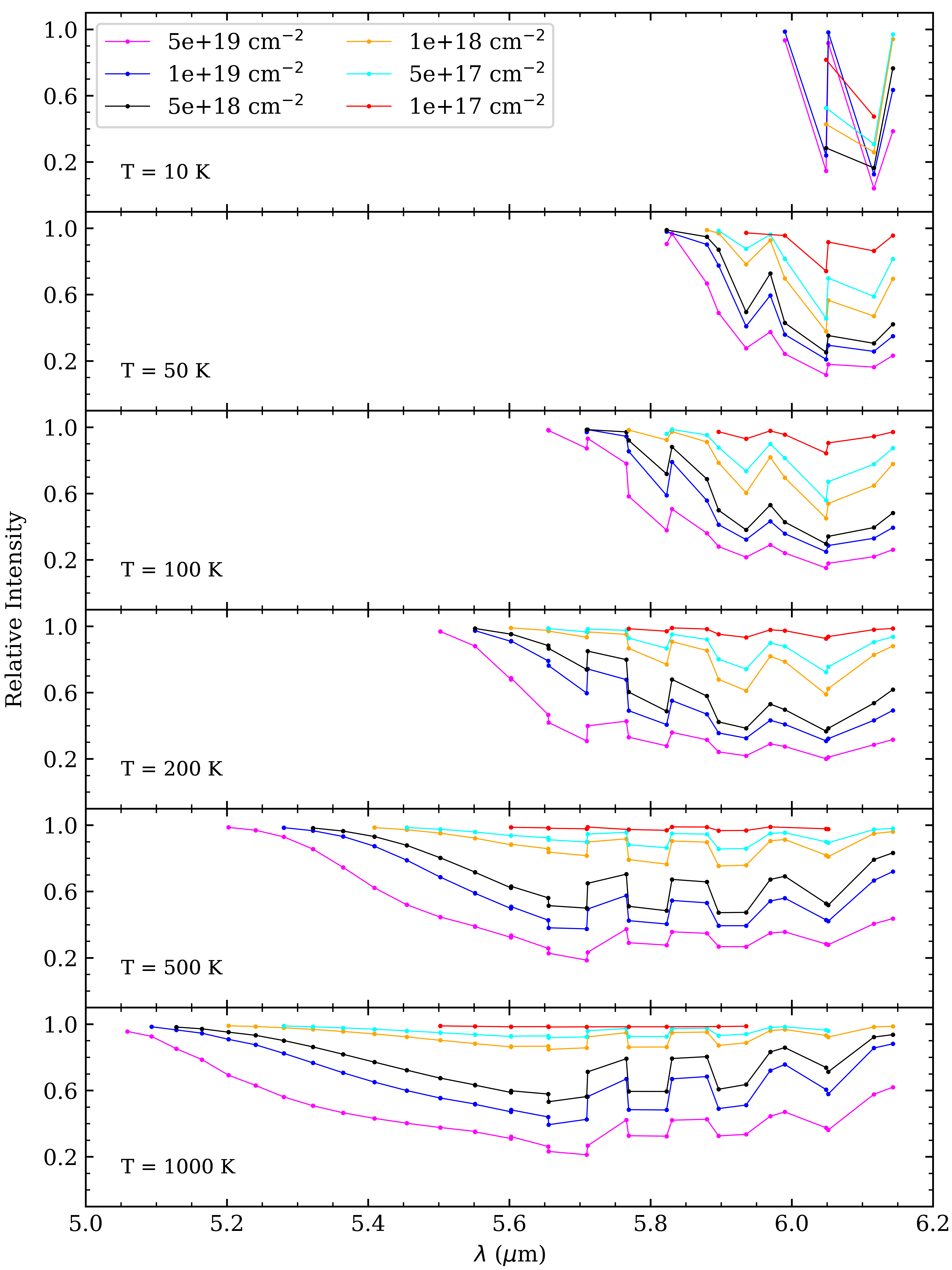}
    \caption{\textit{Left:} The $\nu_2$ band transitions between $J_{0, J}$--$(J-1)_{1, J-1}$ and $J_{1, J}$--$(J-1)_{0, J-1}$. $J_{0, J}$--$(J-1)_{1, J-1}$ and $J_{1, J}$--$(J-1)_{0, J-1}$ lines overlap when $J\geq$9 and are separated when when $J<$9. These transitions in the $R$-branch specifically consist of the envelope of the overall water spectral profile at $\gtrsim 300$~K. \textit{Right}: $J_{0, J}$--$(J-1)_{1, J-1}$ and $J_{1, J}$--$(J-1)_{0, J-1}$ transitions under different column density from $10^{17}$ to $5\times10^{19}$~cm$^{-2}$. Lines that are plotted are stronger than a detection limit of 1\%.}
    \label{fig:r-branch}
\end{figure*}

\section{Reference Grids for the Line Ratio Method}\label{app:grids}

{We present in this section the grids of line ratios of six pairs in different $N_\textrm{beam}, T_\textrm{ex}$, and velocity width ($b$=2.4~\kms). None of the pairs has a universal applicable parameter space. }

\begin{figure*}[!t]
    \centering
    \includegraphics[width=0.32\linewidth]{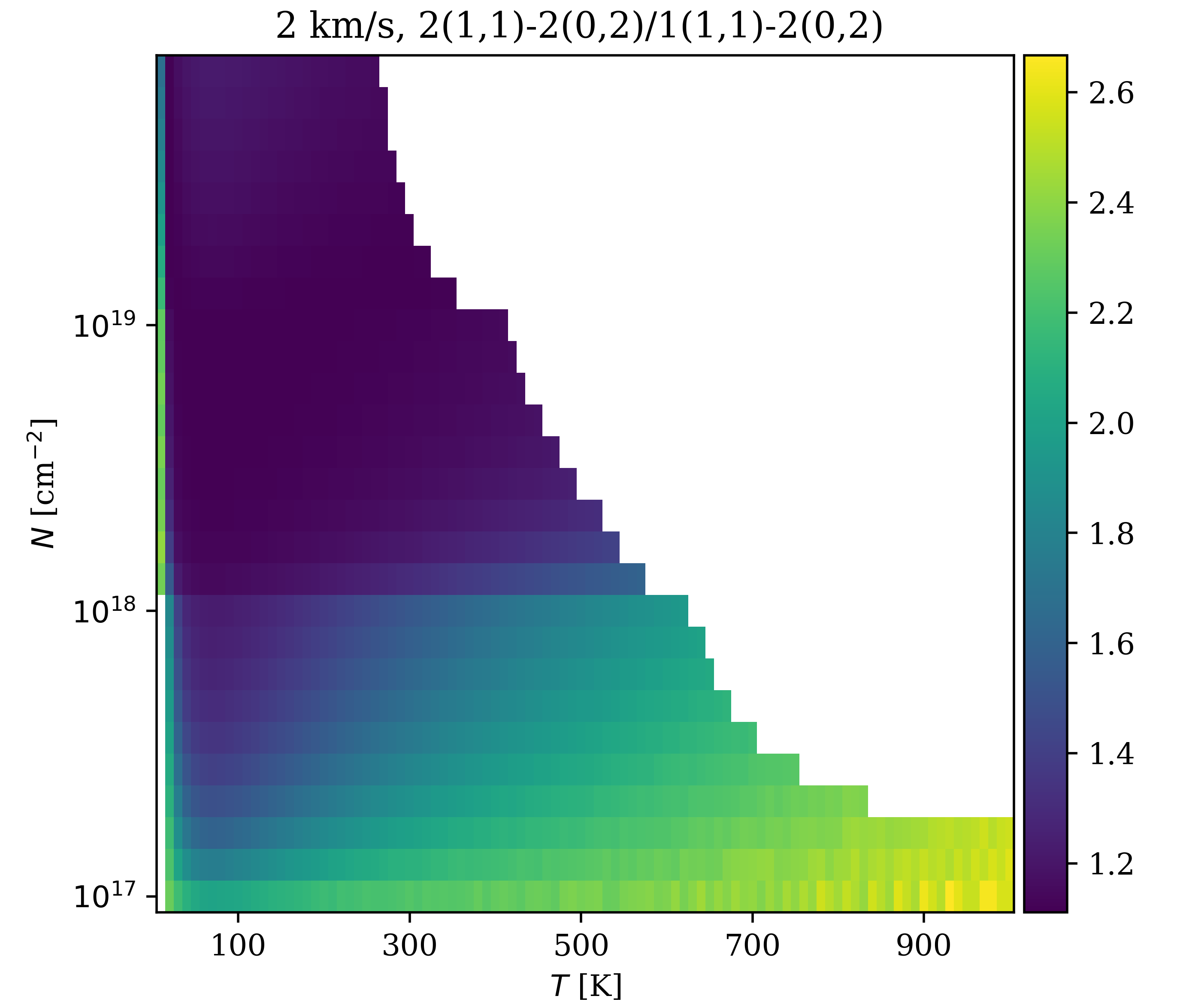}
    \includegraphics[width=0.32\linewidth]{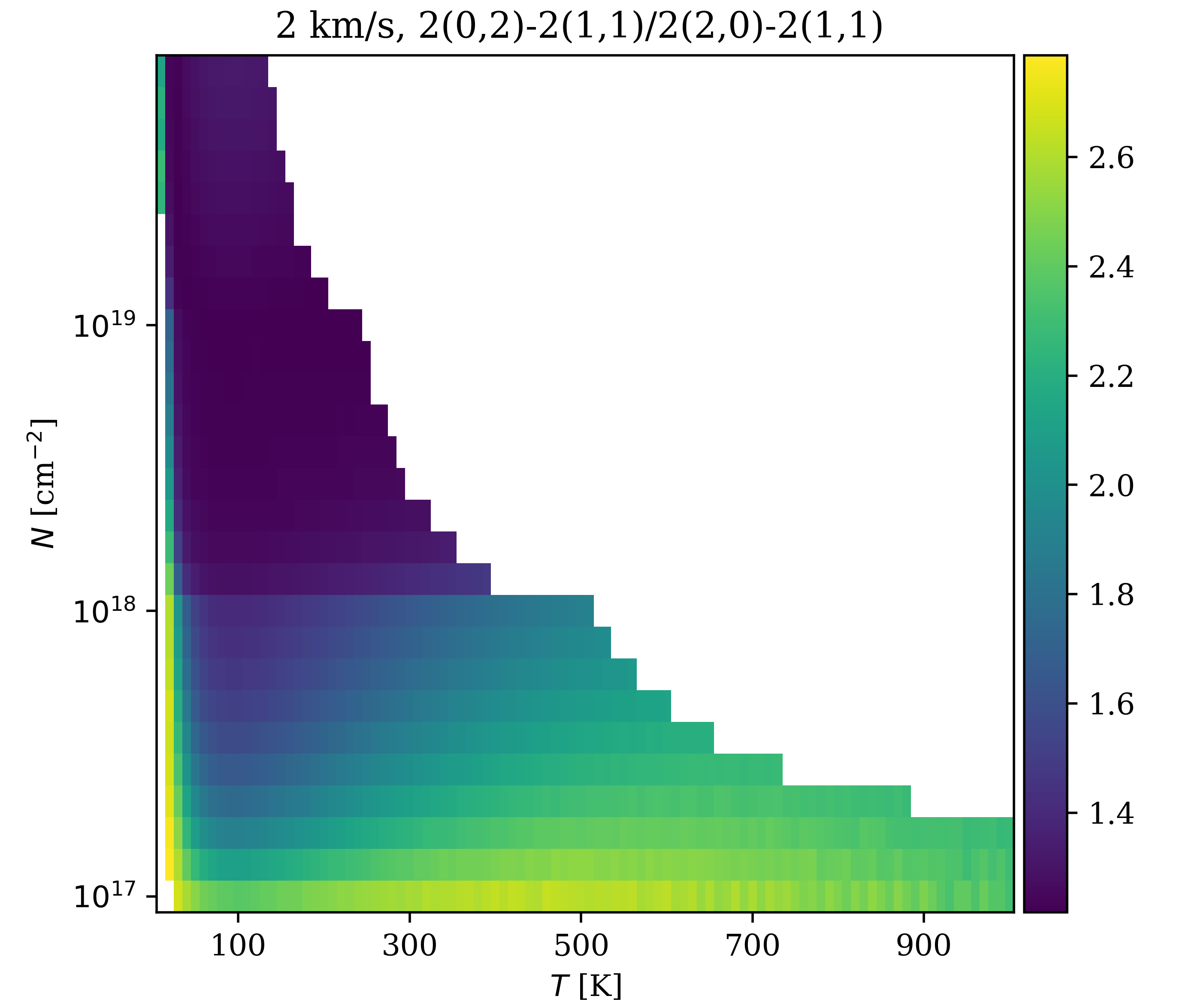}
    \includegraphics[width=0.32\linewidth]{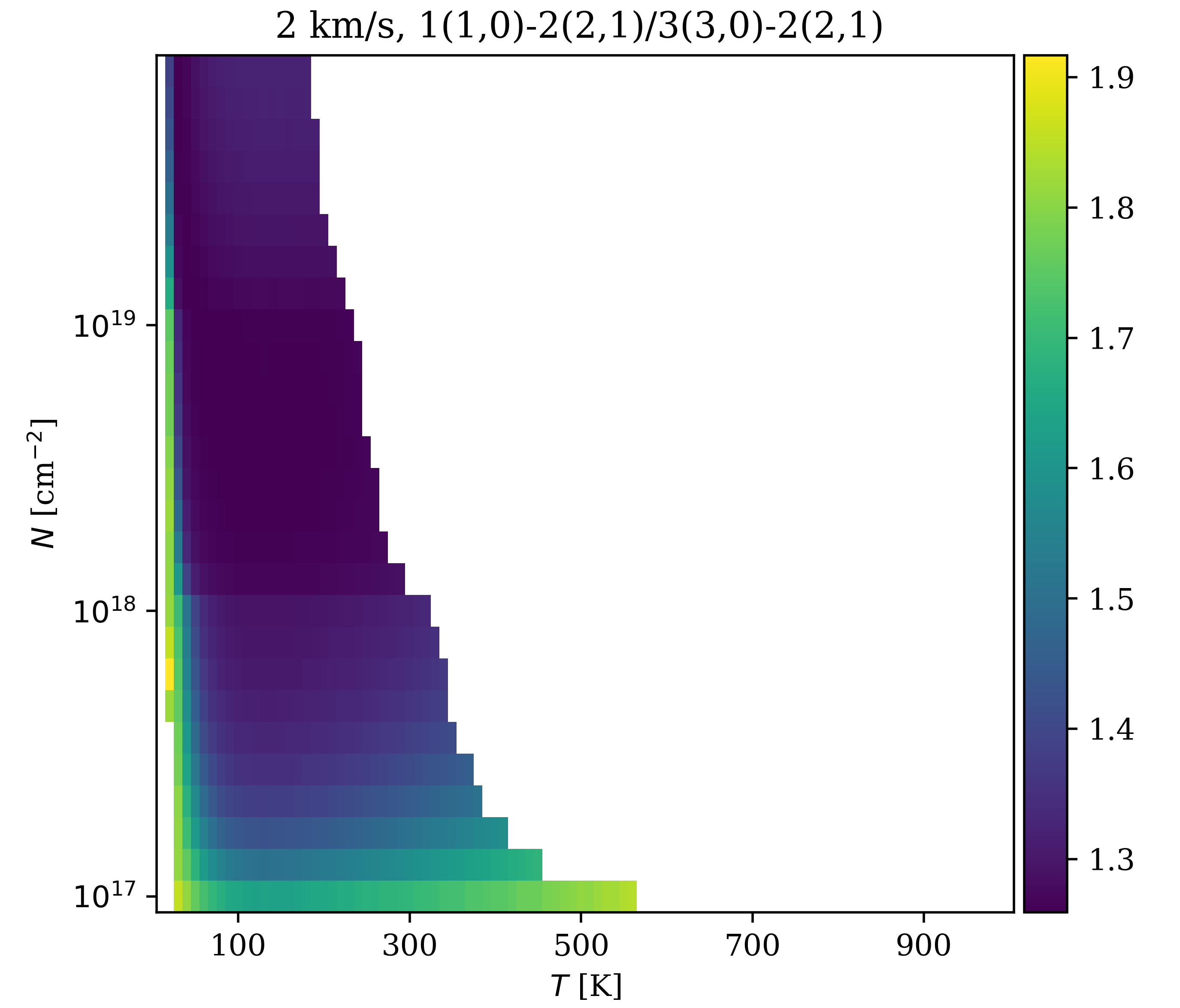}
    \includegraphics[width=0.32\linewidth]{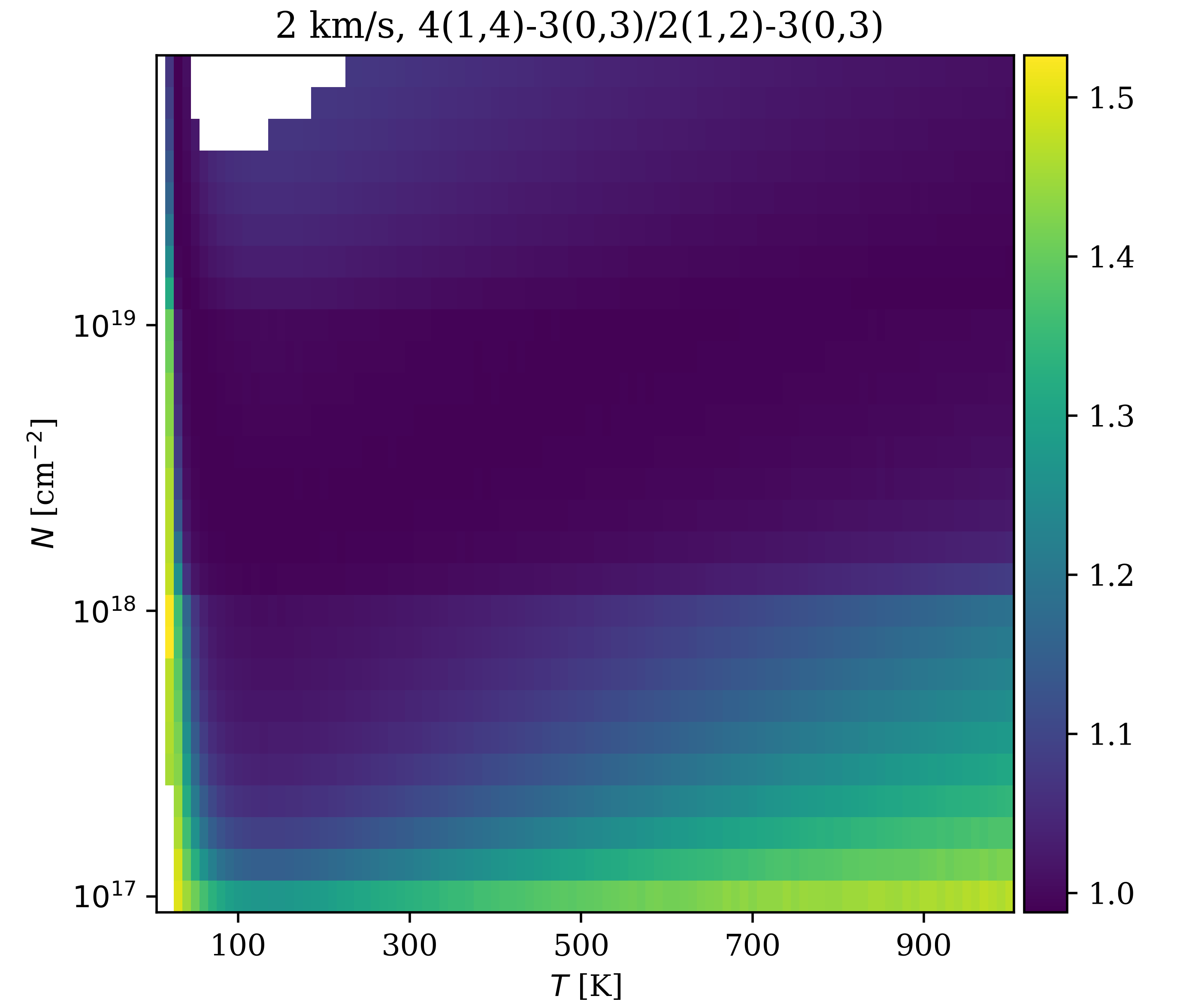}
    \includegraphics[width=0.32\linewidth]{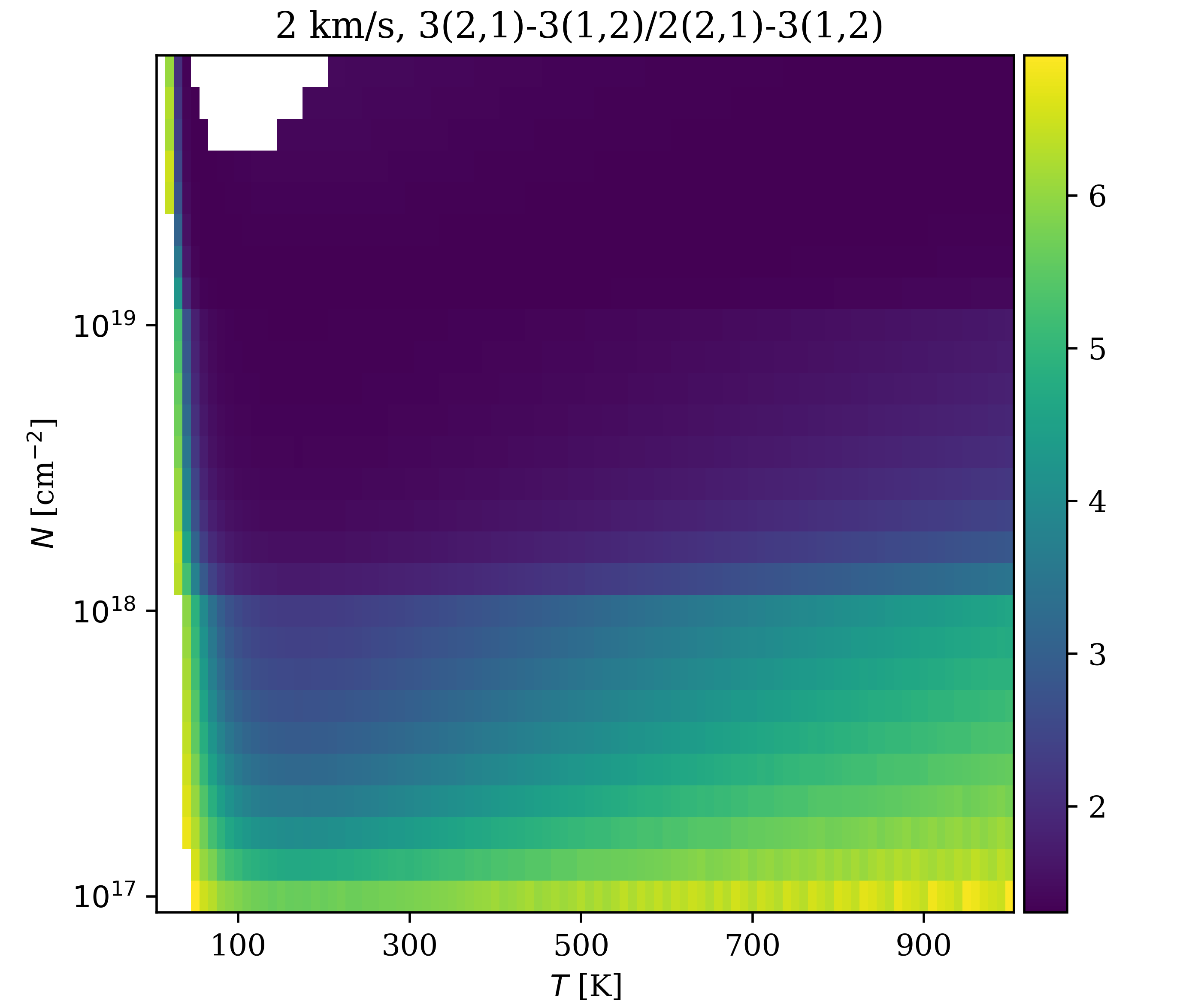}
    \includegraphics[width=0.32\linewidth]{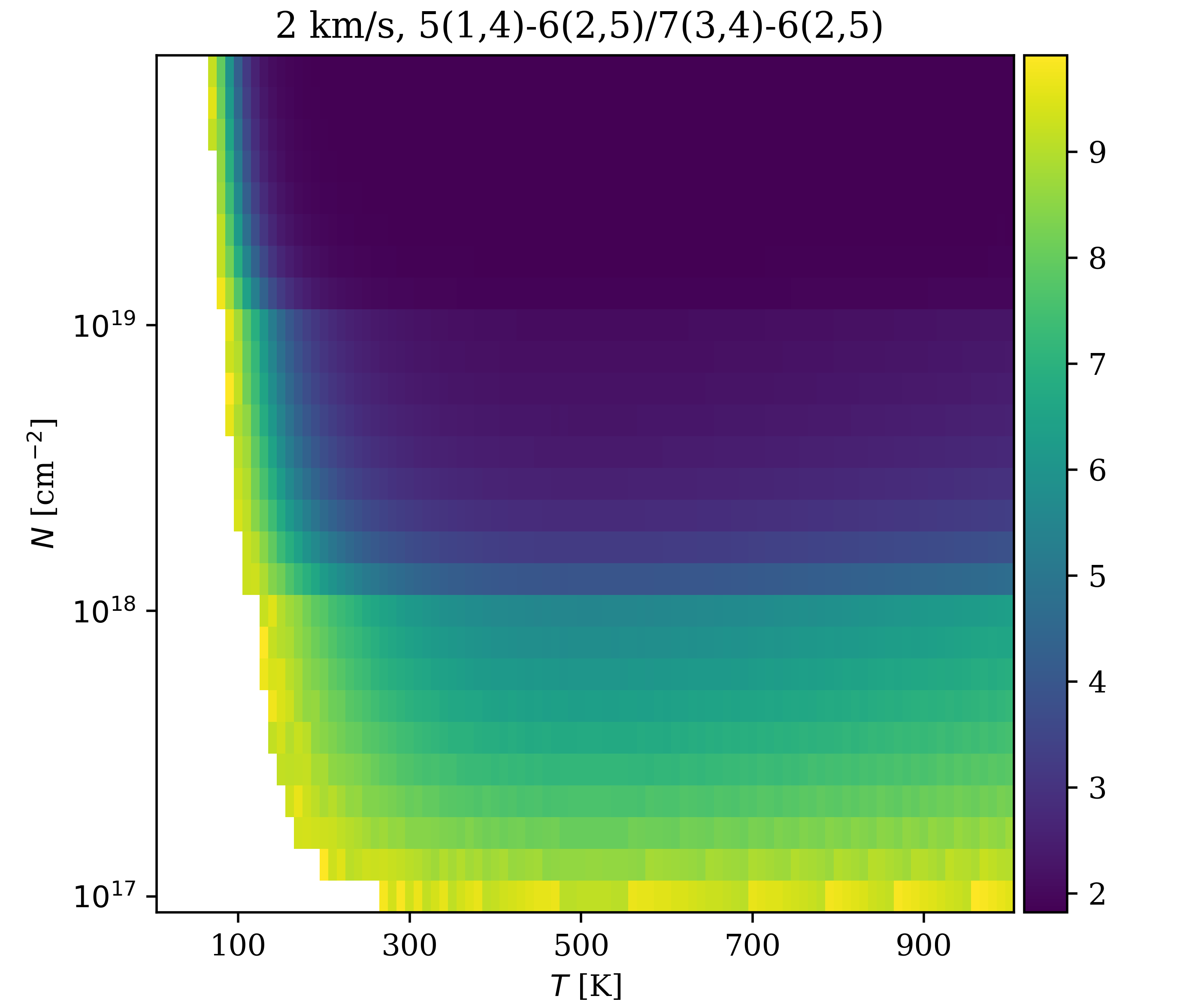}
    \caption{Reference grids of line ratios under ($N_\textrm{beam}$, \tex) with $\sigma_v=2$~\kms~($b$=2.8~\kms). The temperature axis is in an interval of 10~K, and the $N_\textrm{beam}$ axis is in an interval of 10$^{17}$ (from 10$^{17}$ to 10$^{18}$ cm$^{-2}$), 10$^{18}$ (from 10$^{18}$ to 10$^{19}$ cm$^{-2}$), and 10$^{19}$ (from 10$^{19}$ to 10$^{20}$ cm$^{-2}$).}
    \label{fig:grids-2kms}
\end{figure*}

\begin{figure*}[!t]
    \centering
    \includegraphics[width=0.32\linewidth]{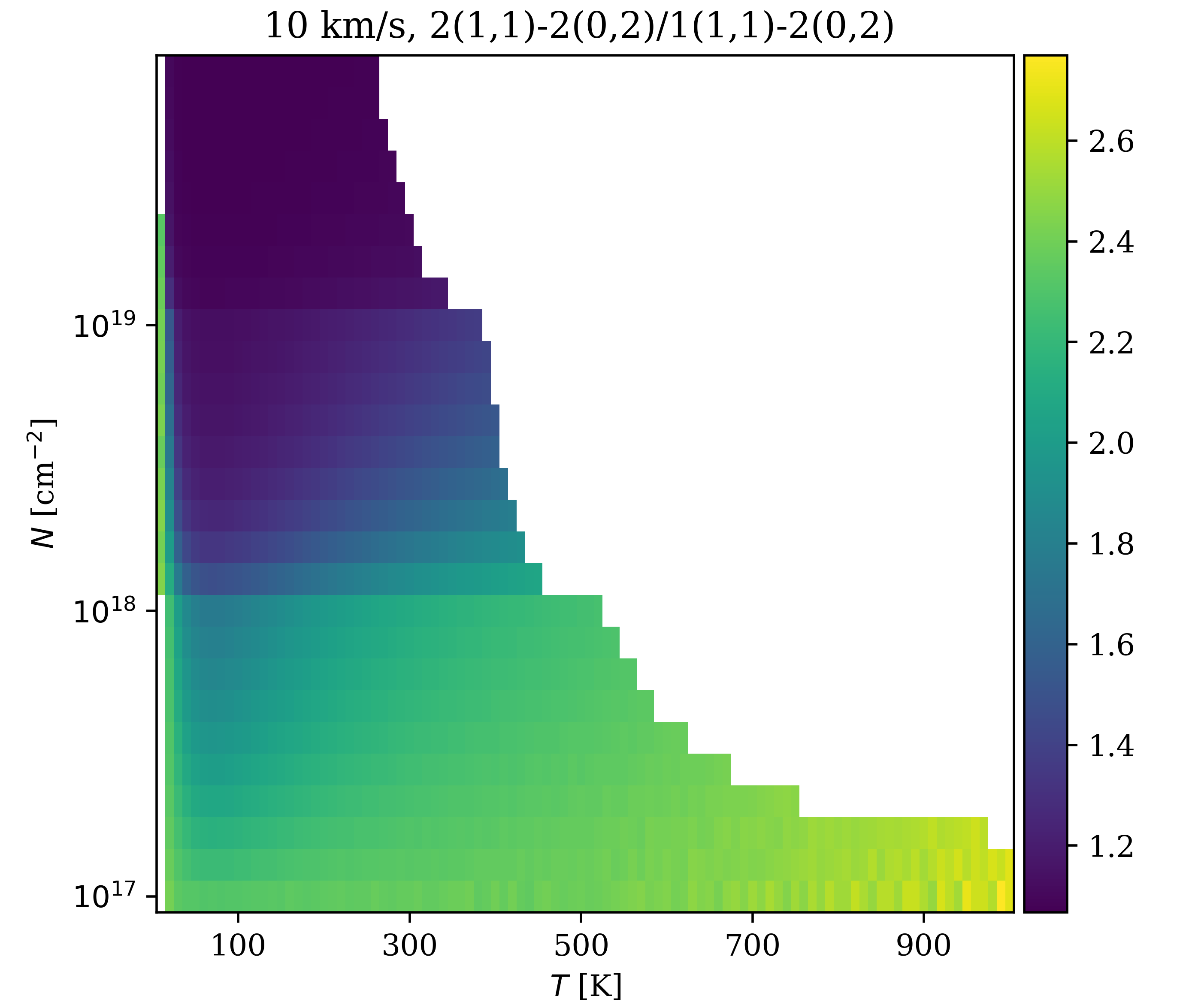}
    \includegraphics[width=0.32\linewidth]{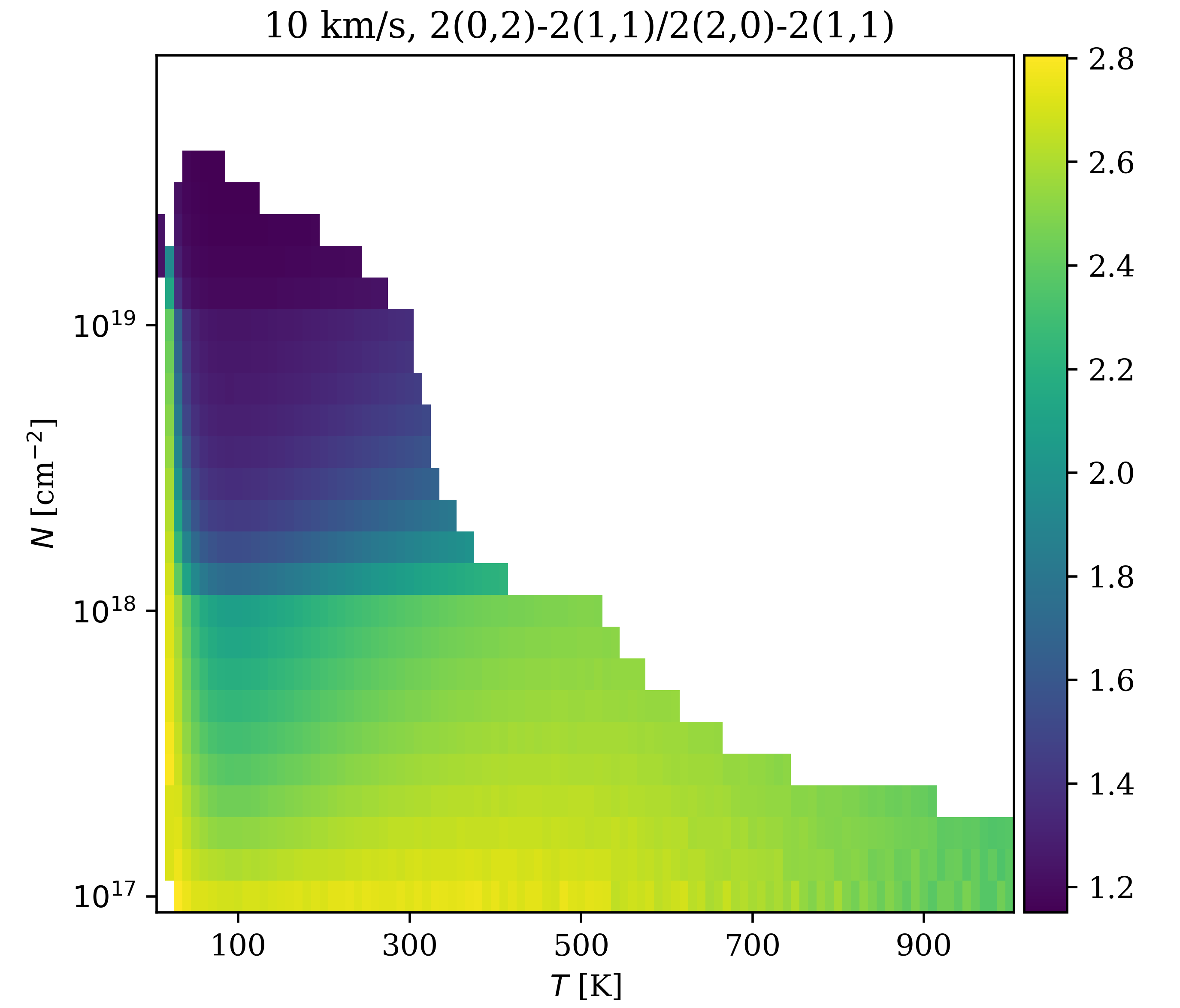}
    \includegraphics[width=0.32\linewidth]{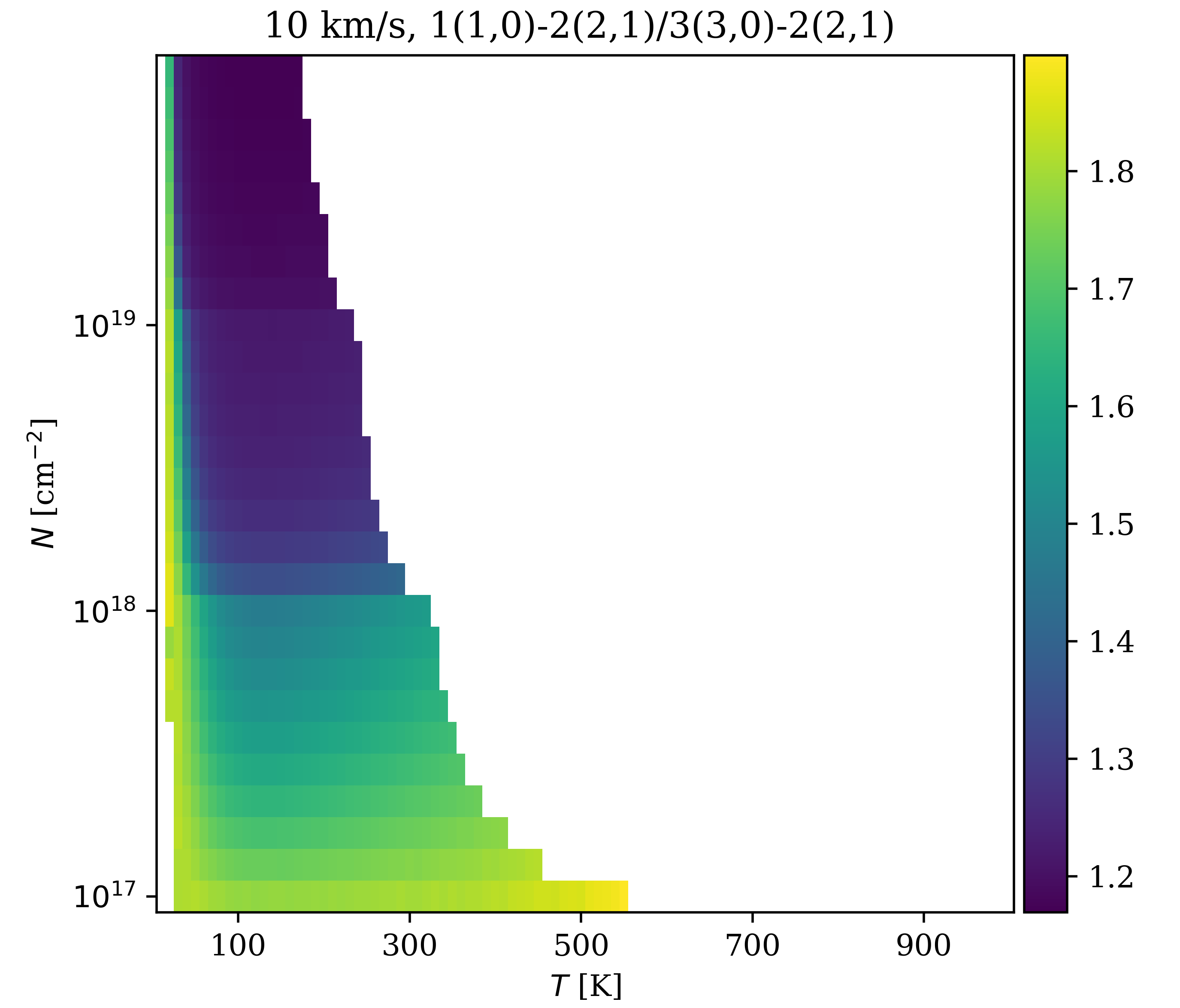}
    \includegraphics[width=0.32\linewidth]{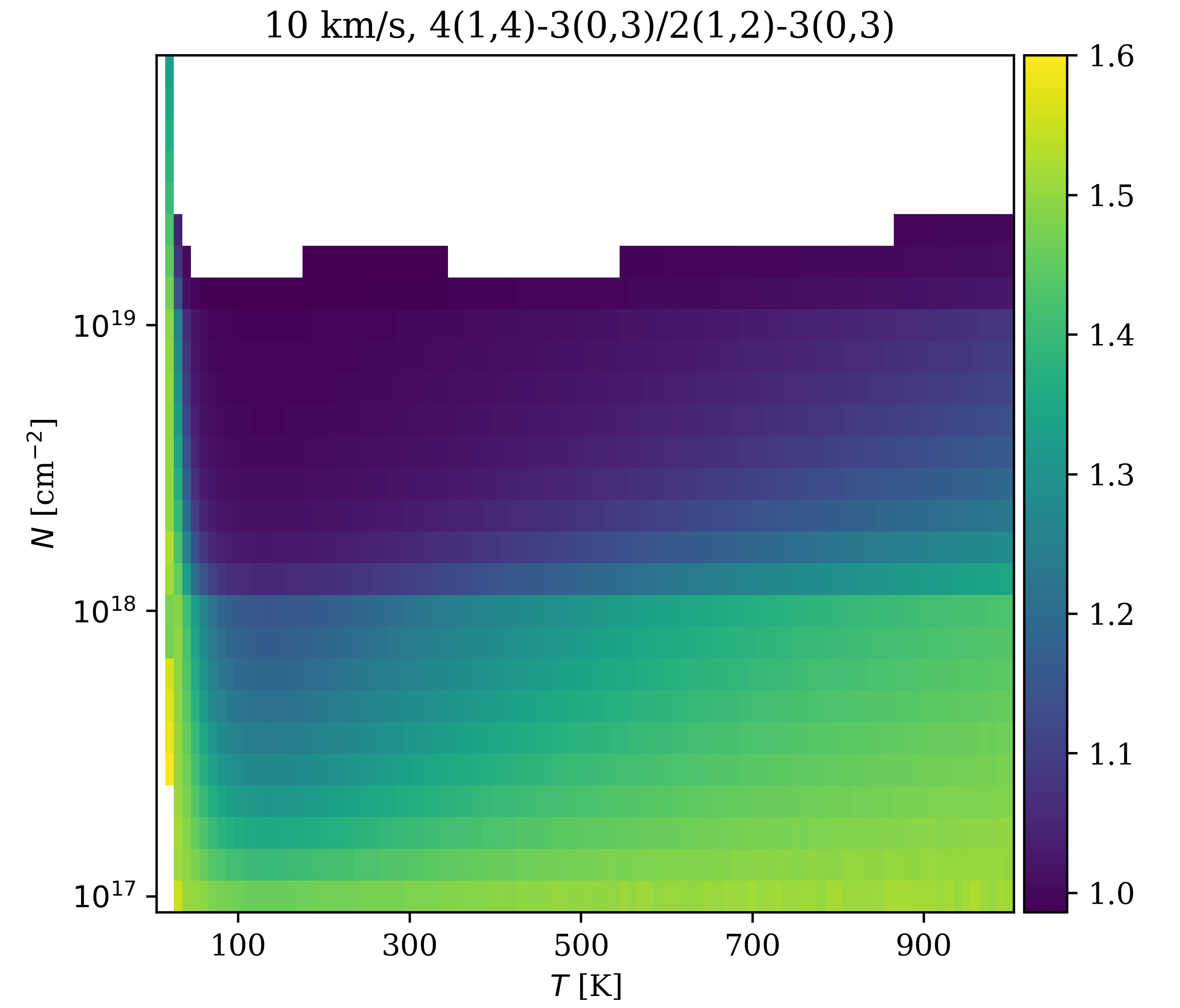}
    \includegraphics[width=0.32\linewidth]{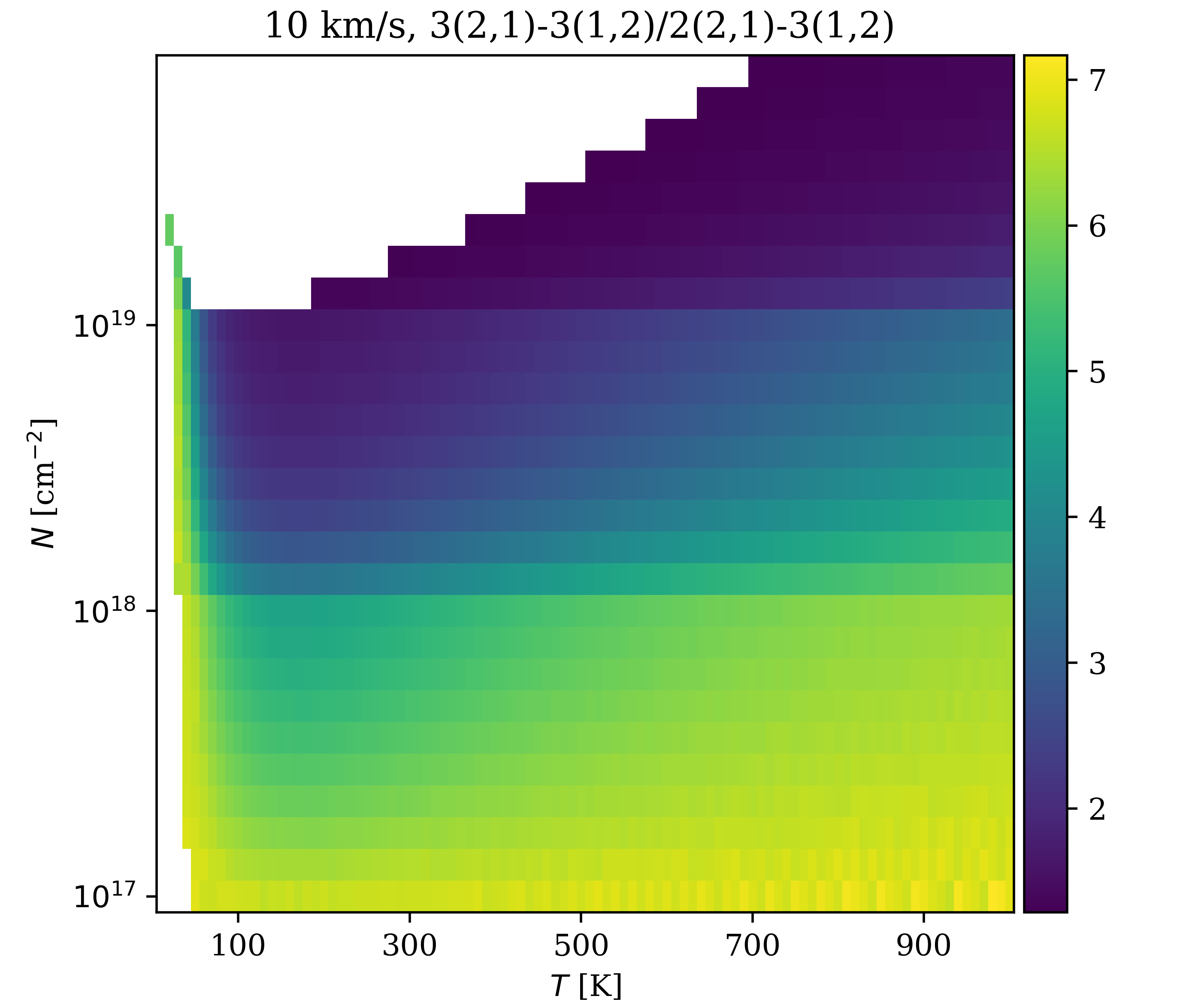}
    \includegraphics[width=0.32\linewidth]{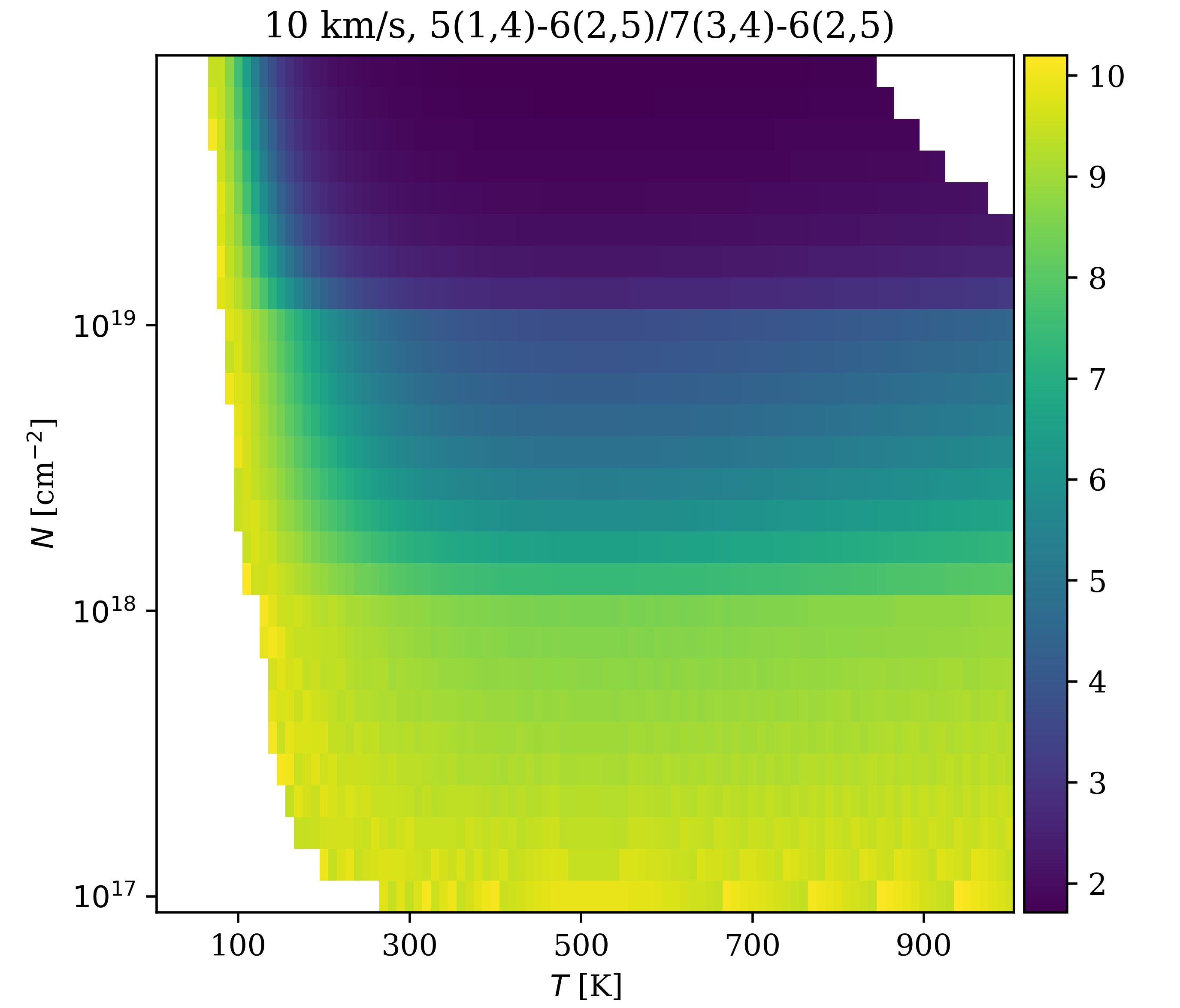}
    \caption{Reference grids similar to those in Figure~\ref{fig:grids-2kms} with $\sigma_v=10$~\kms~($b$=14~\kms).}
    \label{fig:grids-10kms}
\end{figure*}

\begin{figure*}[!t]
    \centering
    \includegraphics[width=0.32\linewidth]{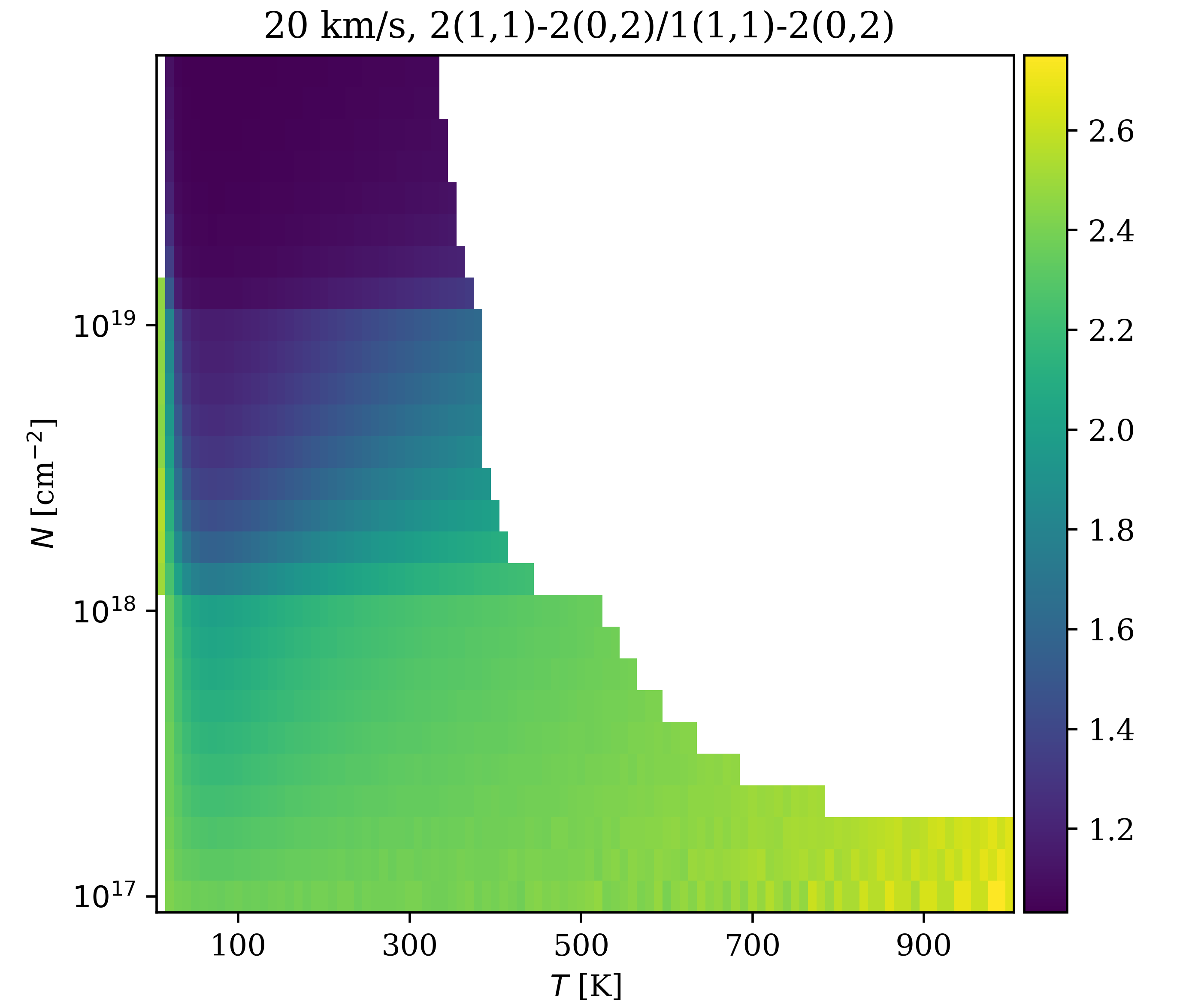}
    \includegraphics[width=0.32\linewidth]{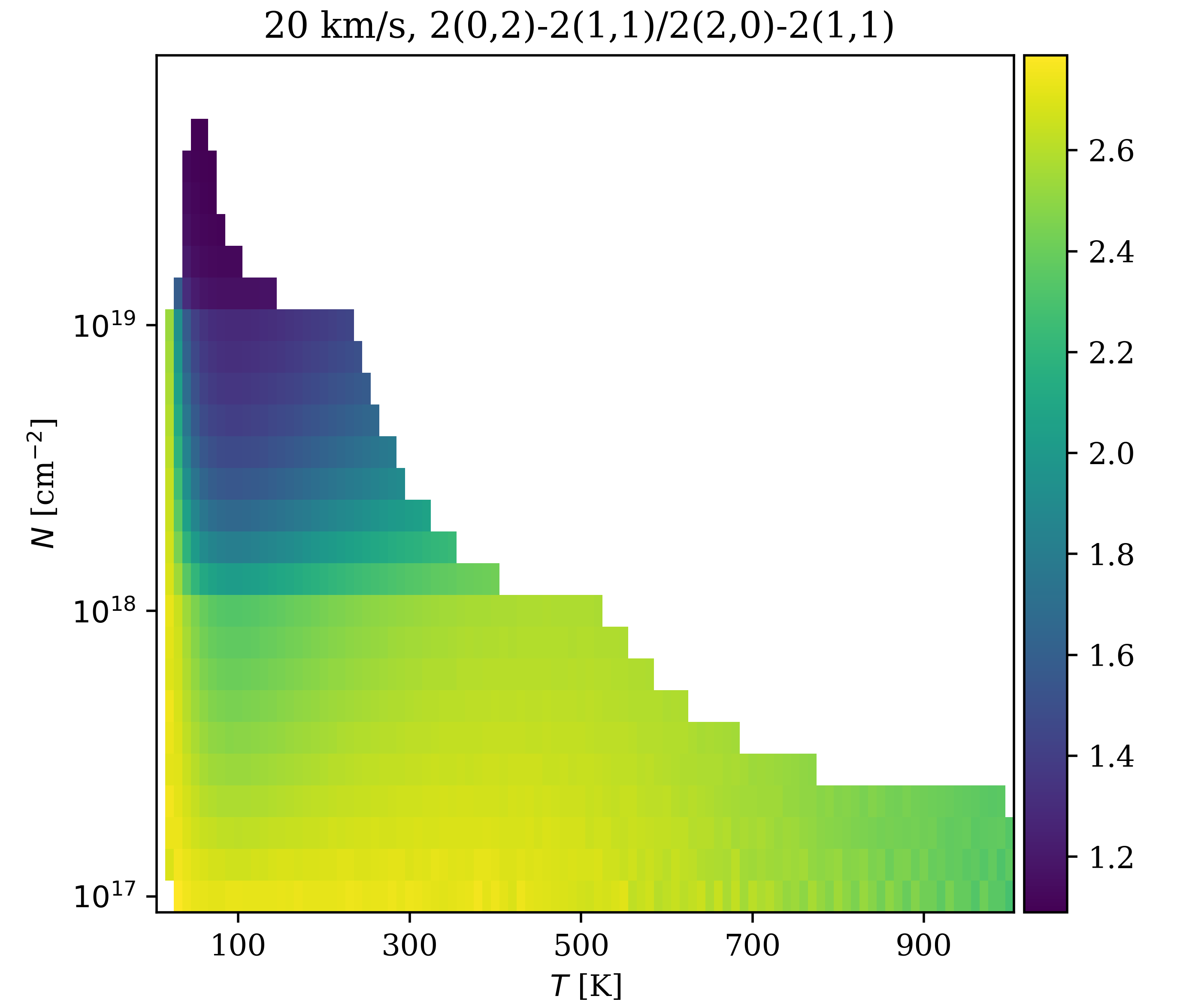}
    \includegraphics[width=0.32\linewidth]{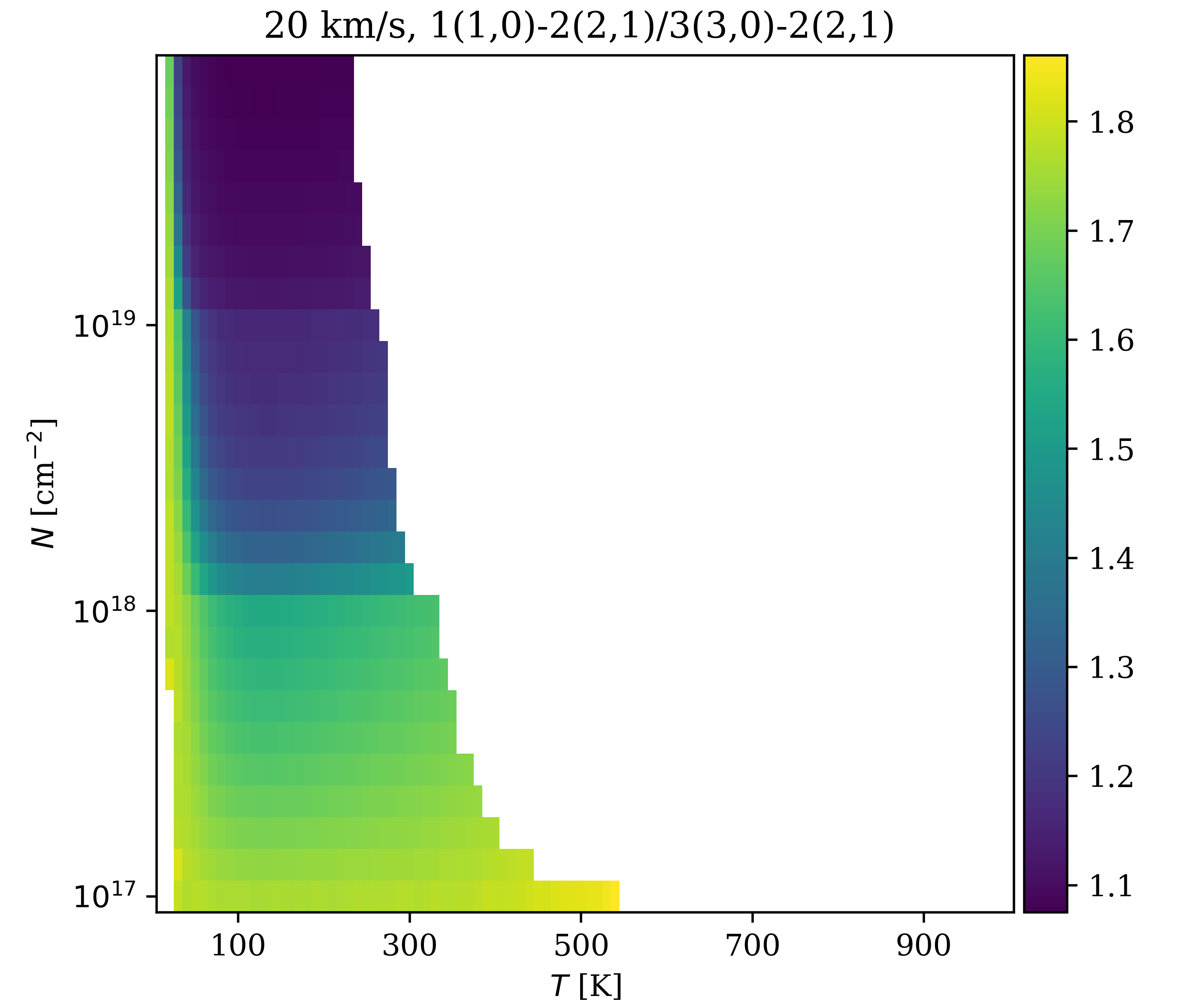}
    \includegraphics[width=0.32\linewidth]{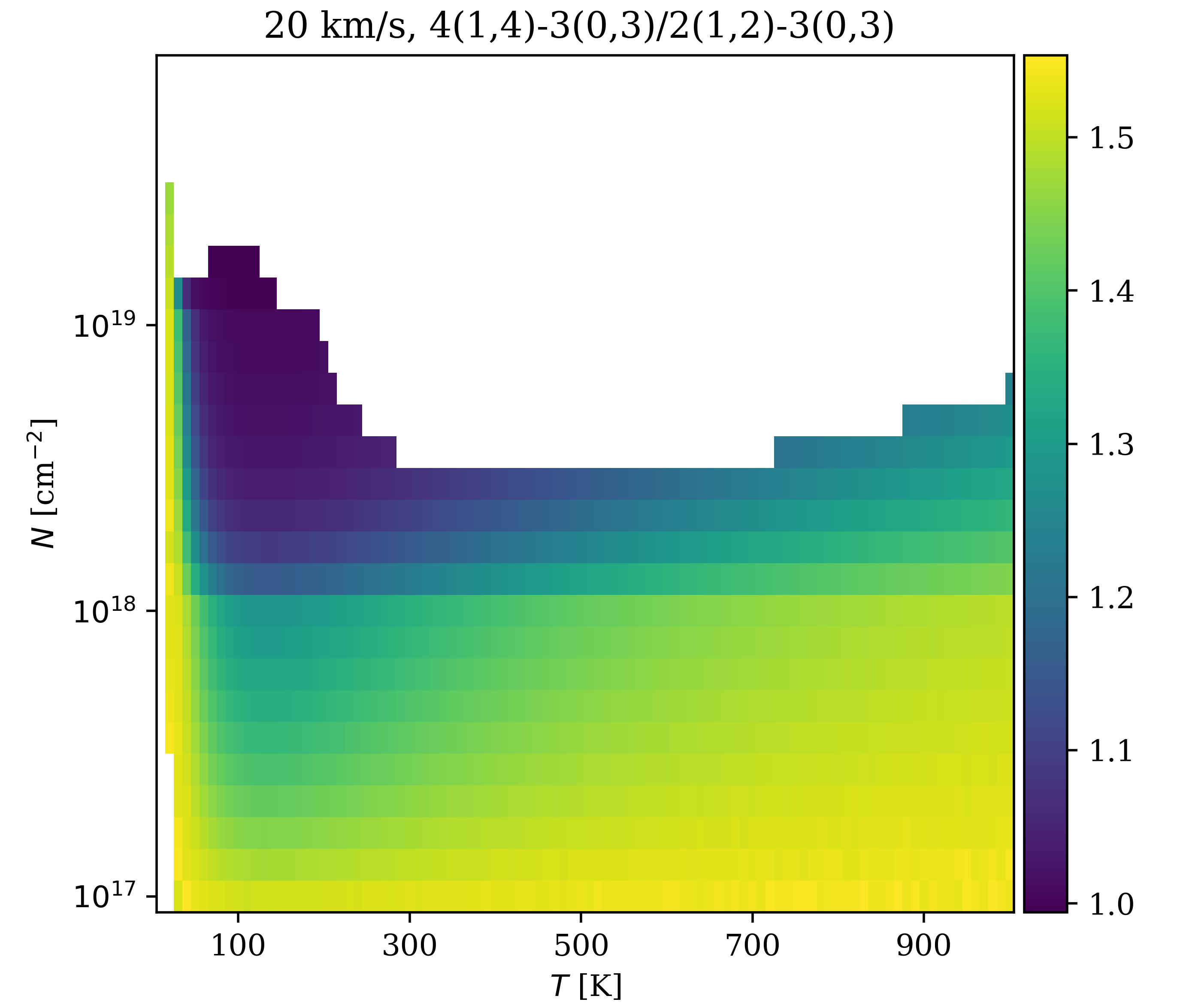}
    \includegraphics[width=0.32\linewidth]{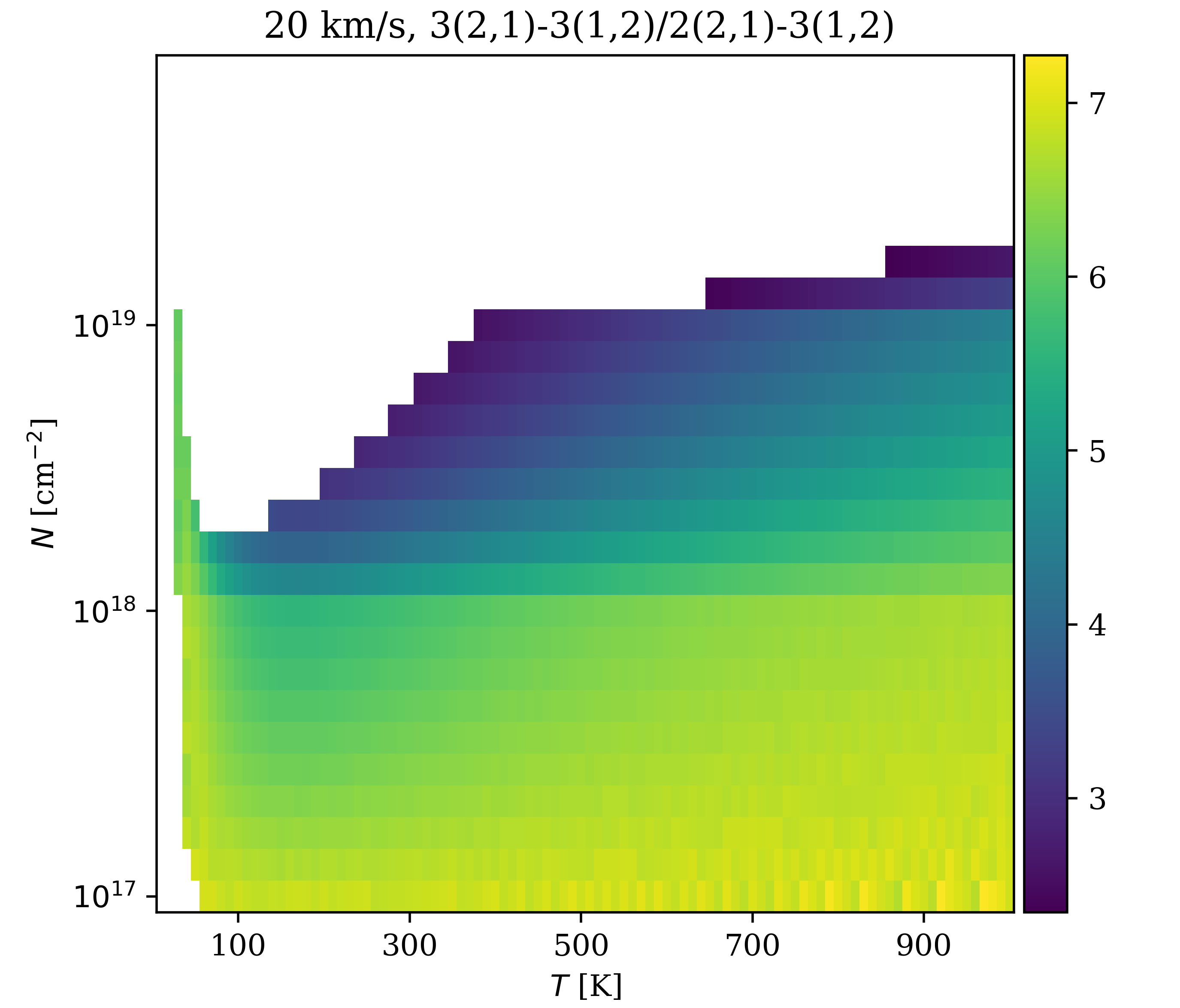}
    \includegraphics[width=0.32\linewidth]{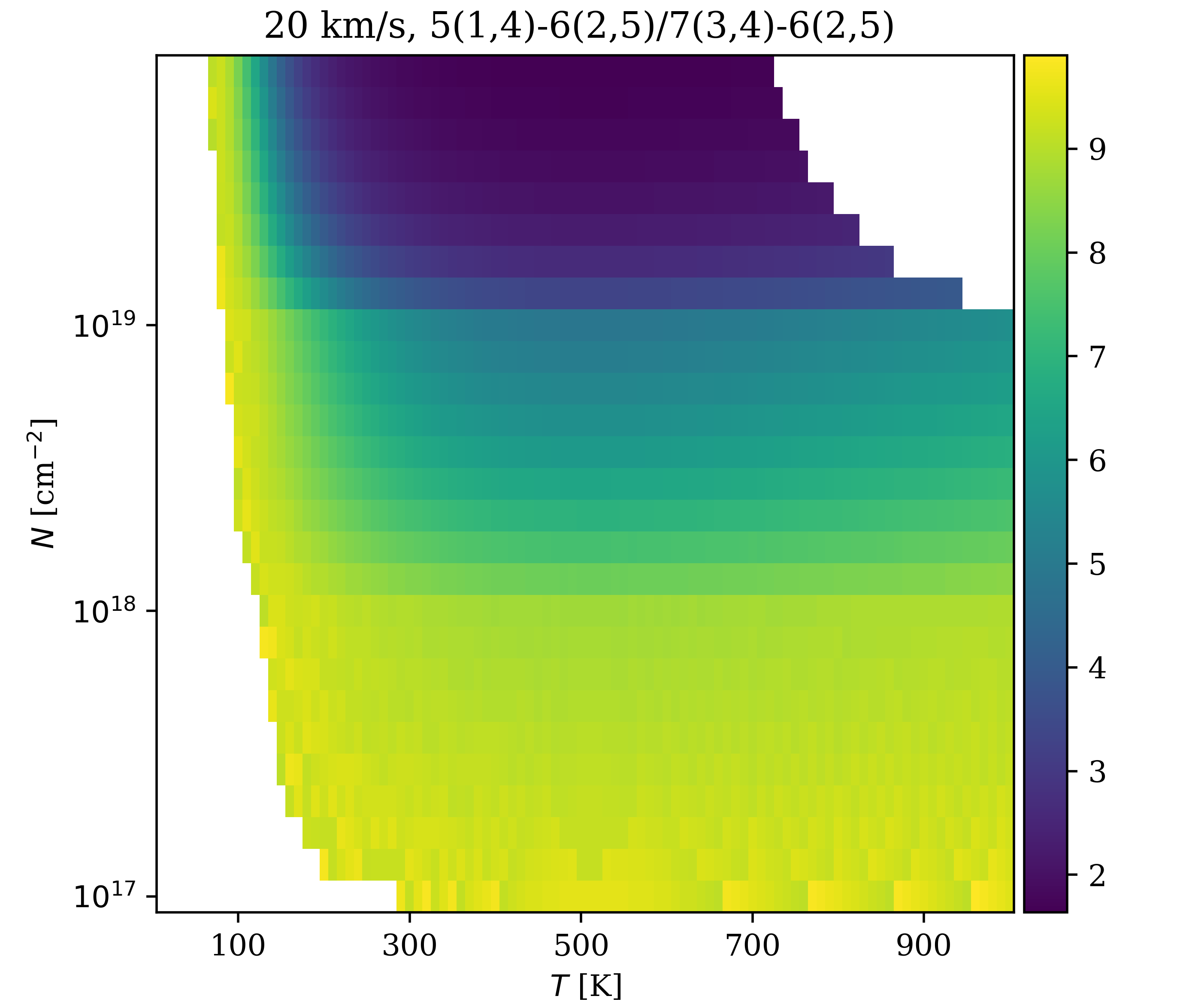}
    \caption{Reference grids similar to those in Figure~\ref{fig:grids-2kms} with $\sigma_v=20$~\kms~($b$=28~\kms).}
    \label{fig:grids-20kms}
\end{figure*}

\begin{figure*}[!t]
    \centering
    \includegraphics[width=0.32\linewidth]{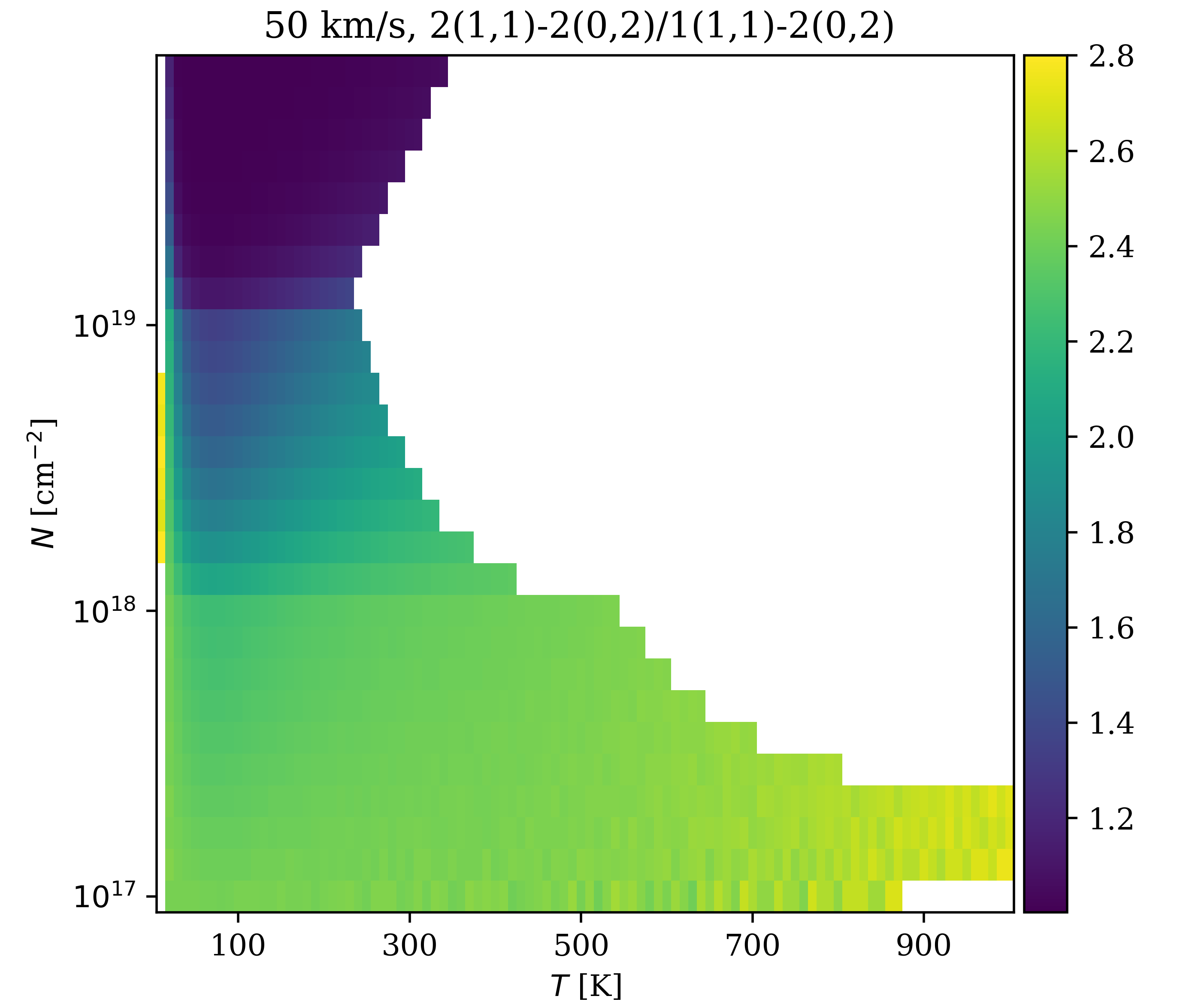}
    \includegraphics[width=0.32\linewidth]{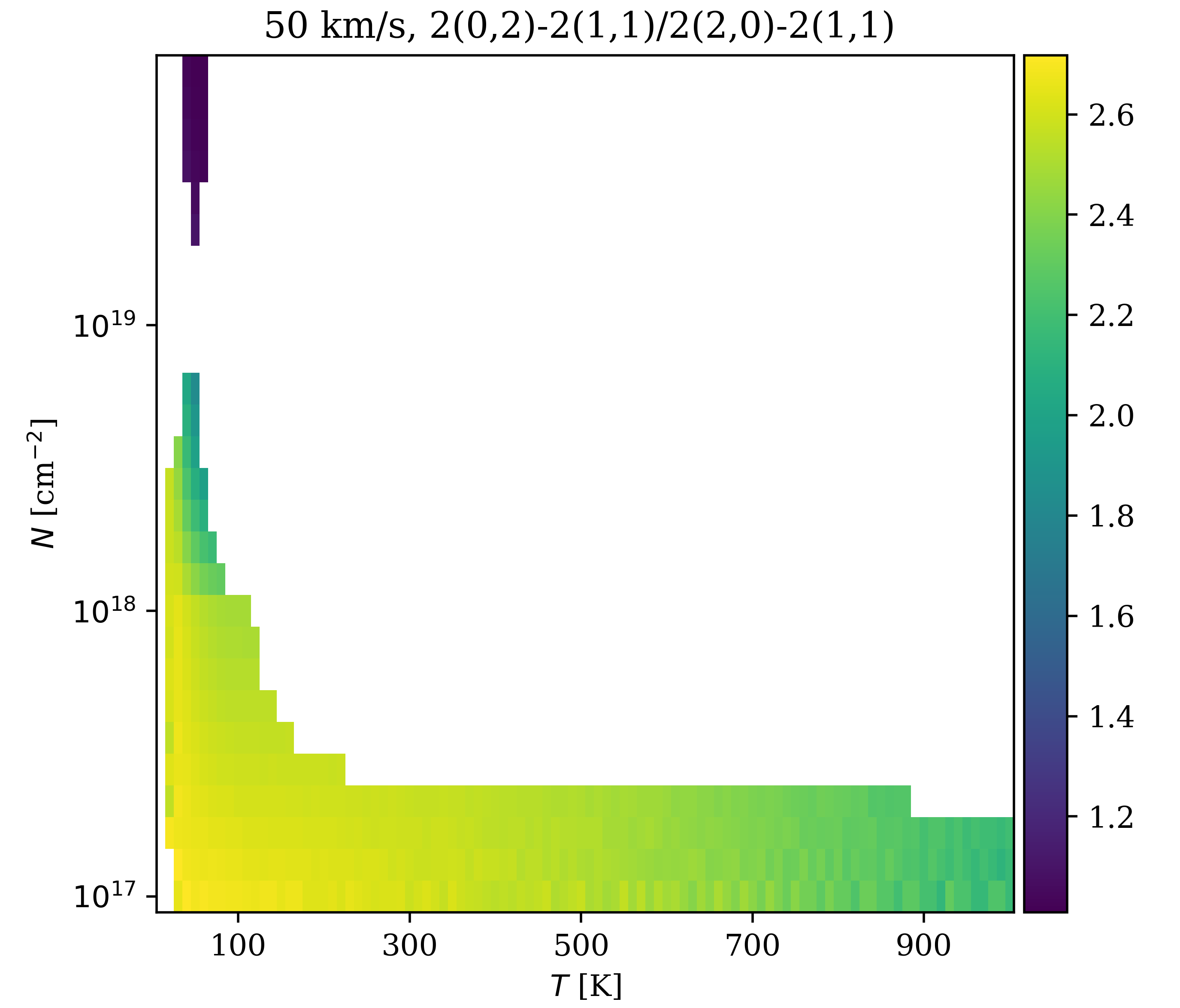}
    \includegraphics[width=0.32\linewidth]{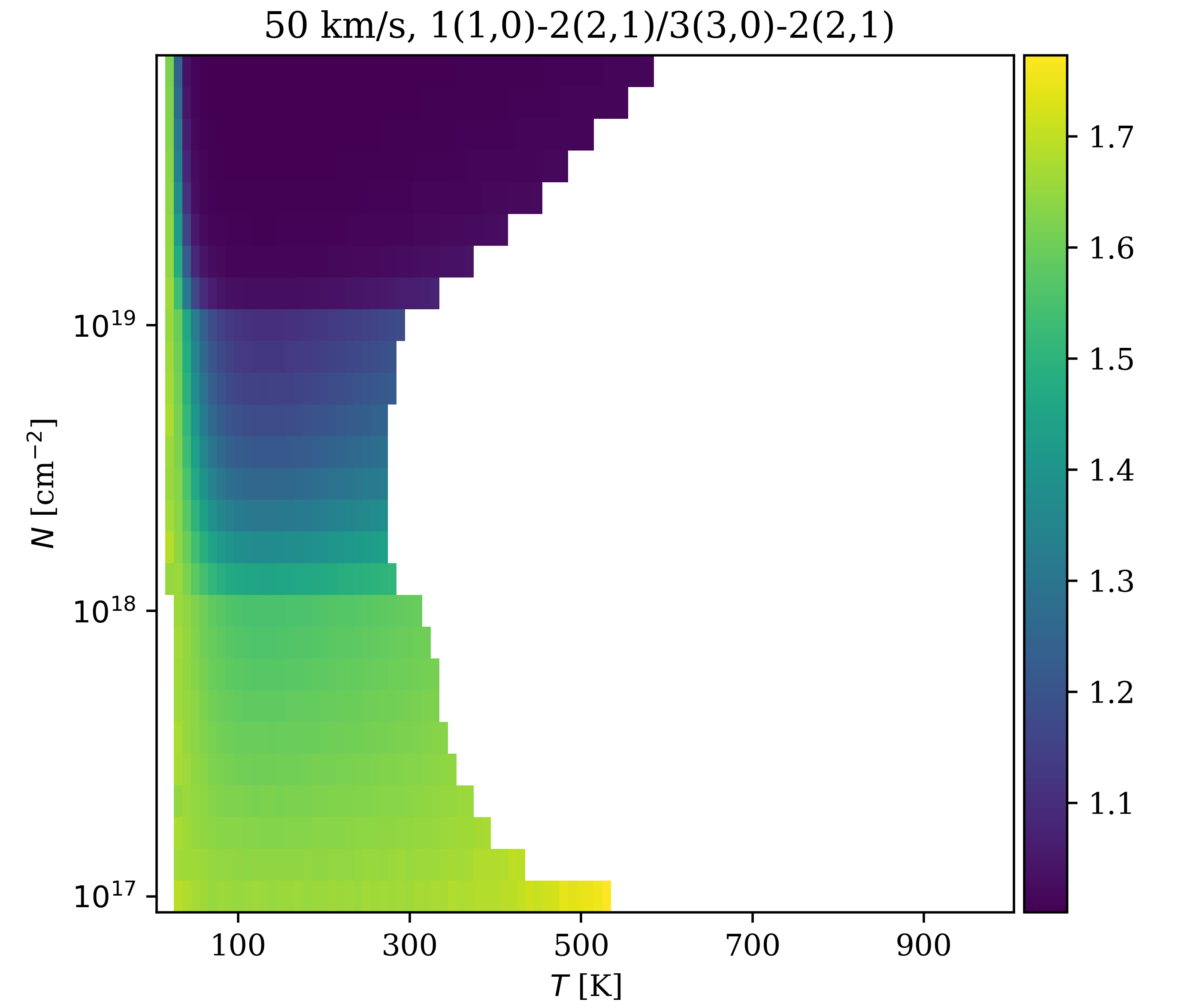}
    \includegraphics[width=0.32\linewidth]{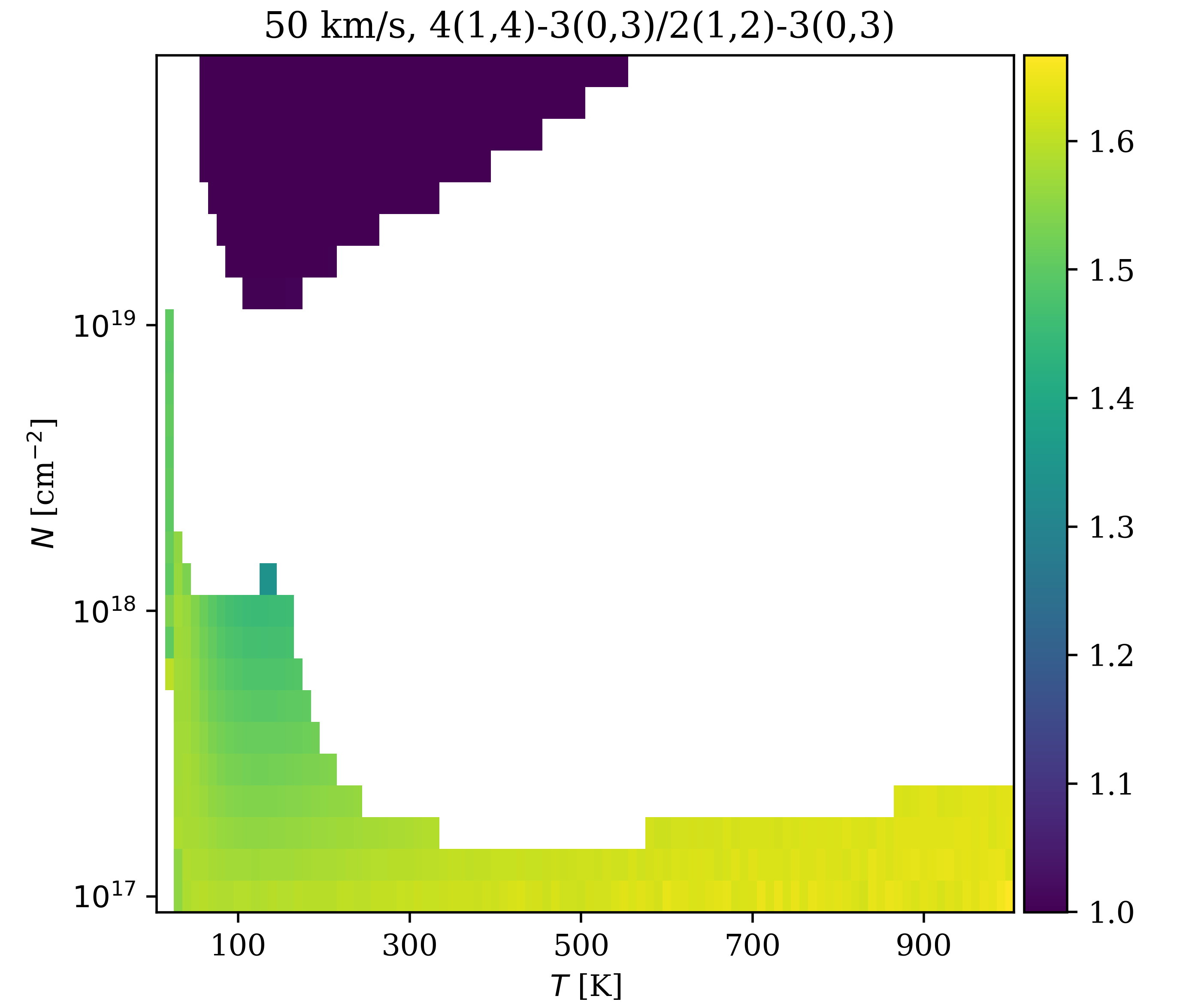}
    \includegraphics[width=0.32\linewidth]{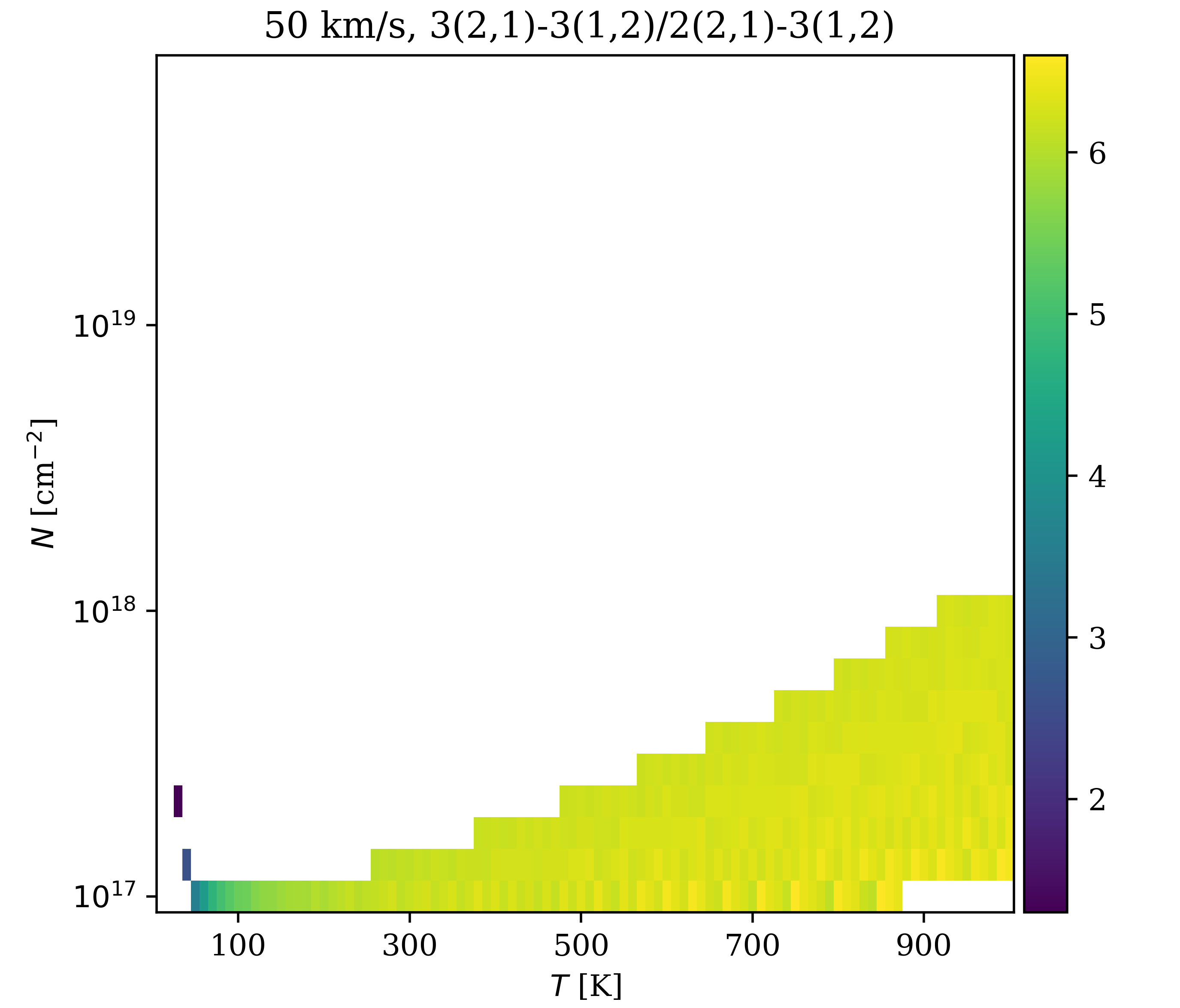}
    \includegraphics[width=0.32\linewidth]{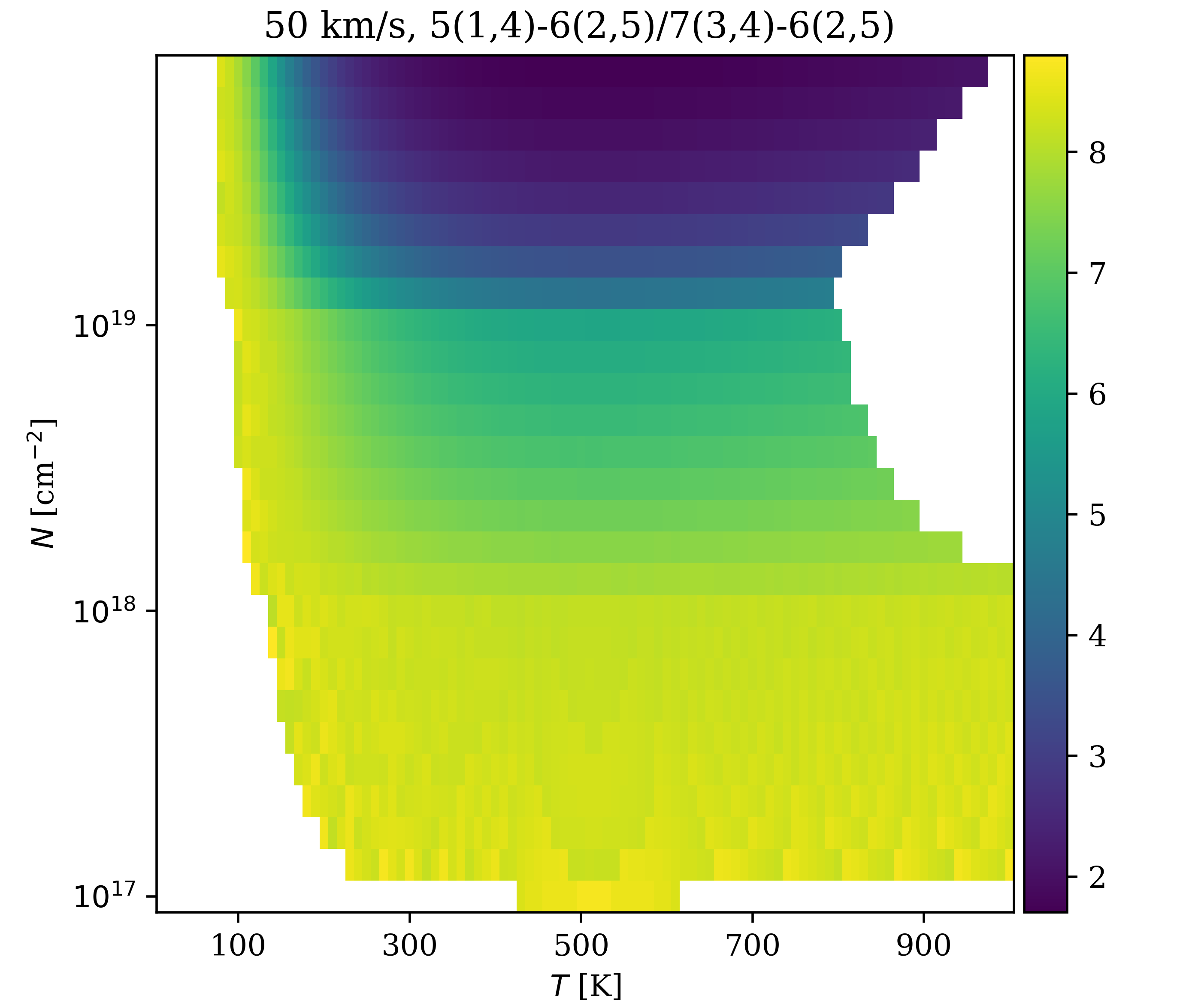}
    \caption{Reference grids similar to those in Figure~\ref{fig:grids-2kms} with $\sigma_v=50$~\kms~($b$=70~\kms).}
    \label{fig:grids-50kms}
\end{figure*}

\begin{figure*}[!t]
    \centering
    \includegraphics[width=0.32\linewidth]{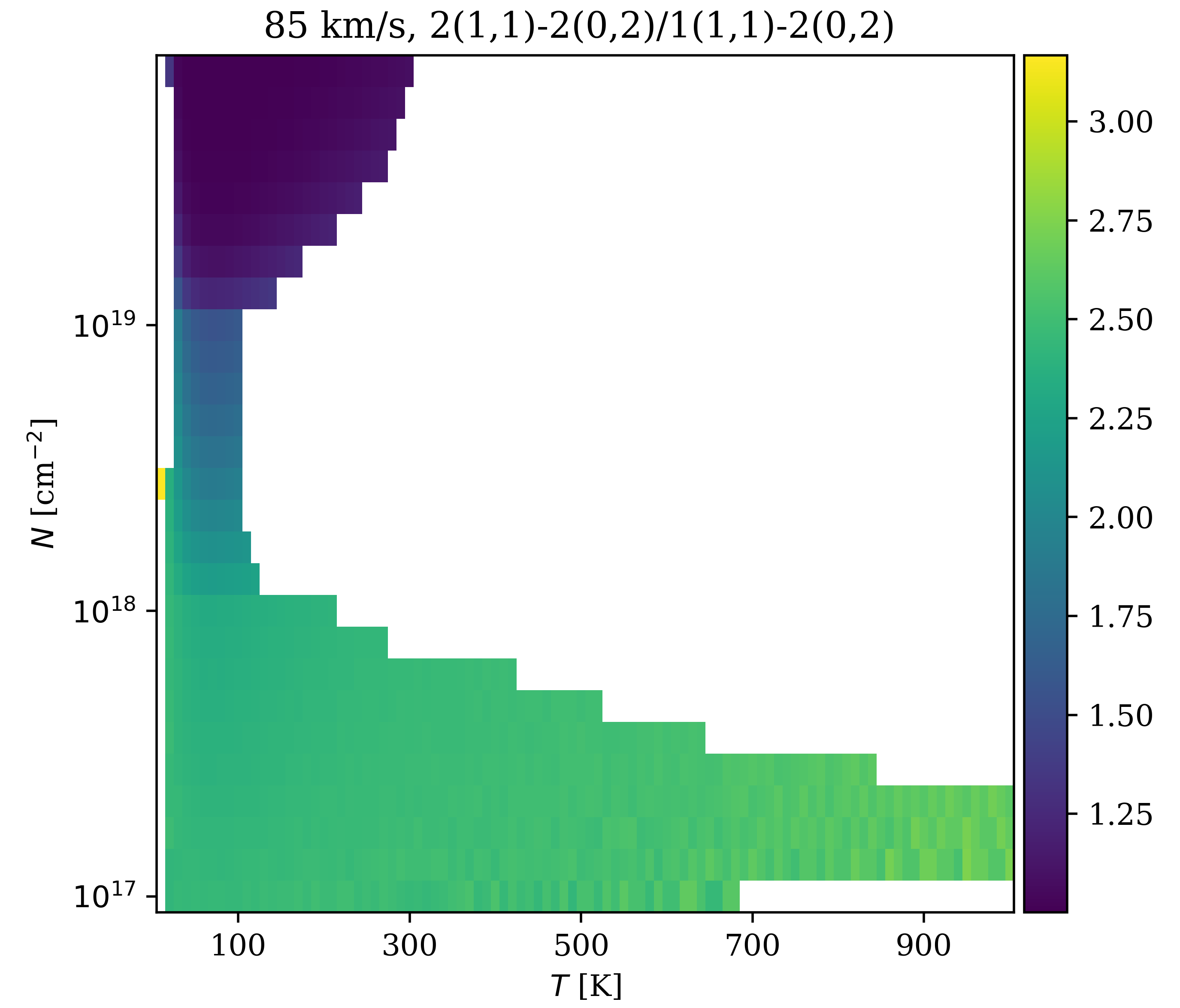}
    \includegraphics[width=0.32\linewidth]{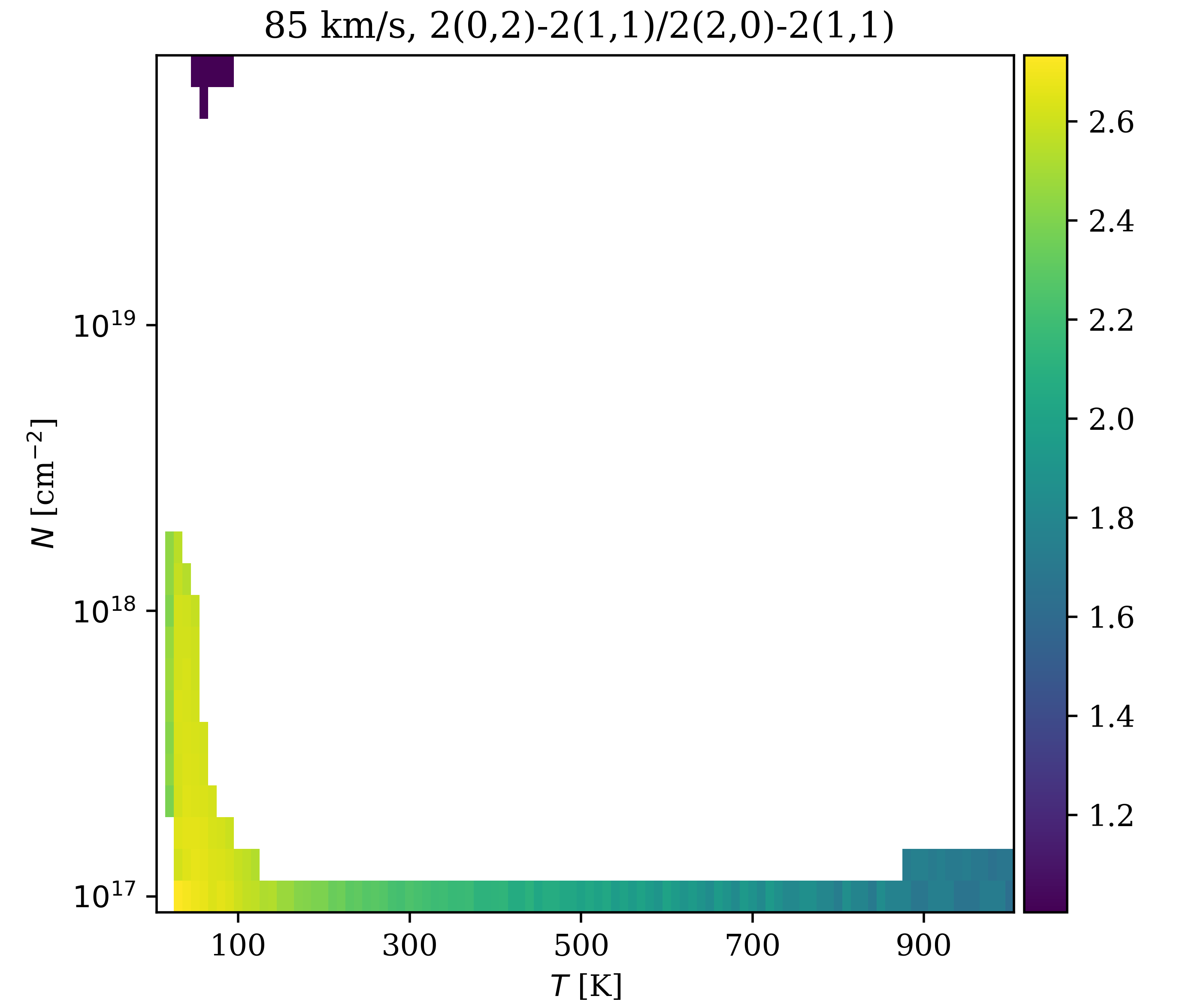}
    \includegraphics[width=0.32\linewidth]{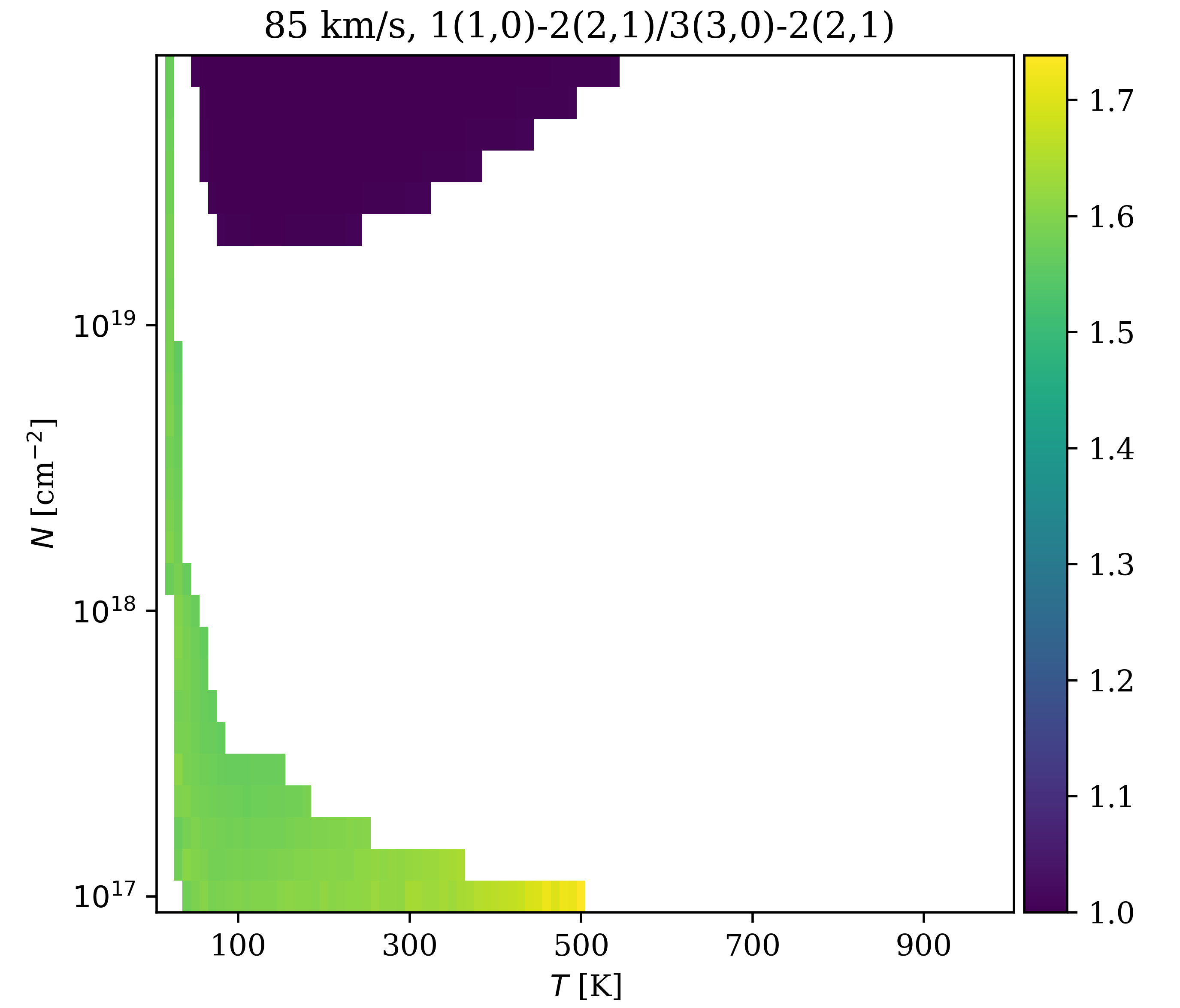}
    \includegraphics[width=0.32\linewidth]{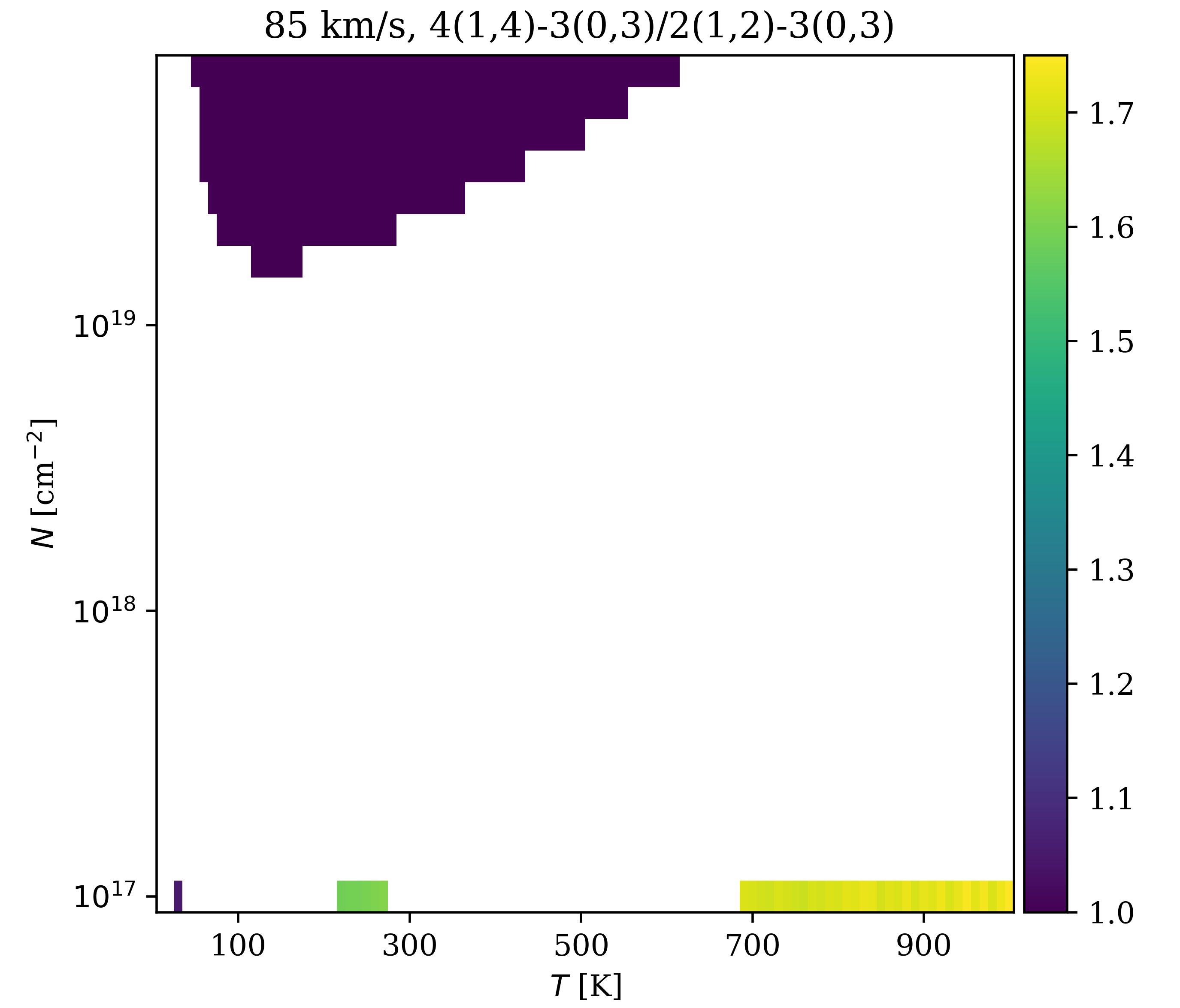}
    \includegraphics[width=0.32\linewidth]{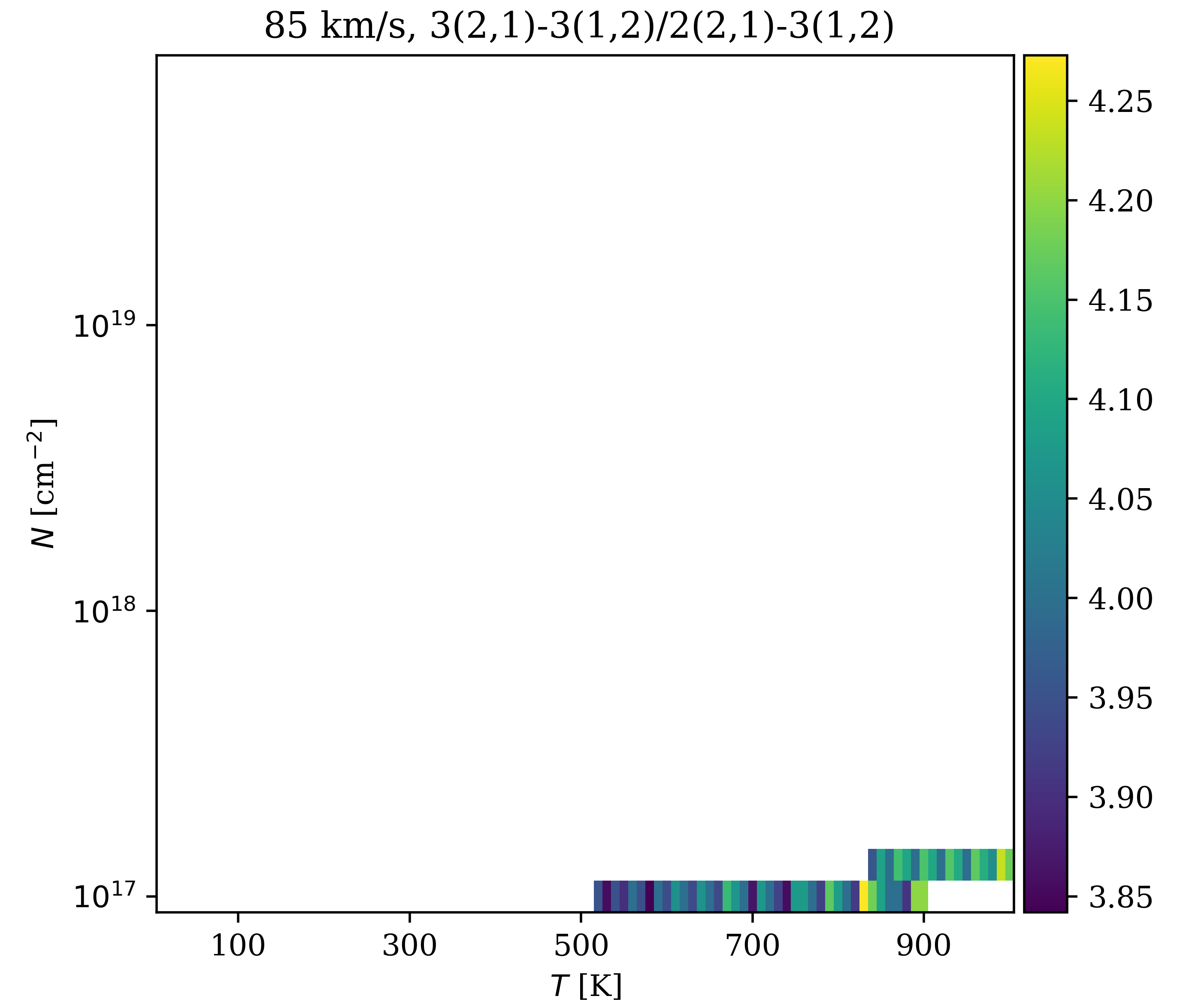}
    \includegraphics[width=0.32\linewidth]{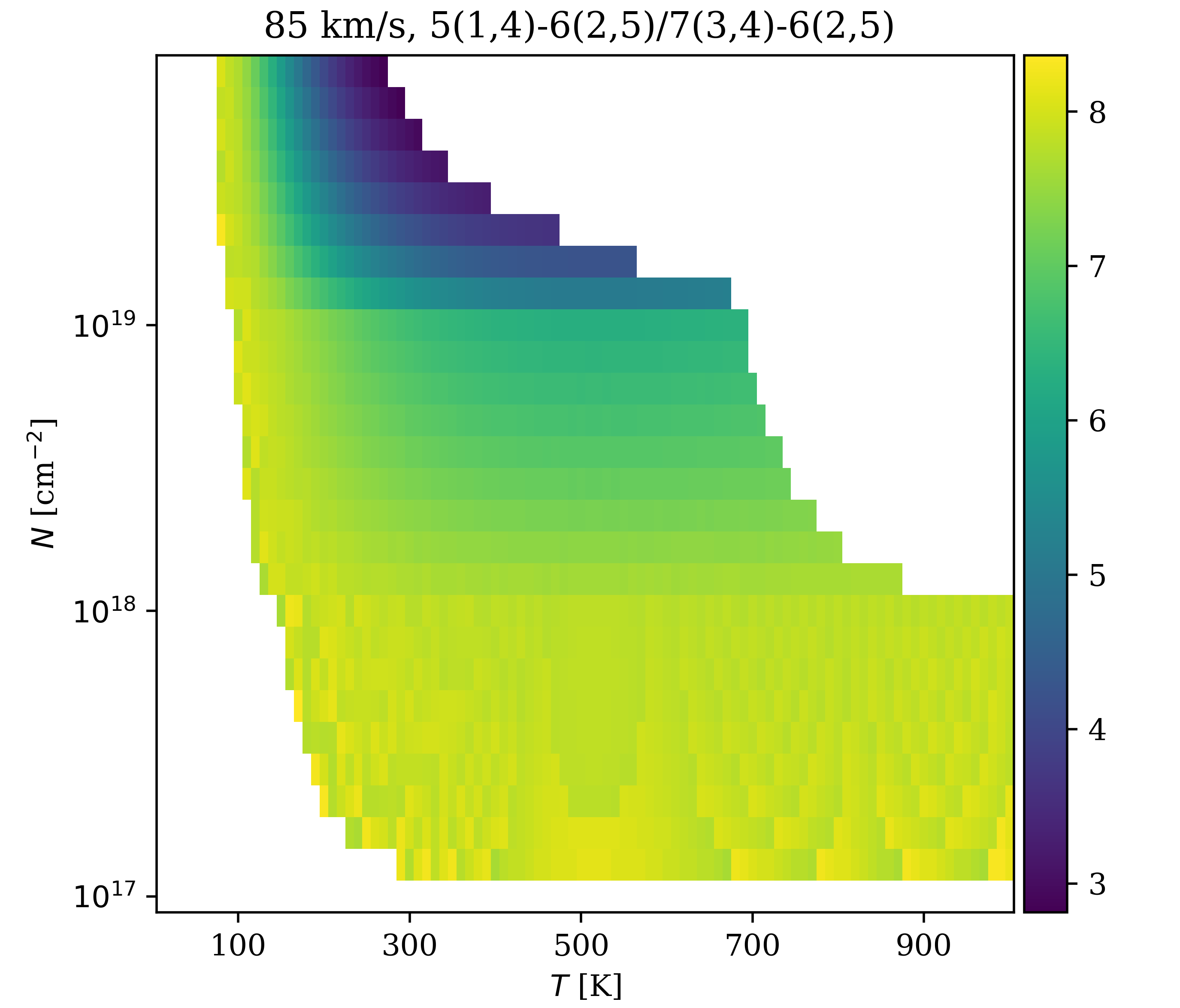}
    \caption{Reference grids similar to those in Figure~\ref{fig:grids-2kms} with $\sigma_v=85$~\kms~($b$=120~\kms).}
    \label{fig:grids-85kms}
\end{figure*}

\begin{figure*}[!t]
    \centering
    \includegraphics[width=0.32\linewidth]{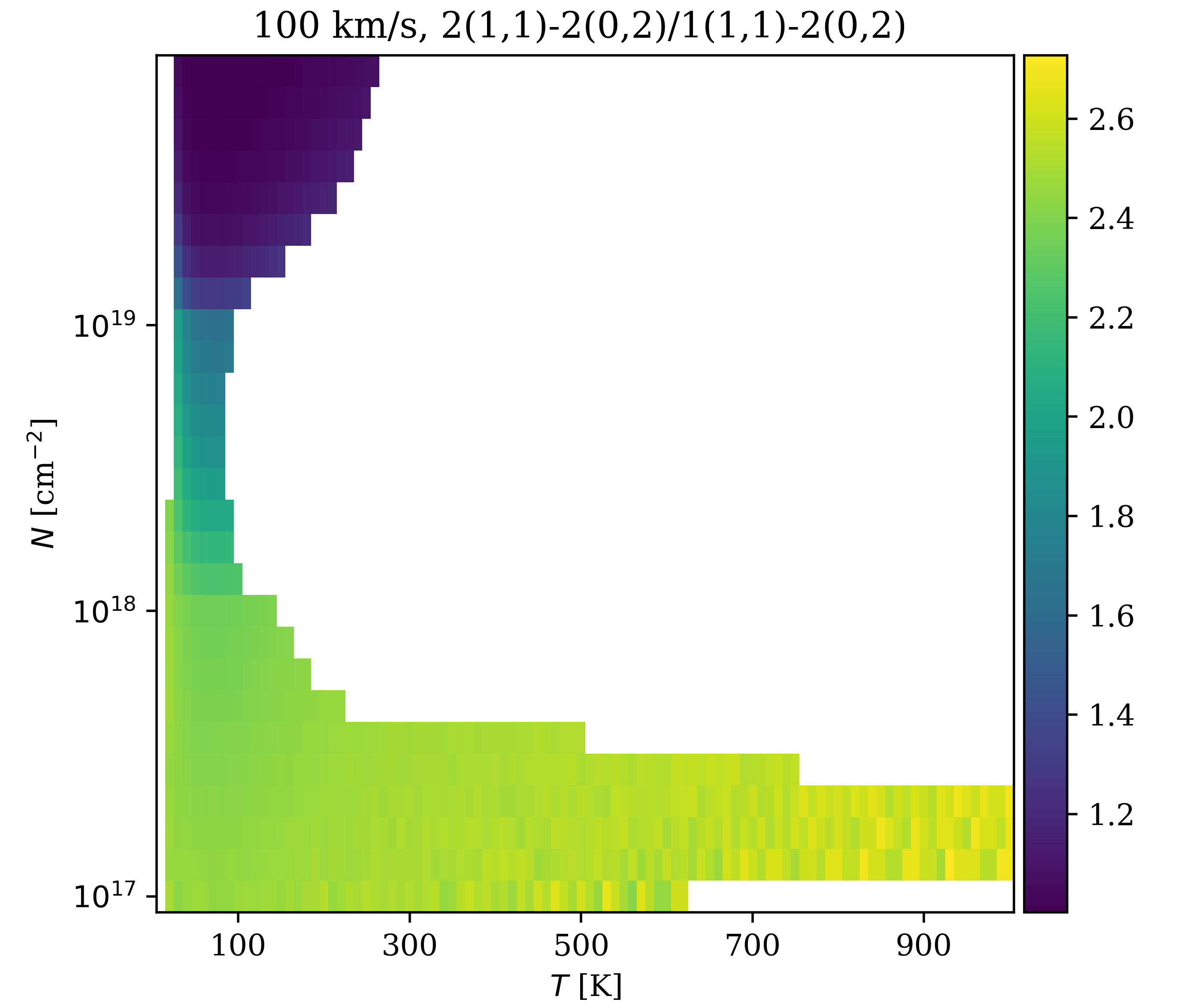}
    \includegraphics[width=0.32\linewidth]{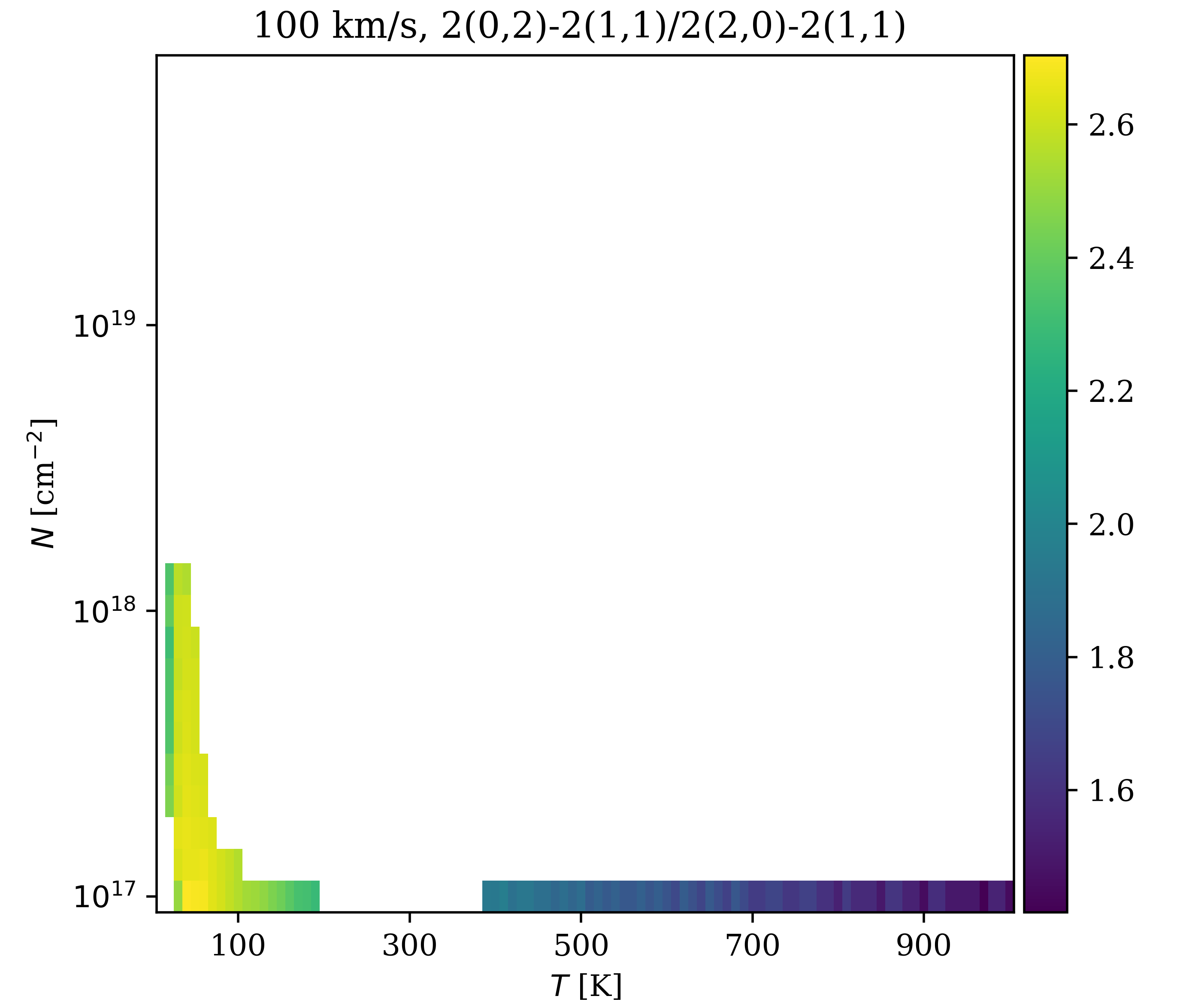}
    \includegraphics[width=0.32\linewidth]{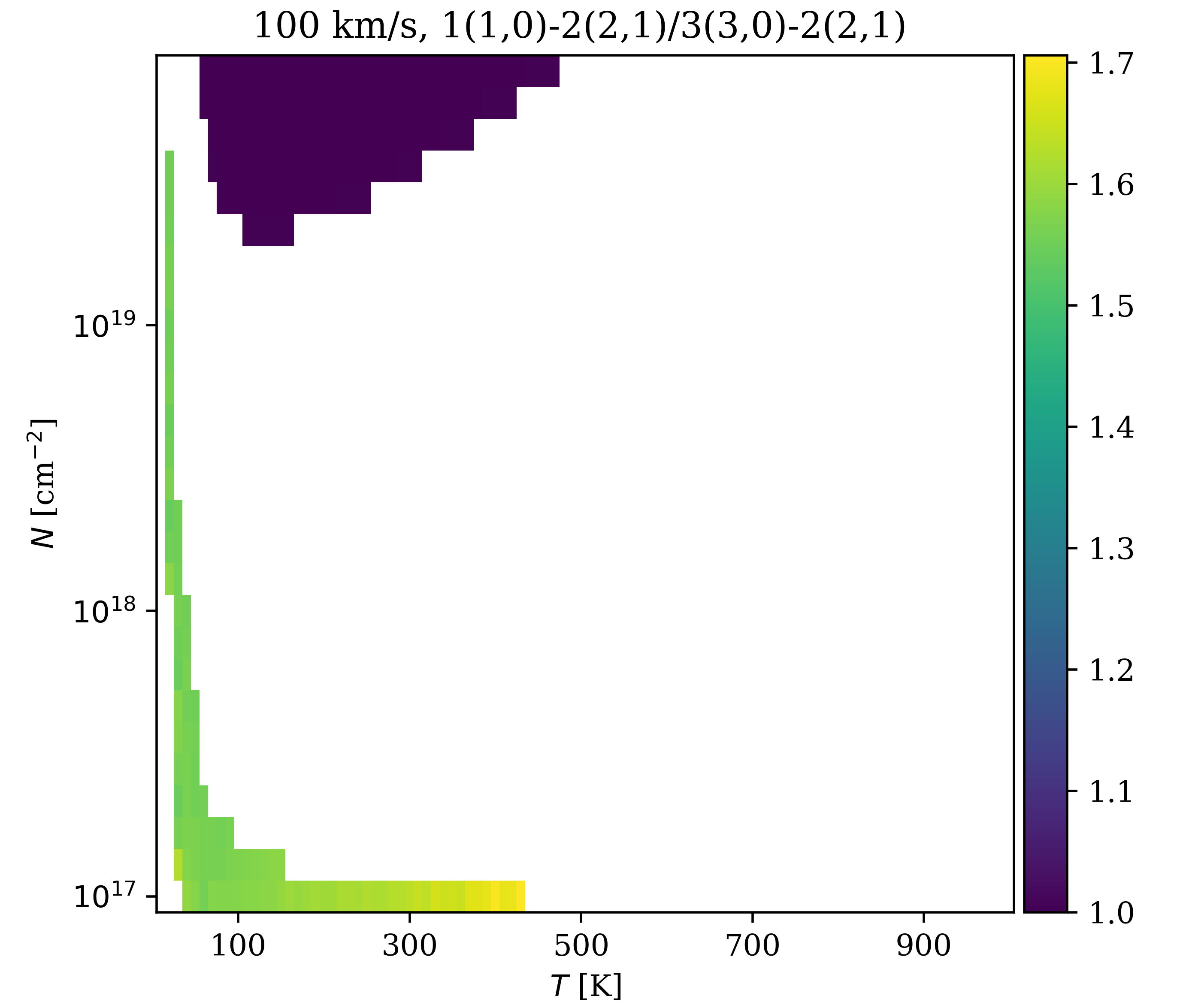}
    \includegraphics[width=0.32\linewidth]{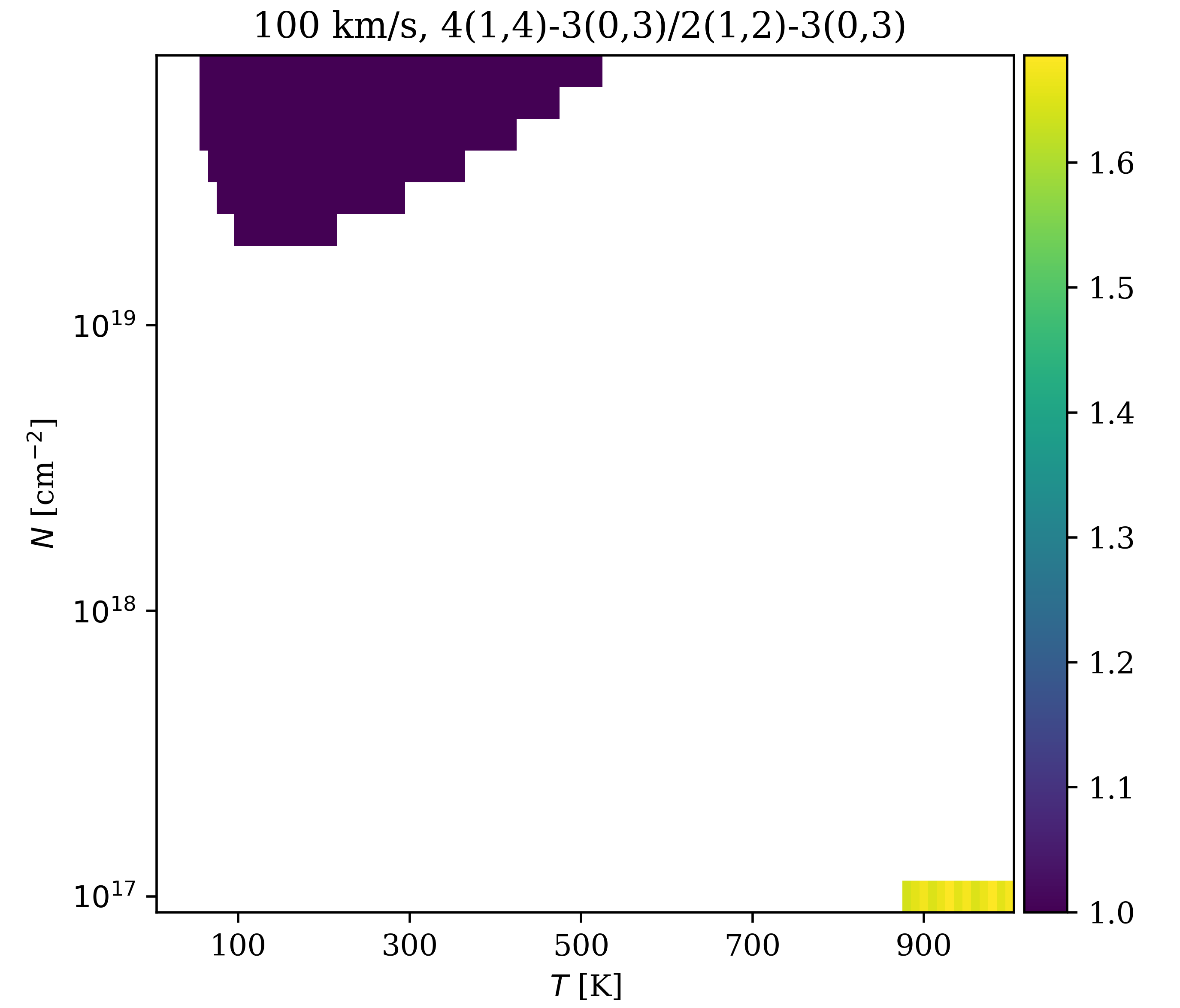}
    \includegraphics[width=0.32\linewidth]{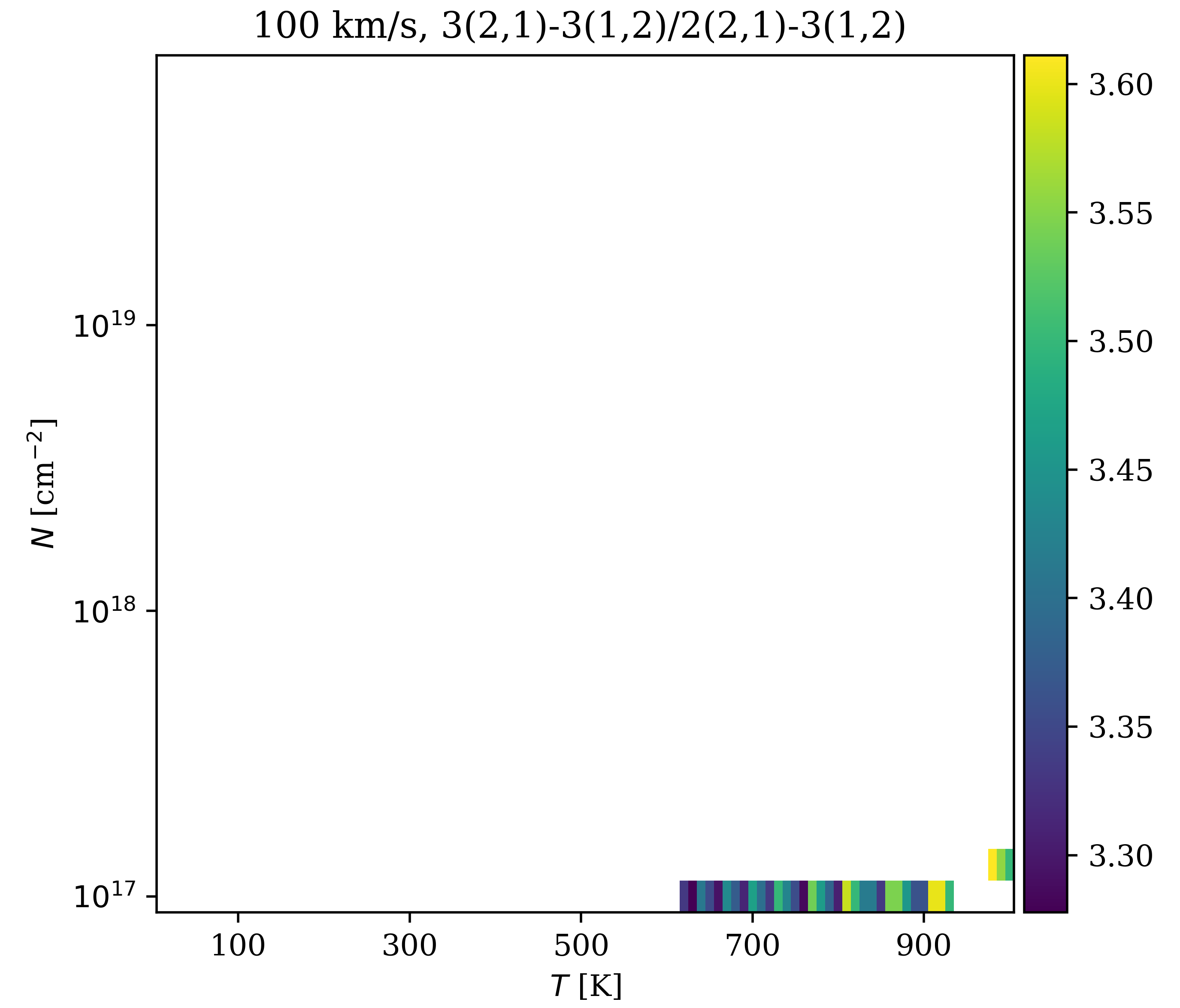}
    \includegraphics[width=0.32\linewidth]{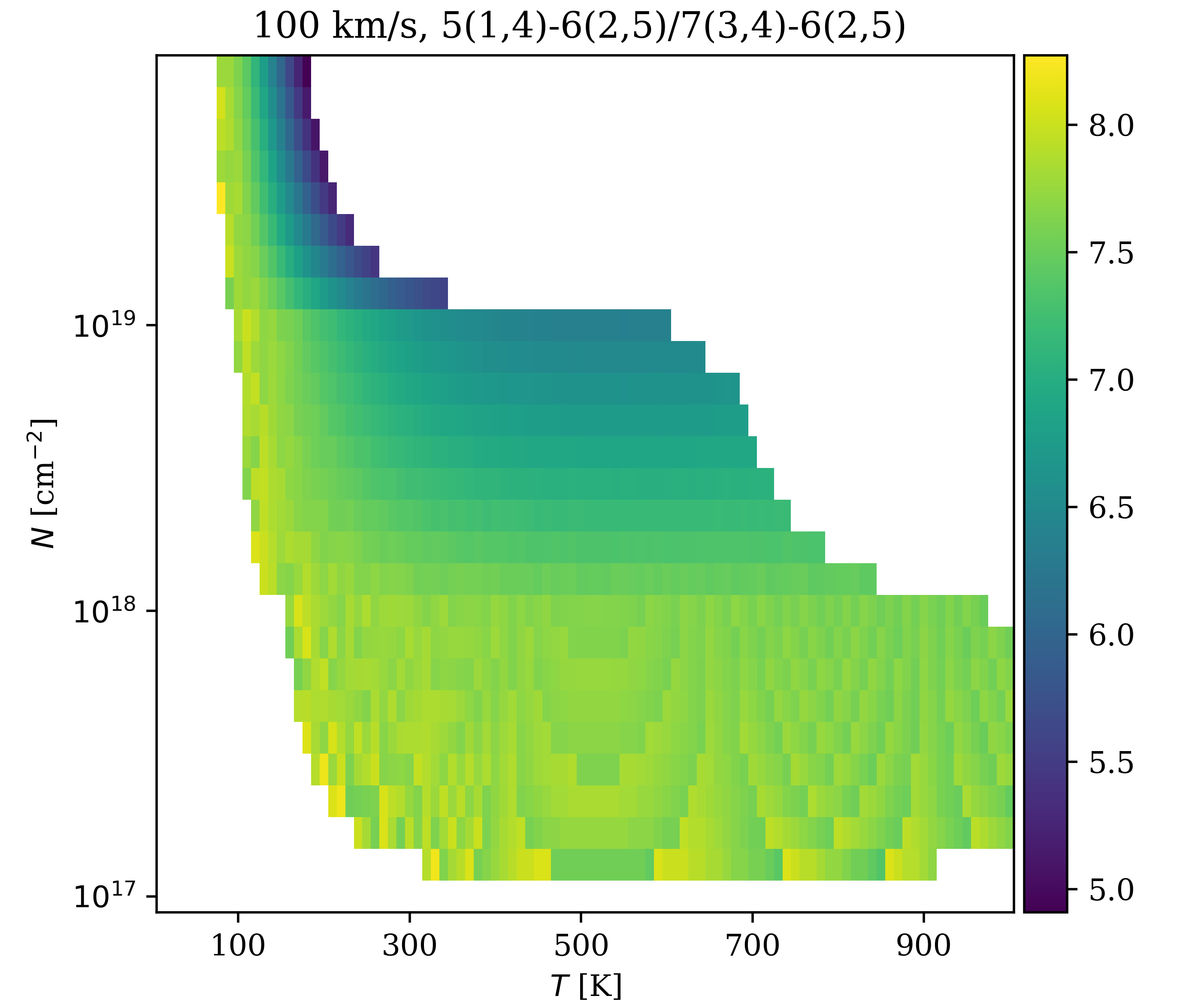}
    \caption{Reference grids similar to those in Figure~\ref{fig:grids-2kms} with $\sigma_v=100$~\kms.}
    \label{fig:grids-100kms}
\end{figure*}

\end{document}